\documentclass[traditabstract]{aa}  % for the abstract without structuration
\usepackage{psfig}

\begin{document}

\title{Unveiling the nature of {\it INTEGRAL} objects through optical 
spectroscopy\thanks{Based on observations collected at the following 
observatories: Cerro Tololo Interamerican Observatory (Chile);
Observatorio del Roque de los Muchachos of the Instituto de 
Astrof\'{\i}sica de Canarias (Canary Islands, Spain); Astronomical 
Observatory of Bologna in Loiano (Italy); Astronomical Observatory of 
Asiago (Italy); Observatorio Astron\'omico Nacional (San Pedro M\'artir, 
M\'exico); Anglo-Australian Observatory (Siding Spring, Australia).}}

\subtitle{IX. 22 more identifications, and a glance into the far 
hard X--ray Universe}

\author{N. Masetti\inst{1},
P. Parisi\inst{1,2,3},
E. Jim\'enez-Bail\'on\inst{4},
E. Palazzi\inst{1},
V. Chavushyan\inst{5},
L. Bassani\inst{1},
A. Bazzano\inst{3}, 
A.J. Bird\inst{6},
A.J. Dean\inst{6},
G. Galaz\inst{7},
R. Landi\inst{1},
A. Malizia\inst{1},
D. Minniti\inst{7,8,9},
L. Morelli\inst{10,11},
F. Schiavone\inst{1},
J.B. Stephen\inst{1} and
P. Ubertini\inst{3}
}

\institute{
INAF -- Istituto di Astrofisica Spaziale e Fisica Cosmica di 
Bologna, Via Gobetti 101, I-40129 Bologna, Italy
\and
Dipartimento di Astronomia, Universit\`a di Bologna,
Via Ranzani 1, I-40127 Bologna, Italy
\and
INAF -- Istituto di Astrofisica Spaziale e Fisica Cosmica di
Roma, Via Fosso del Cavaliere 100, I-00133 Rome, Italy
\and
Instituto de Astronom\'{\i}a, Universidad Nacional Aut\'onoma de M\'exico,
Apartado Postal 70-264, 04510 M\'exico D.F., M\'exico
\and
Instituto Nacional de Astrof\'{i}sica, \'Optica y Electr\'onica,
Apartado Postal 51-216, 72000 Puebla, M\'exico
\and
Physics \& Astronomy, University of Southampton, Southampton, 
Hampshire, SO17 1BJ, United Kingdom  
\and
Departamento de Astronom\'{i}a y Astrof\'{i}sica, Pontificia Universidad 
Cat\'olica de Chile, Casilla 306, Santiago 22, Chile
\and
Specola Vaticana, V-00120 Citt\`a del Vaticano
\and
Department of Astrophysical Sciences, University of Princeton, Princeton, 
NJ 08544-1001, USA
\and
Dipartimento di Astronomia, Universit\`a di Padova,
Vicolo dell'Osservatorio 3, I-35122 Padua, Italy
\and
INAF-Osservatorio Astronomico di Padova, Vicolo dell'Osservatorio 5, 
I-35122 Padua, Italy
}

\offprints{N. Masetti (\texttt{masetti@iasfbo.inaf.it)}}
\date{Received 1 December 2011; accepted 29 December 2011}

\abstract{Since its launch in October 2002, the {\it INTEGRAL} satellite 
has revolutionized our knowledge of the hard X--ray sky thanks to its 
unprecedented imaging capabilities and source detection positional 
accuracy above 20 keV. Nevertheless, many of the newly-detected sources in 
the {\it INTEGRAL} sky surveys are of unknown nature. The combined use of 
available information at longer wavelengths (mainly soft X--rays and 
radio) and of optical spectroscopy on the putative counterparts of these 
new hard X--ray objects allows us to pinpoint their exact nature. Continuing 
our long-standing program that has been running since 2004, and using 6 different 
telescopes of various sizes together with data from an online 
spectroscopic survey, here we report the classification through optical 
spectroscopy of 22 more unidentified or poorly studied high-energy sources 
detected with the IBIS instrument onboard {\it INTEGRAL}. We found that 16 
of them are active galactic nuclei (AGNs), while the remaining 6 objects 
are within our Galaxy. Among the identified extragalactic sources, the 
large majority (14) is made up of Type 1 AGNs (i.e. with broad emission 
lines); of these, 6 lie at redshift larger than 0.5 and one (IGR J12319$-$0749) 
has $z$ = 3.12, which makes it the second farthest object detected in the 
{\it INTEGRAL} surveys up to now. The remaining AGNs are of type 2 (that is, 
with narrow emission lines only), and one of the two cases is confirmed as a pair 
of interacting Seyfert 2 galaxies. The Galactic objects are identified as two 
cataclysmic variables, one high-mass X--ray binary, one symbiotic binary 
and two chromospherically active stars, possibly of RS CVn type. The main 
physical parameters of these hard X--ray sources were also determined 
using the multiwavelength information available in the literature. We thus 
still find that AGNs are the most abundant population among hard X--ray 
objects identified through optical spectroscopy. Moreover, we note that 
the higher sensitivity of the more recent {\it INTEGRAL} surveys is now enabling
the detection of high-redshift AGNs, thus allowing the exploration 
of the most distant hard X--ray emitting sources and possibly of the 
most extreme blazars.}


\titlerunning{The nature of 22 more {\it INTEGRAL} sources}
\authorrunning{N. Masetti et al.}

\maketitle

\section{Introduction}

\begin{figure*}[th!]
%\begin{center}
\hspace{-.1cm}
\centering{\mbox{\psfig{file=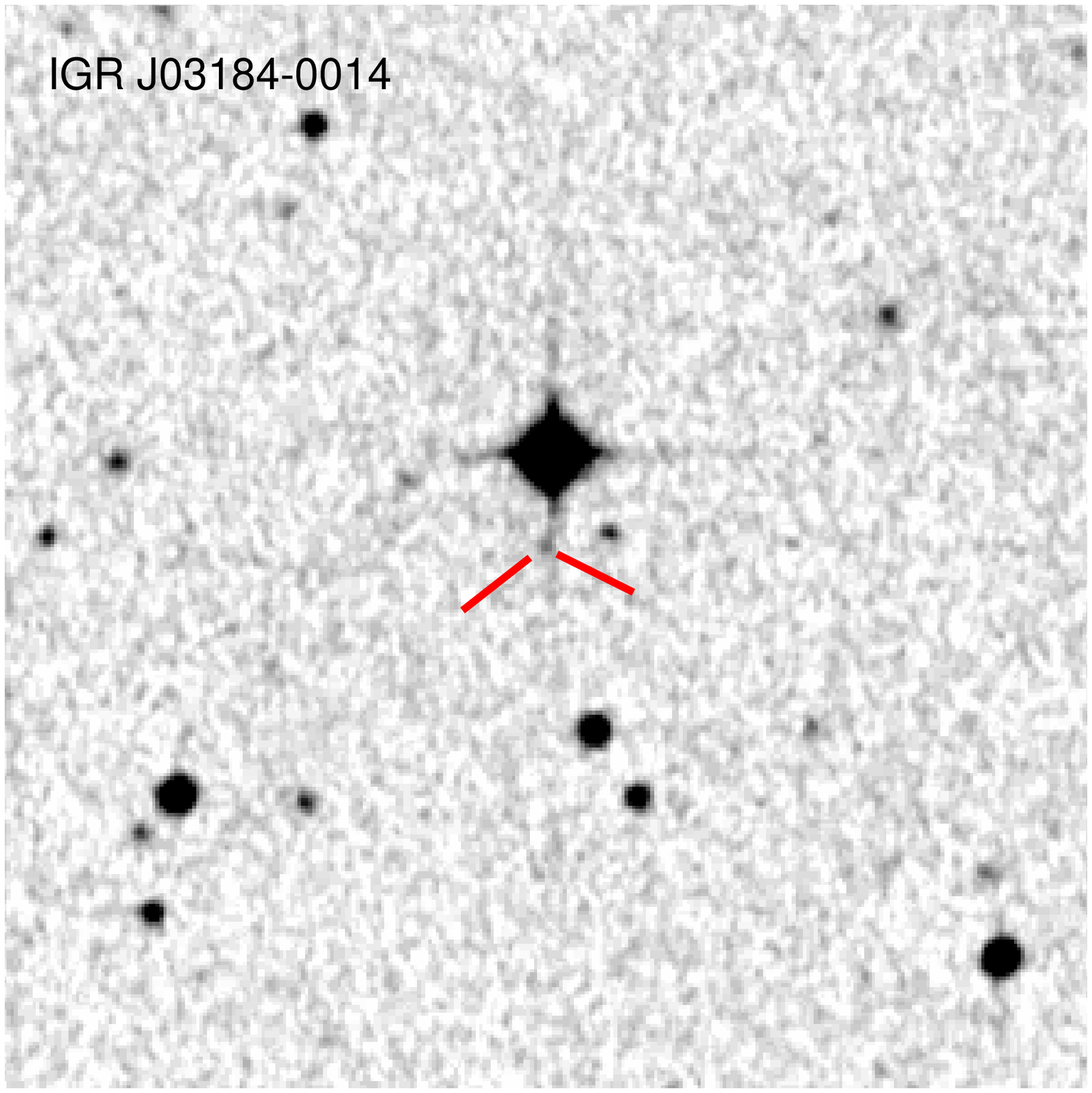,width=5.9cm}}}
%\vspace{-.3cm}
\centering{\mbox{\psfig{file=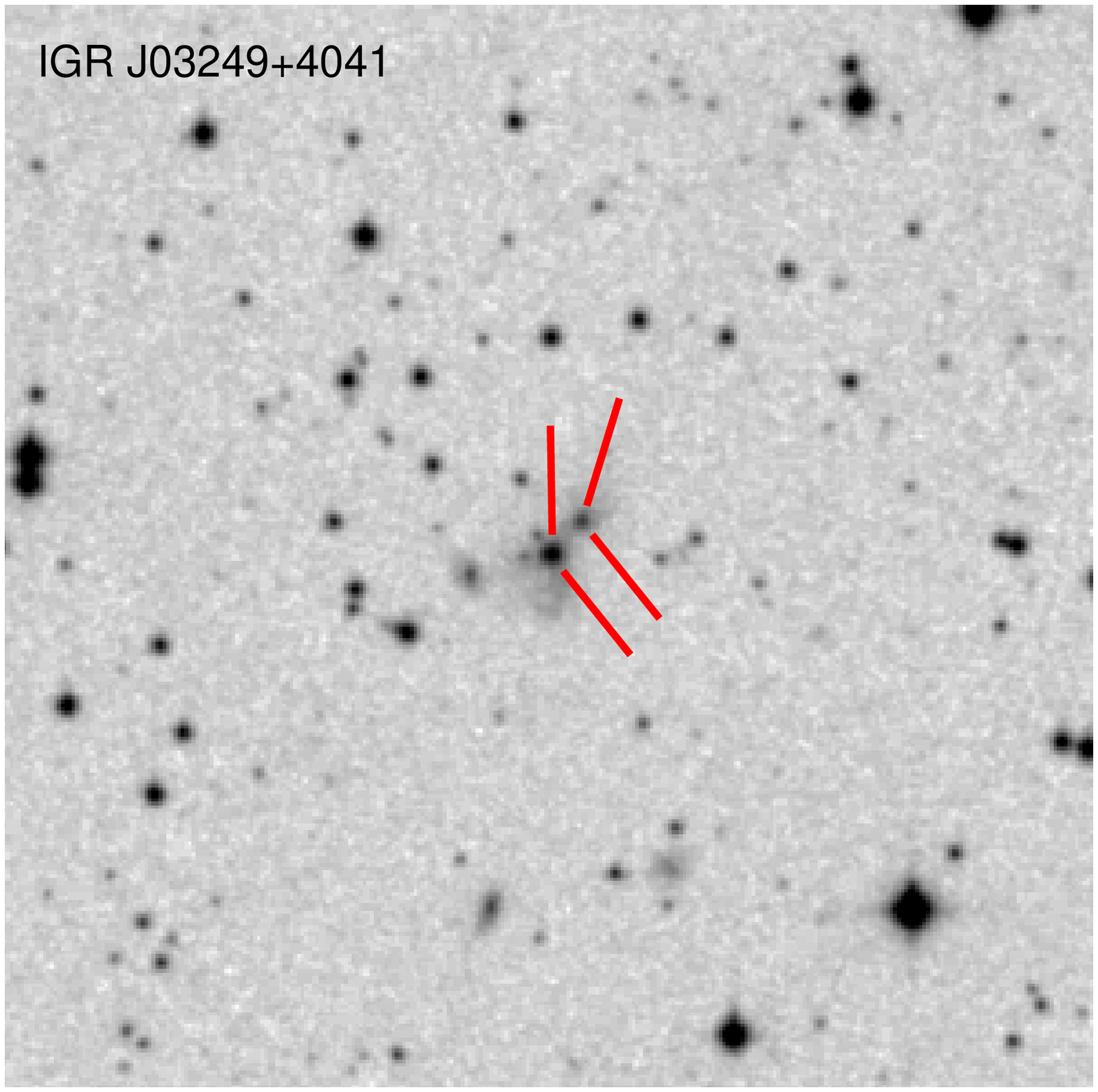,width=5.9cm}}}
%\vspace{-.3cm}
\centering{\mbox{\psfig{file=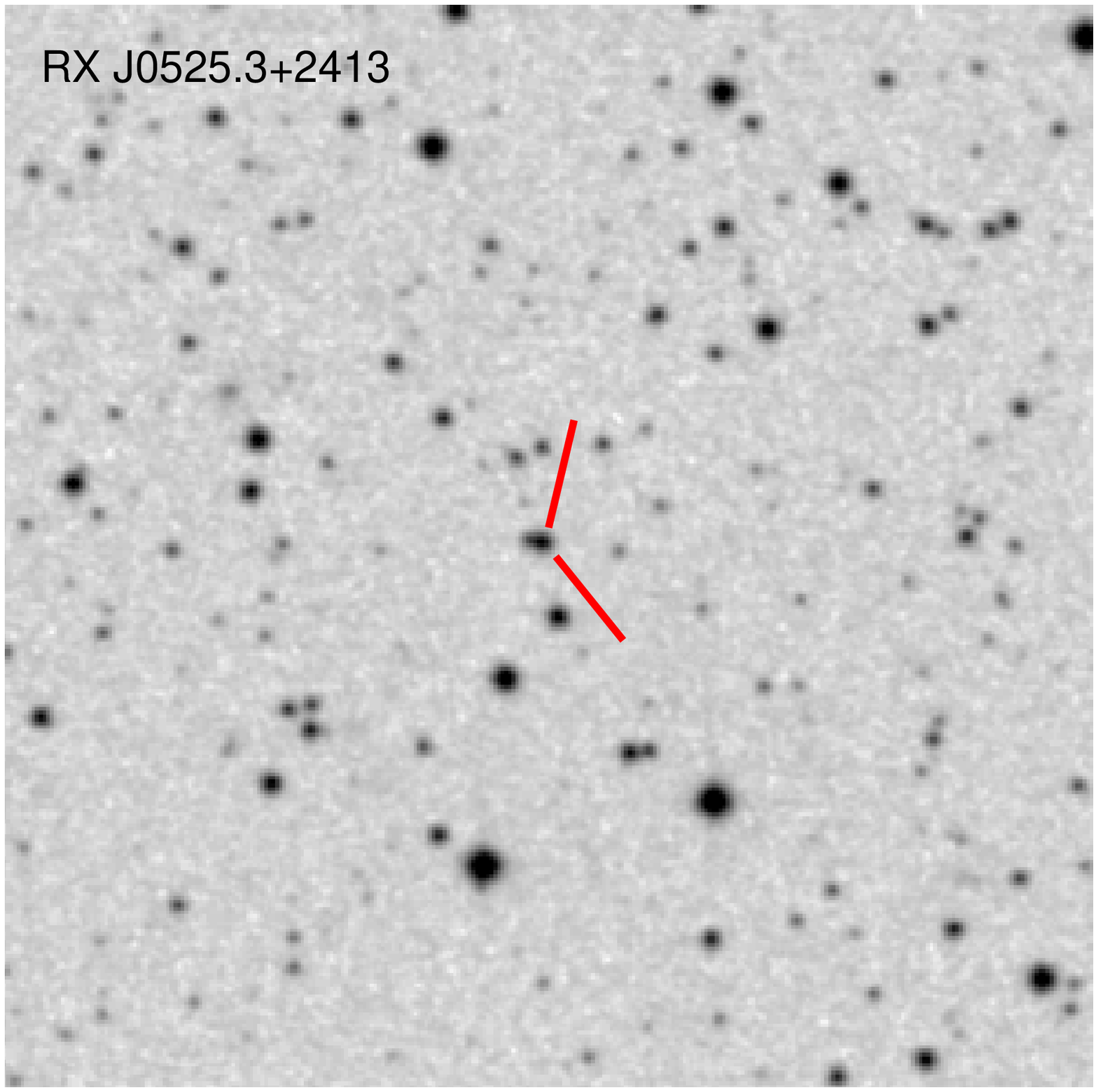,width=5.9cm}}}
%\vspace{-.3cm}
\centering{\mbox{\psfig{file=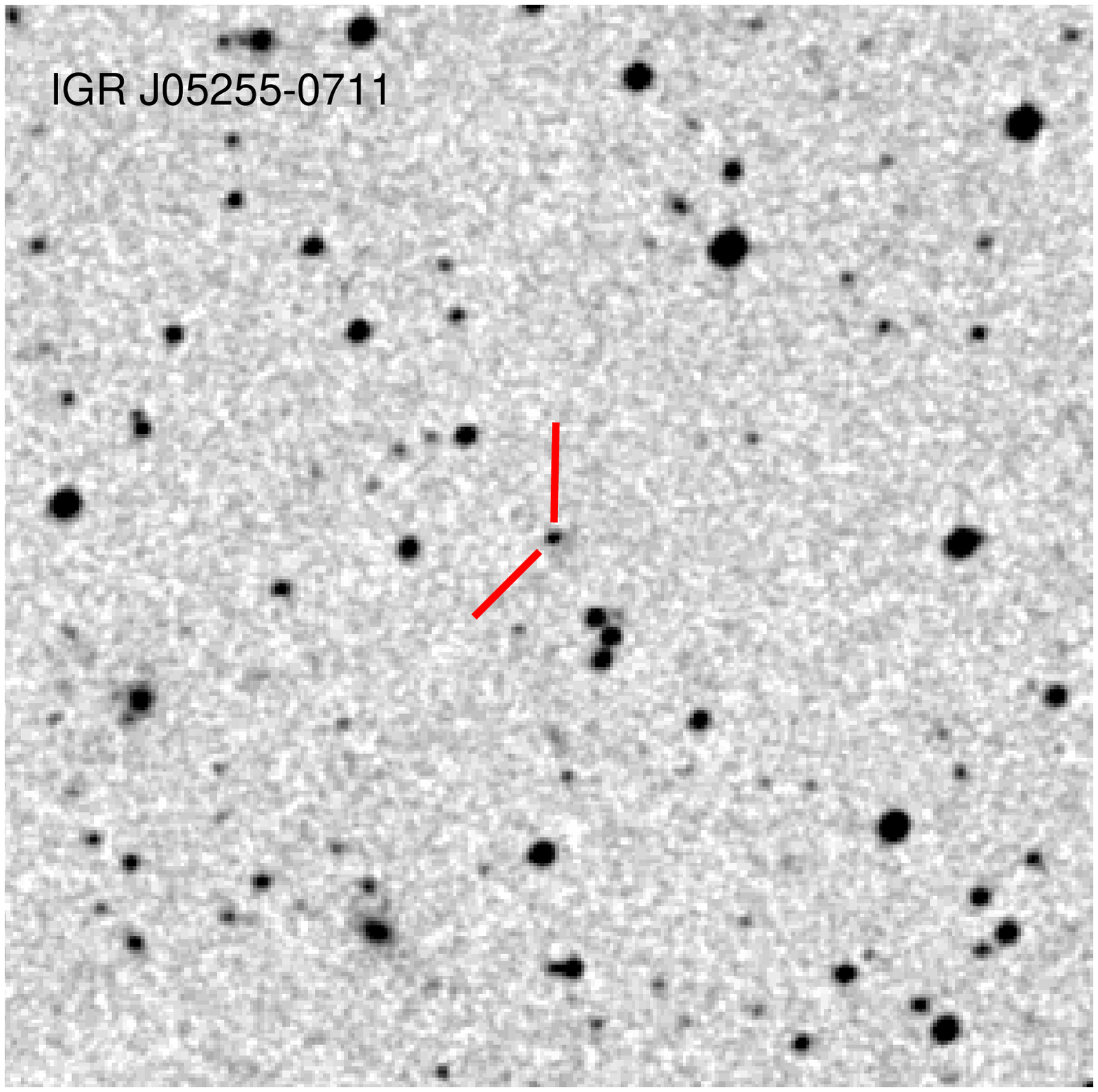,width=5.9cm}}}
%\vspace{-.3cm}
\centering{\mbox{\psfig{file=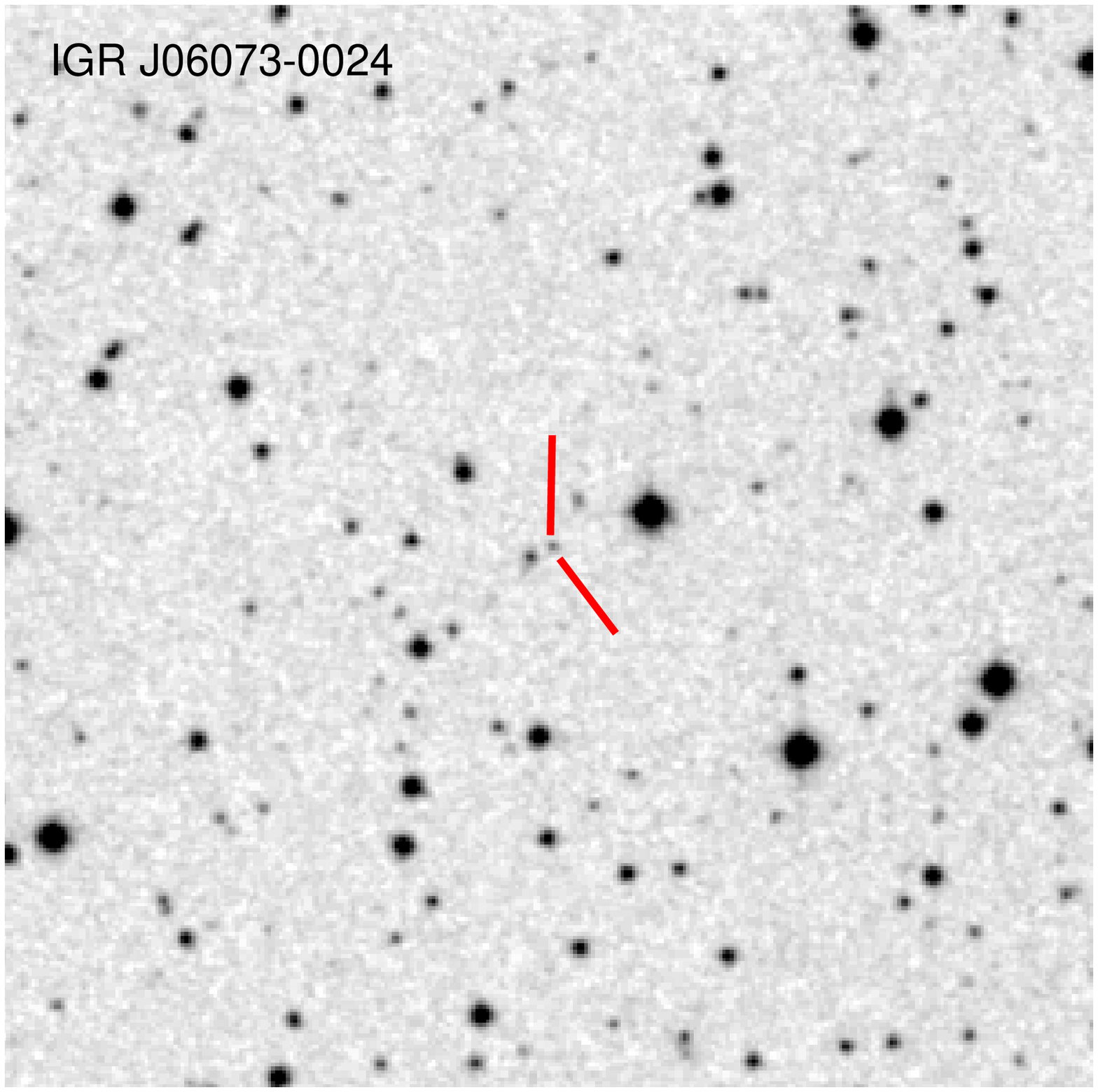,width=5.9cm}}}
%\vspace{-.3cm}
\centering{\mbox{\psfig{file=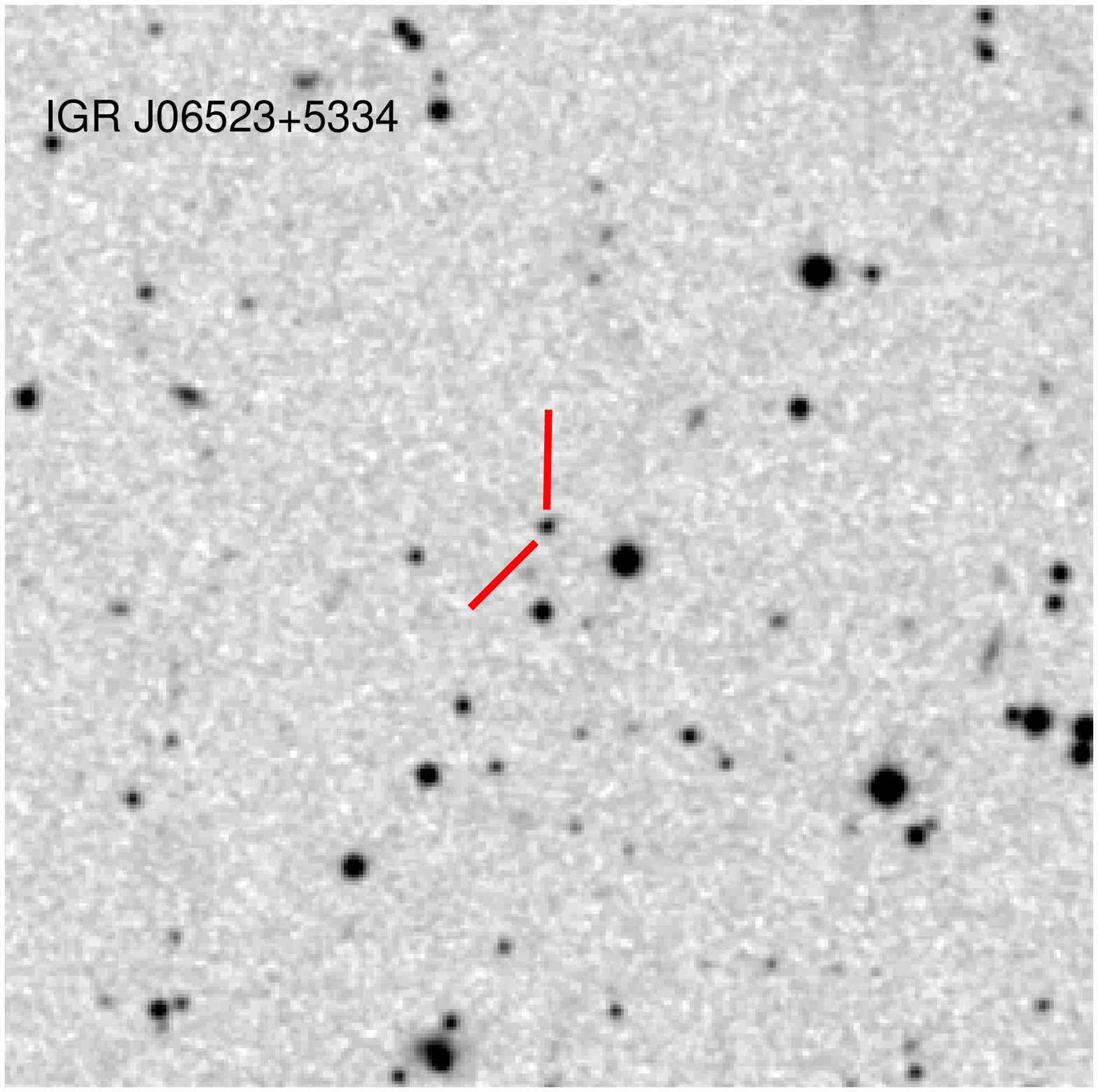,width=5.9cm}}}
%\vspace{-.3cm}
\parbox{6cm}{
\psfig{file=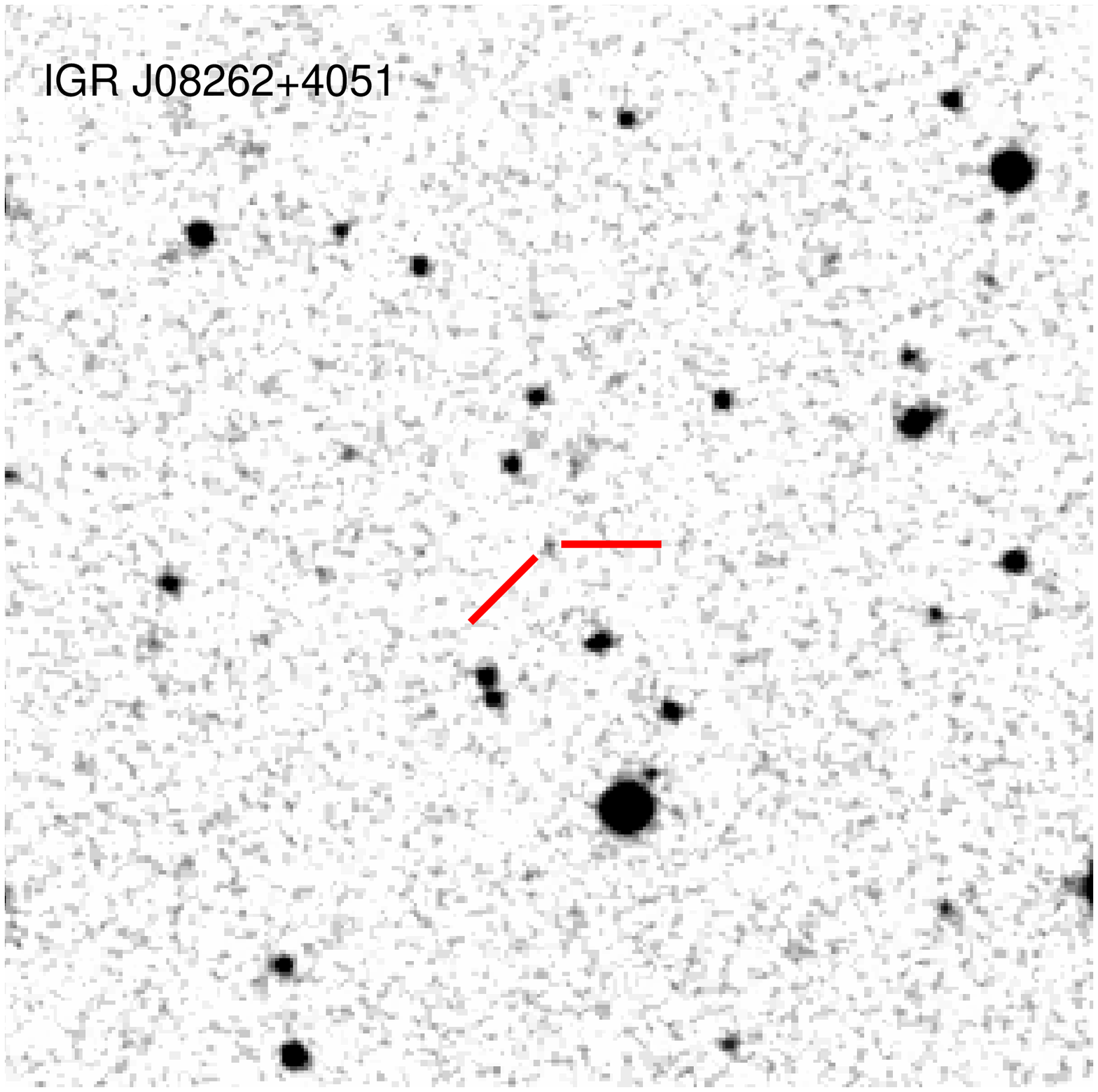,width=5.9cm}
}
\hspace{0.8cm}
\parbox{11cm}{
\vspace{-.5cm}
\hspace{-.3cm}
\caption{Optical images of the fields of 7 of the {\it INTEGRAL} hard 
X--ray sources selected in this paper for optical spectroscopic 
follow-up (see Table 1). The object name is indicated in each panel.
The proposed optical counterparts are indicated with tick marks. Field 
sizes are 5$'$$\times$5$'$ and are extracted from the DSS-II-Red survey.
In all cases, north is up and east to the left.}}
%\end{center}
\end{figure*}

%\addtocounter{figure}{-1}
\begin{figure*}[th!]
%\begin{center}
\hspace{-.1cm}
\centering{\mbox{\psfig{file=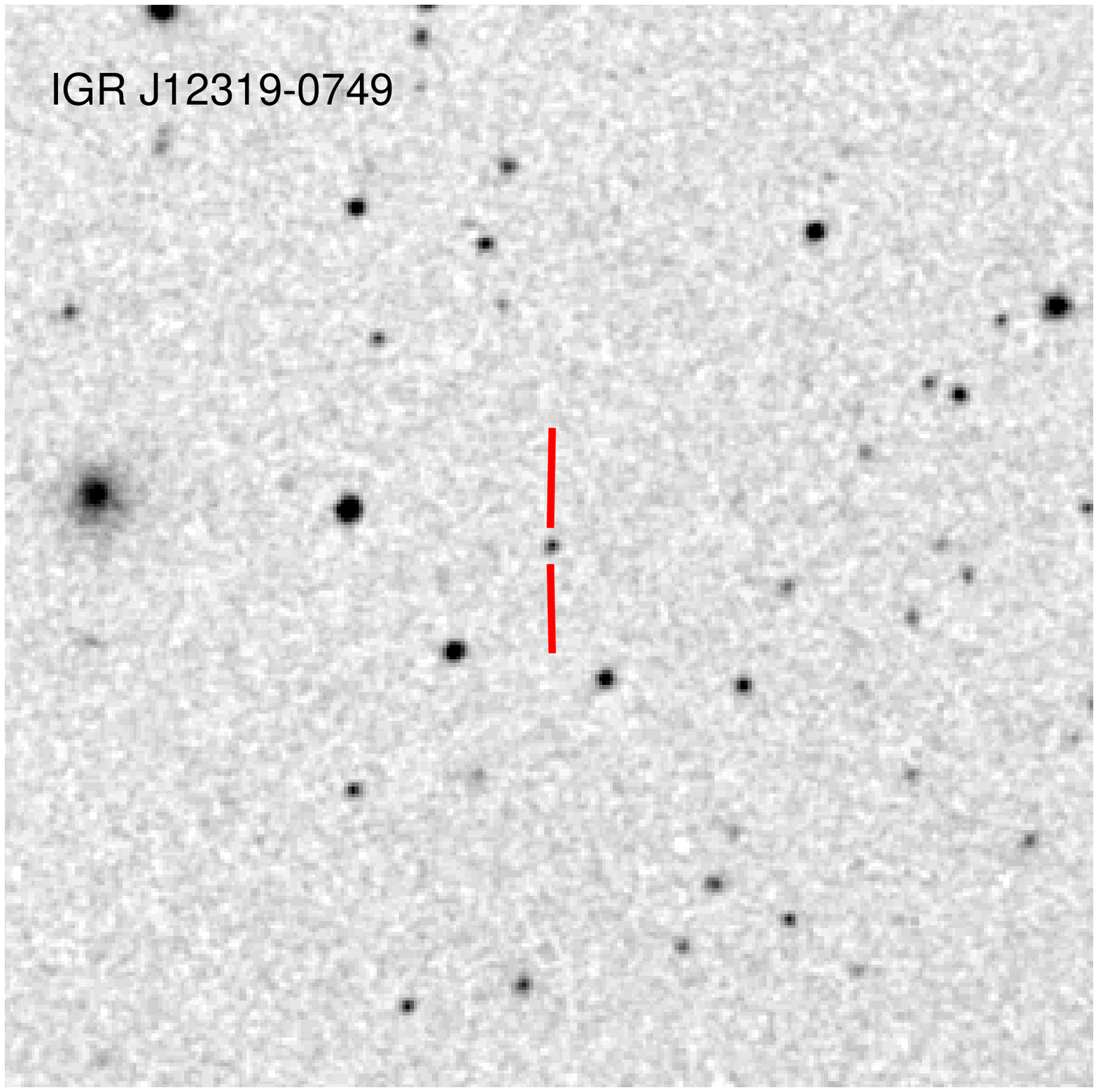,width=5.9cm}}}
%\vspace{-.3cm}
\centering{\mbox{\psfig{file=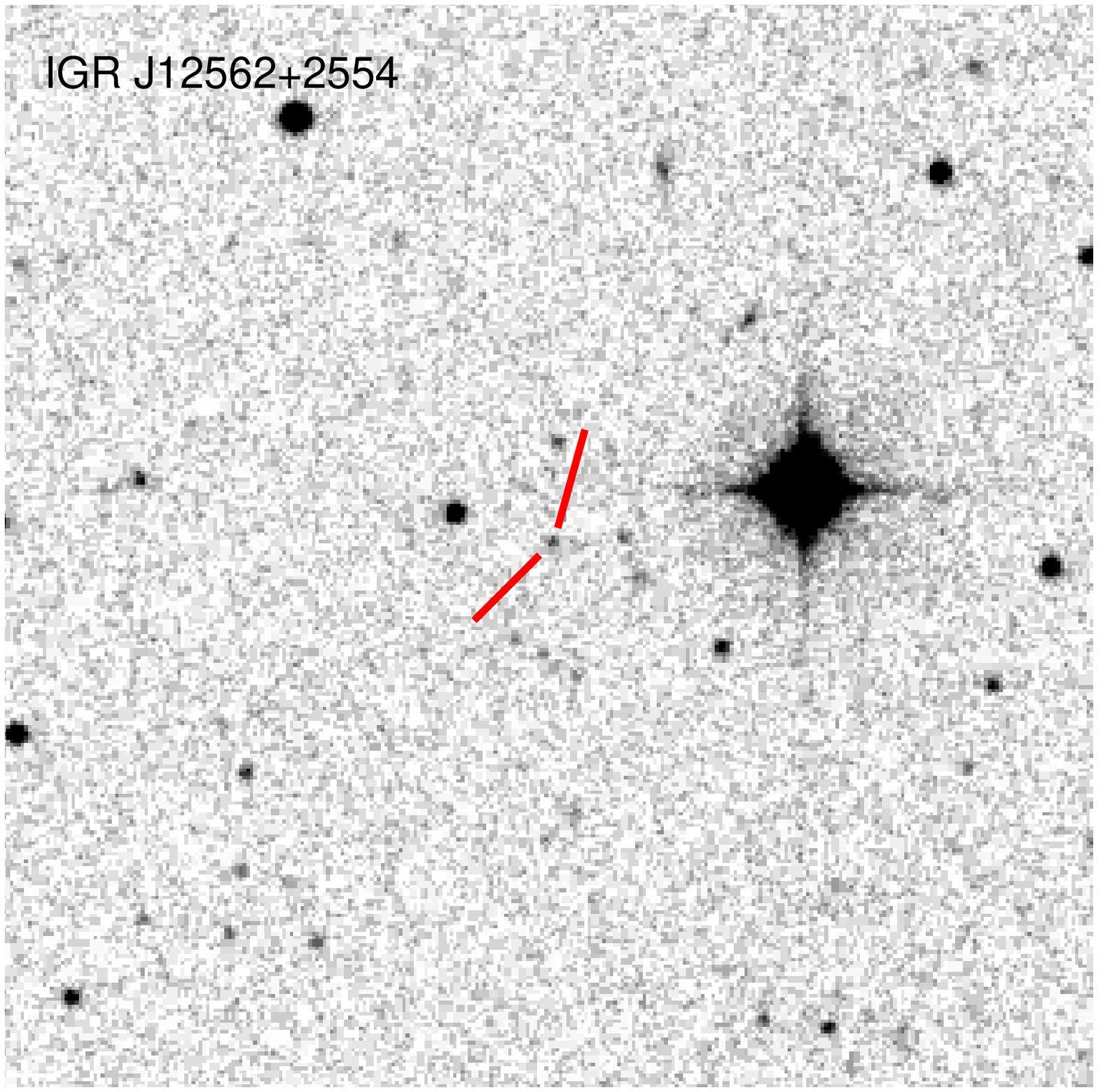,width=5.9cm}}}
%\vspace{-.3cm}
\centering{\mbox{\psfig{file=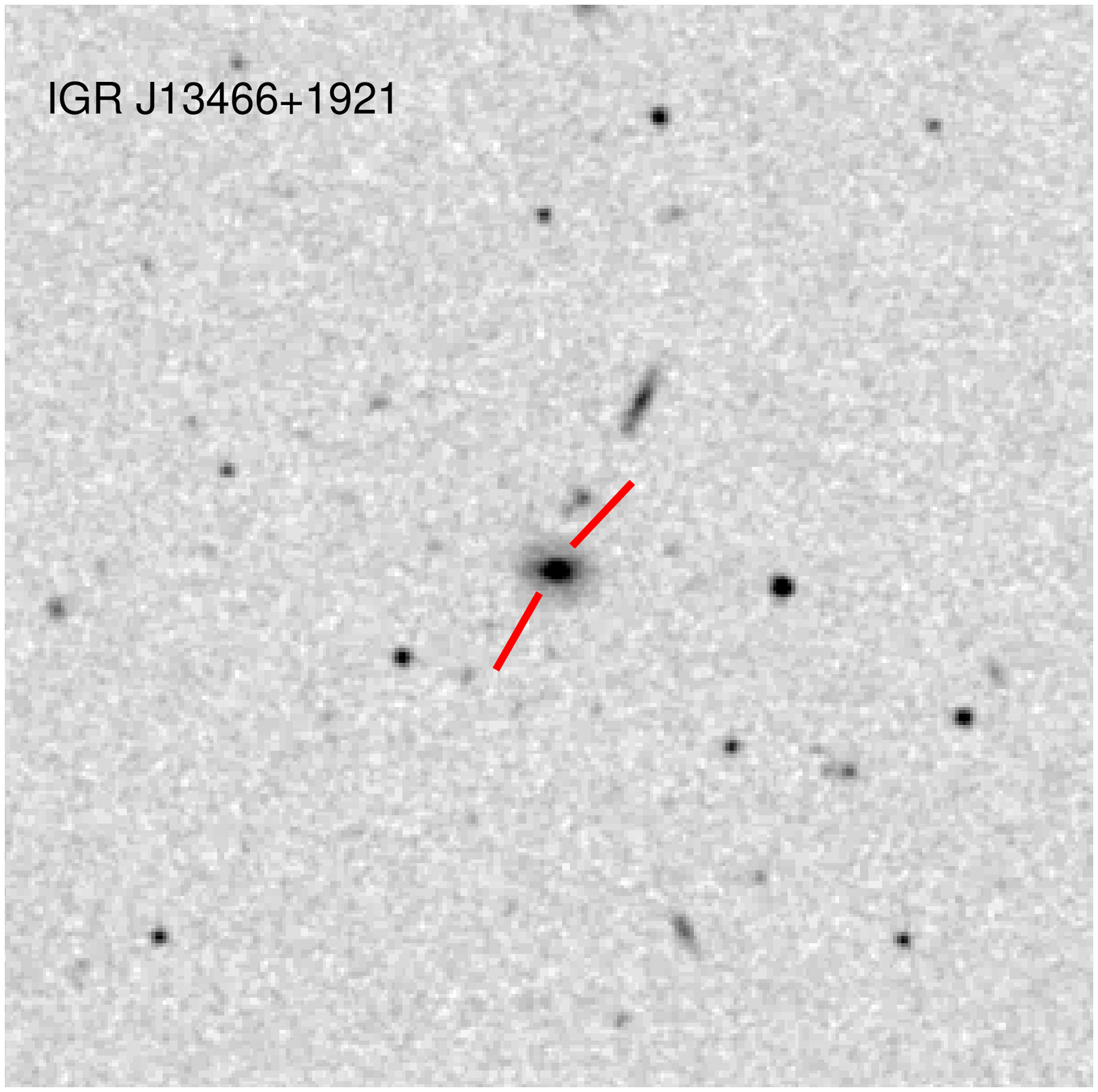,width=5.9cm}}}
%\vspace{-.3cm}
\centering{\mbox{\psfig{file=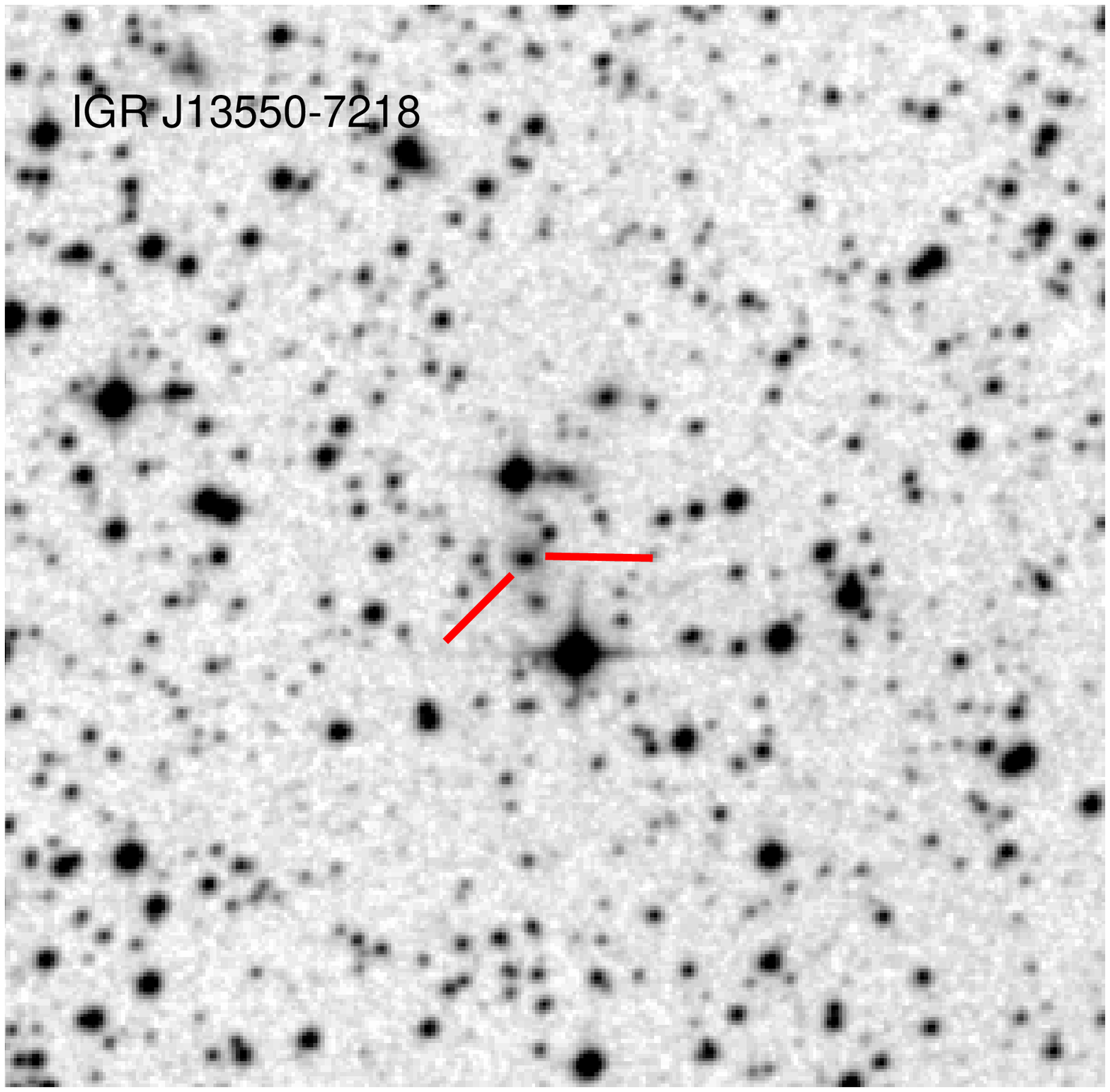,width=5.9cm}}}
%\vspace{-.3cm}
\centering{\mbox{\psfig{file=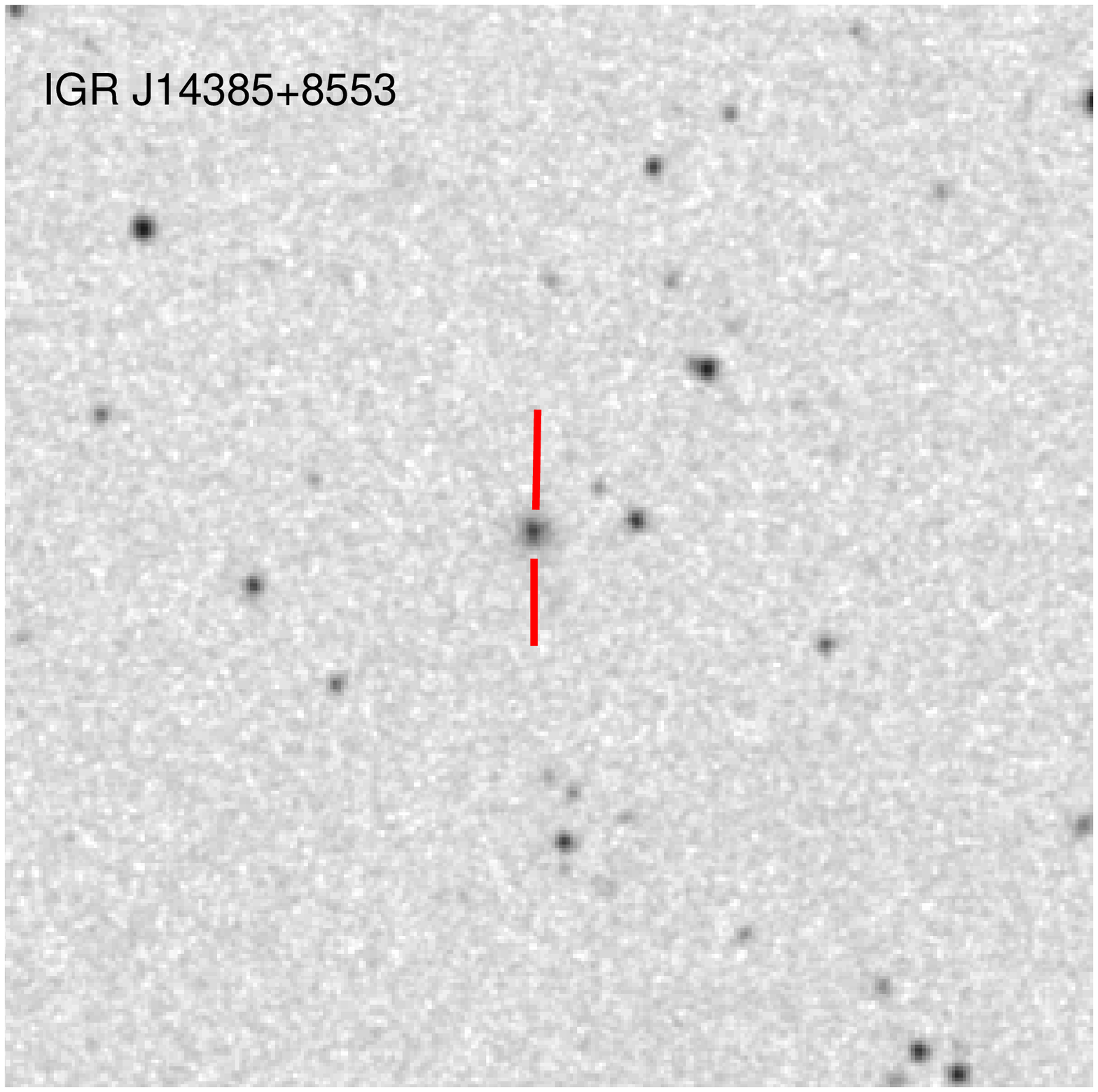,width=5.9cm}}}
%\vspace{-.3cm}
\centering{\mbox{\psfig{file=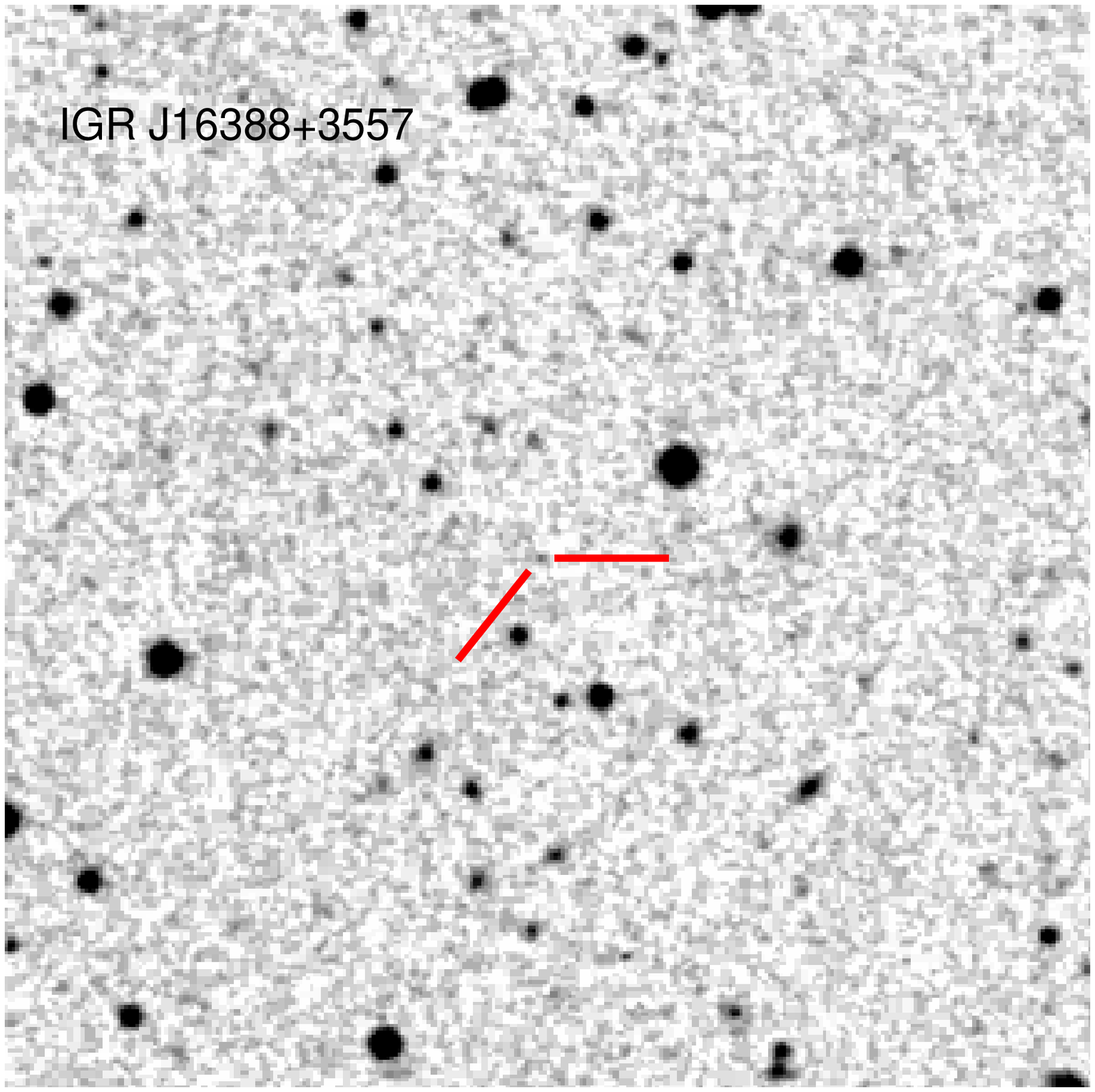,width=5.9cm}}}
%\vspace{-.3cm}
\parbox{12cm}{
\centering{\mbox{\psfig{file=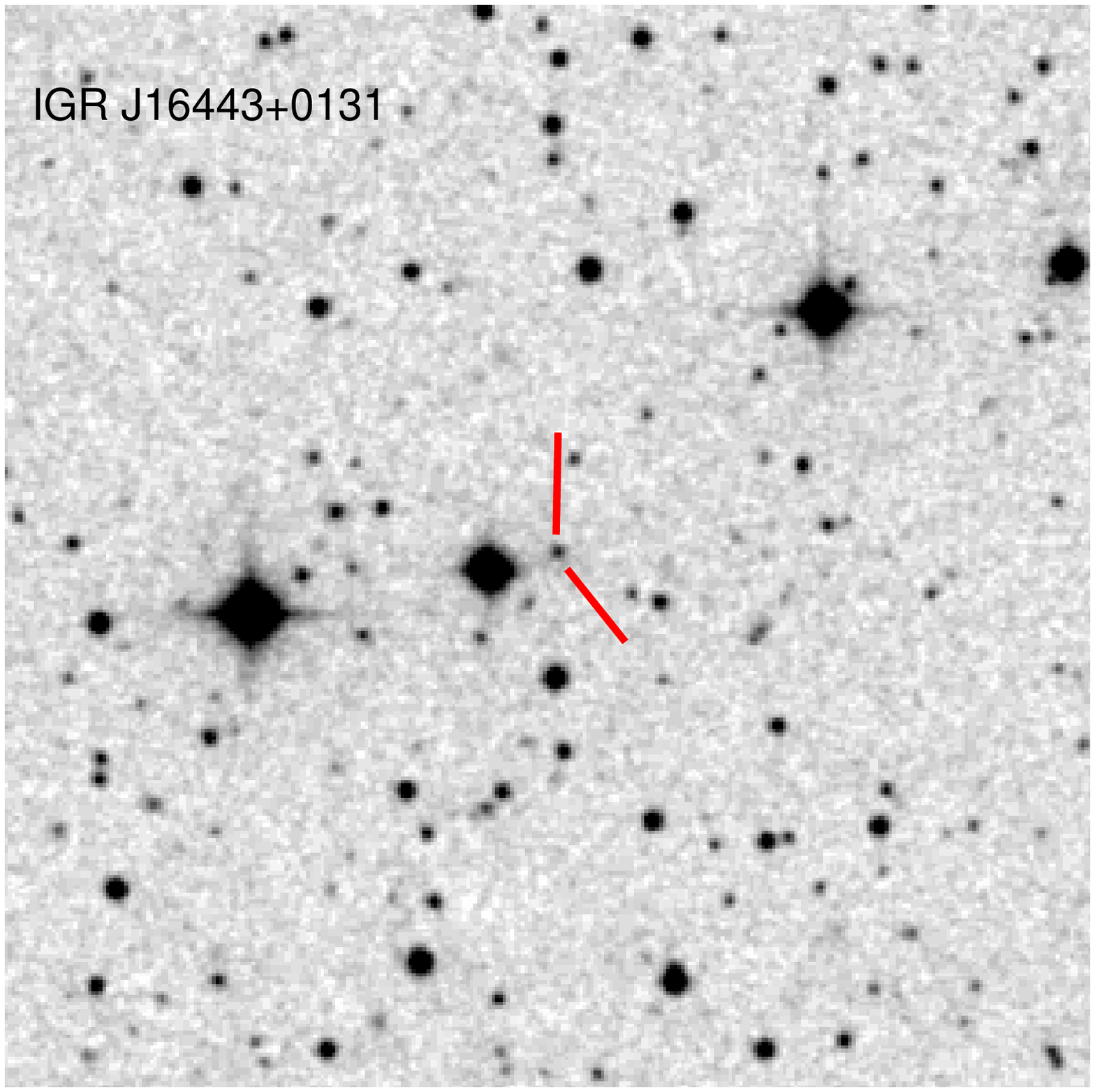,width=5.9cm}}}
\centering{\mbox{\psfig{file=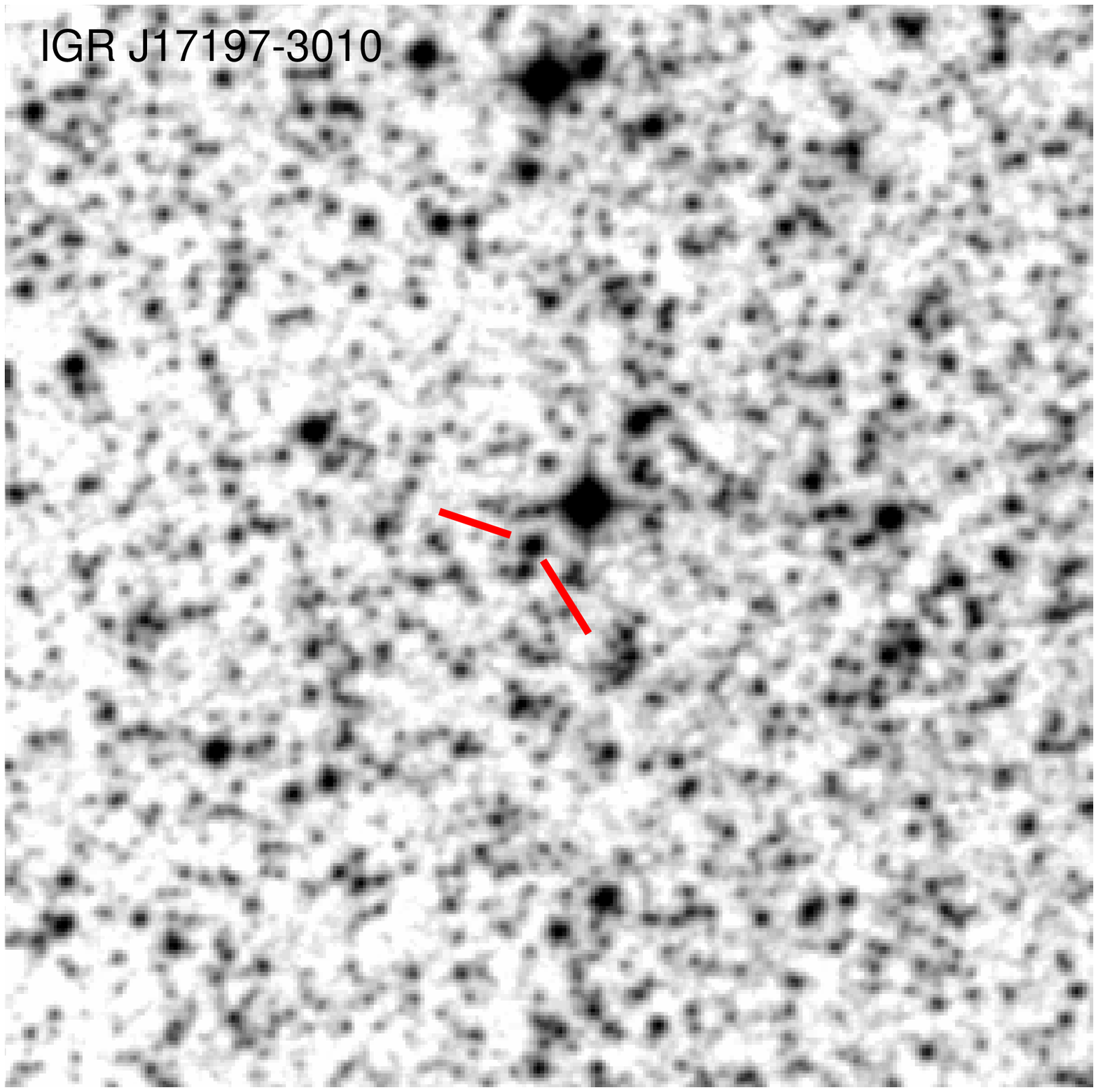,width=5.9cm}}}
}
\hspace{0.8cm}
\parbox{5cm}{
\vspace{-.5cm}
\hspace{-.3cm}
\caption{As Fig. 1, but for 8 more {\it INTEGRAL} sources of our
sample (see Table 1).}}
%\end{center}
\end{figure*}

%\addtocounter{figure}{-1}
\begin{figure*}[th!]
%\begin{center}
\hspace{-.1cm}
\centering{\mbox{\psfig{file=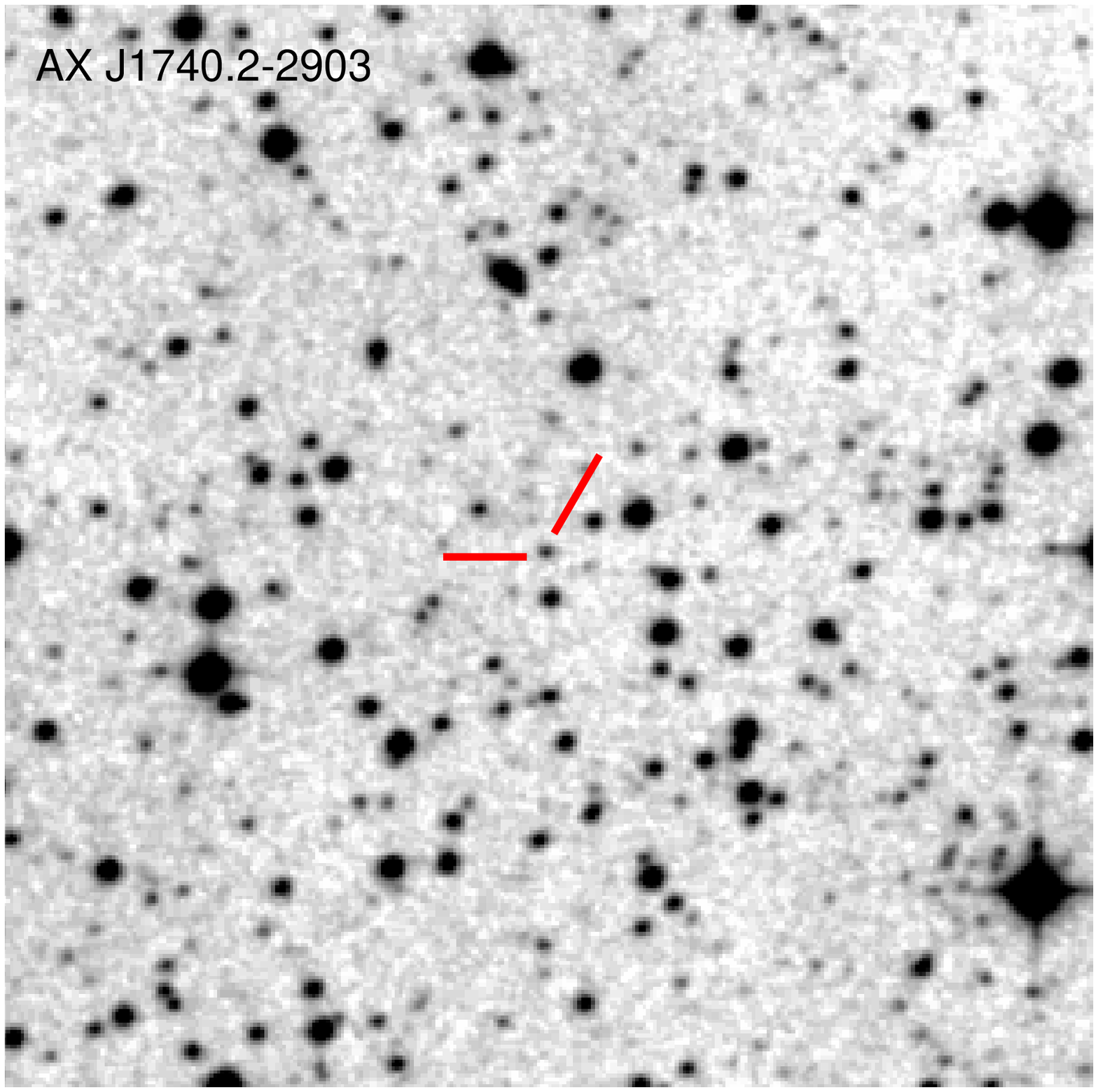,width=5.9cm}}}
%\vspace{-.3cm}
\centering{\mbox{\psfig{file=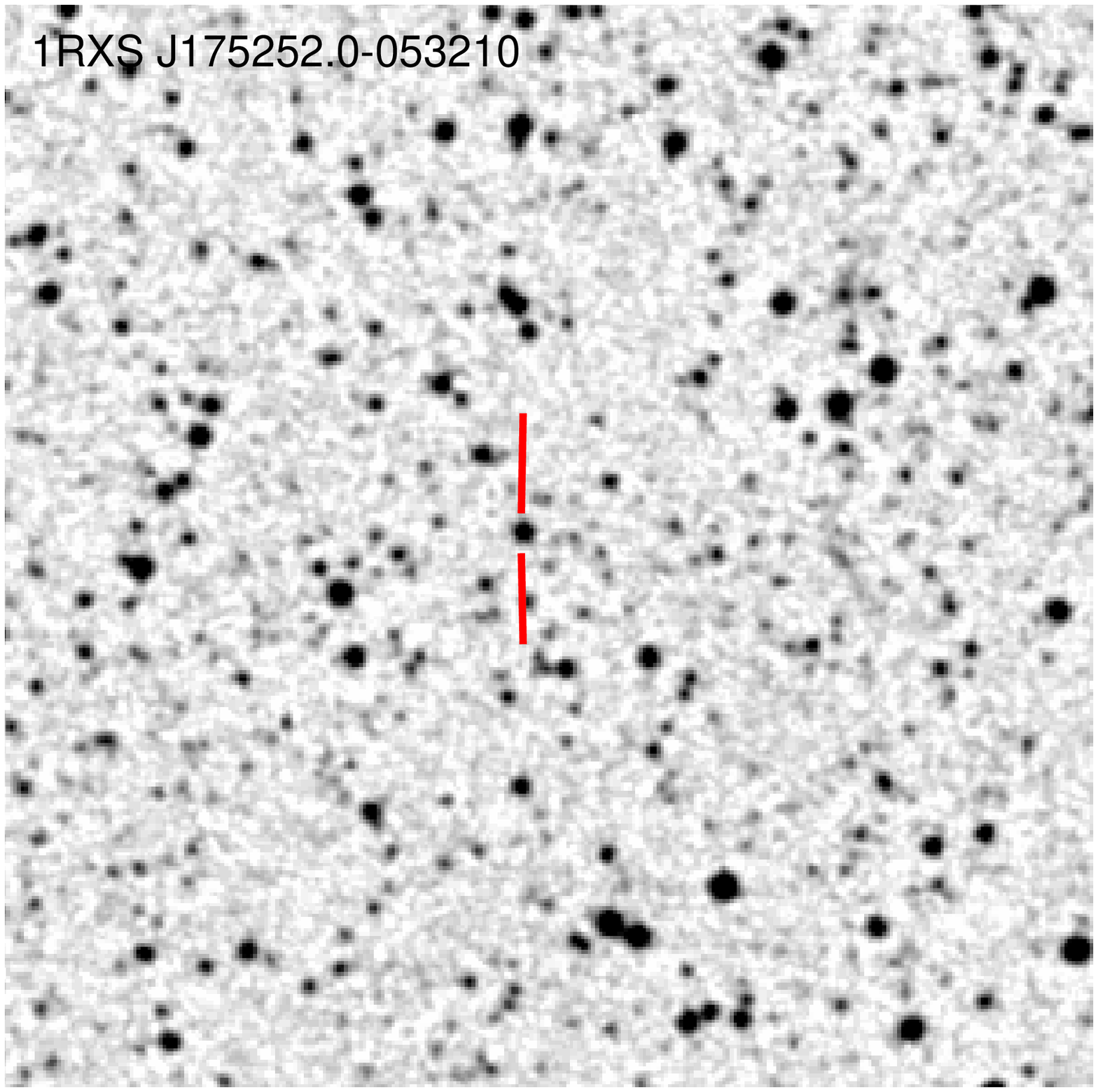,width=5.9cm}}}
%\vspace{-.3cm}
\centering{\mbox{\psfig{file=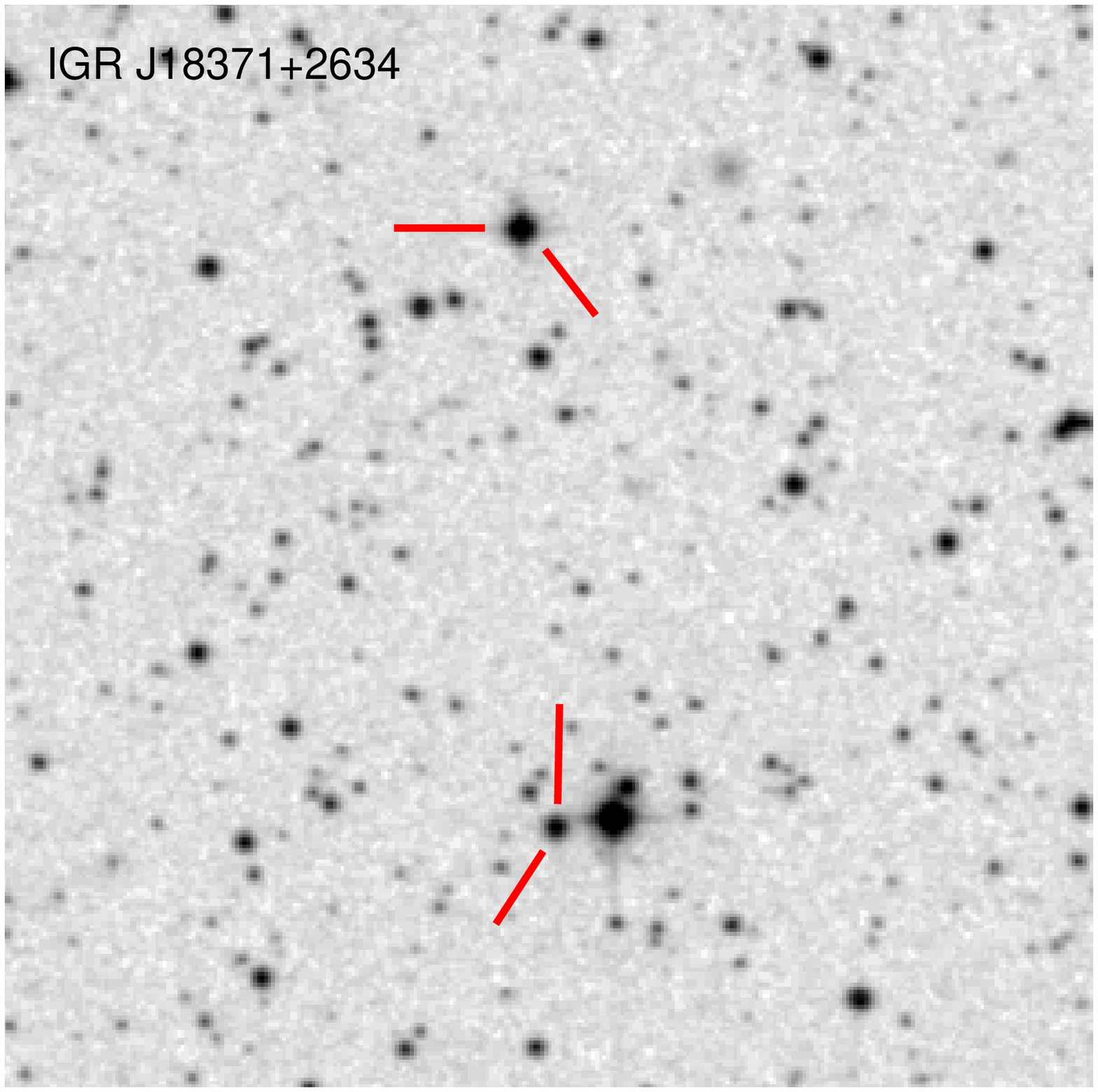,width=5.9cm}}}
%\vspace{-.3cm}
\centering{\mbox{\psfig{file=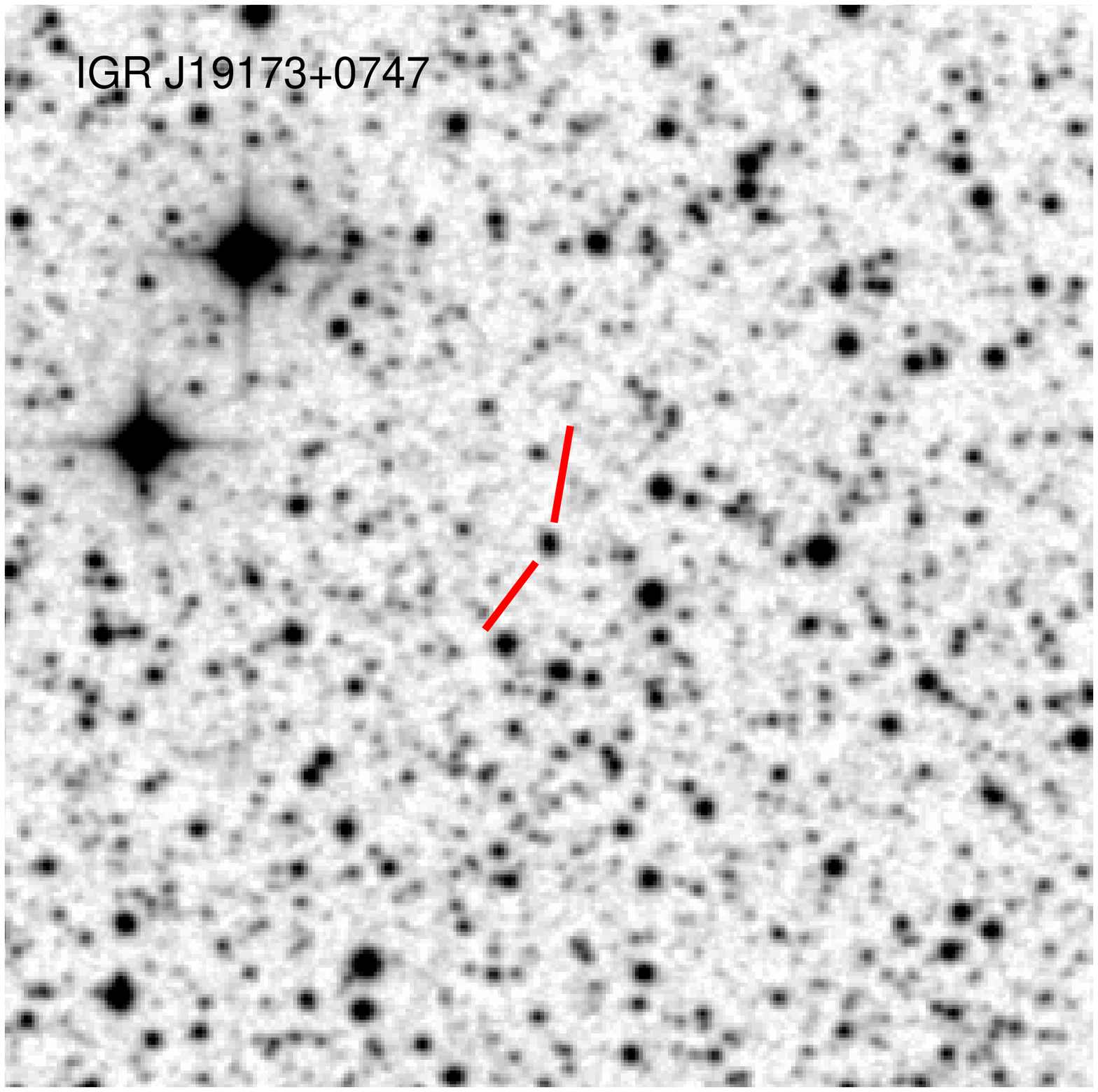,width=5.9cm}}}
%\vspace{-.3cm}
\centering{\mbox{\psfig{file=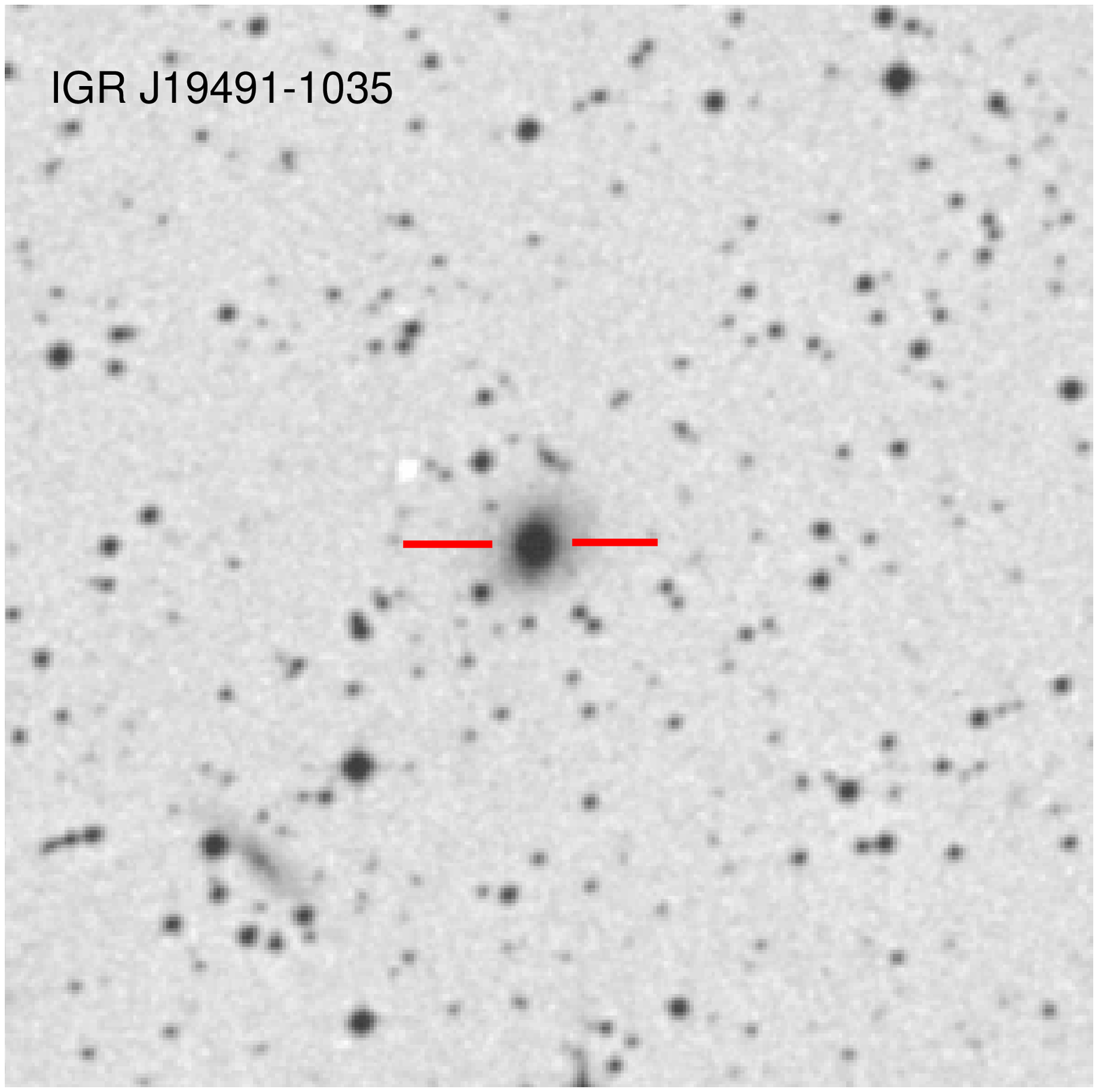,width=5.9cm}}}
%\vspace{-.3cm}
\centering{\mbox{\psfig{file=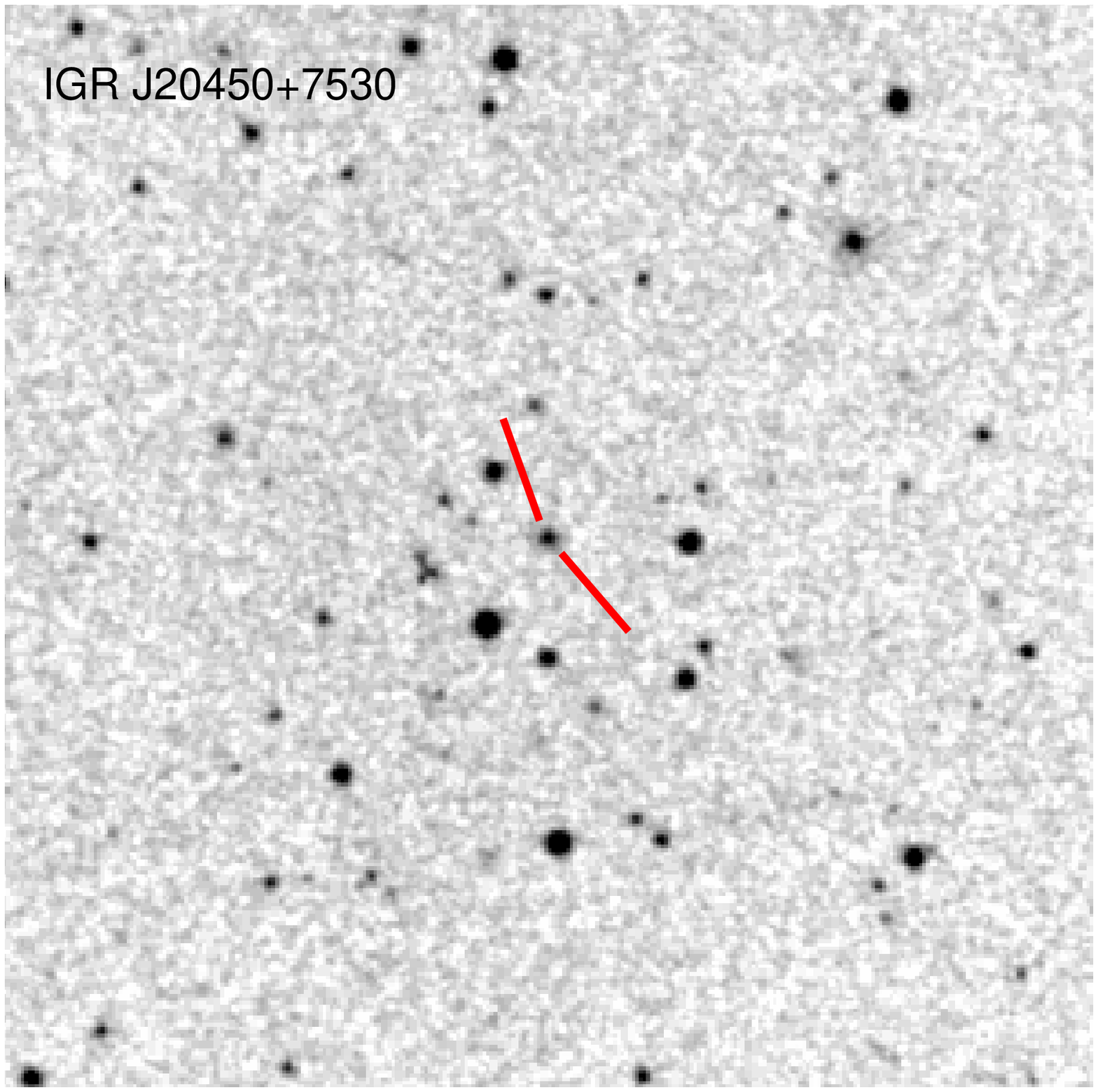,width=5.9cm}}}
%\vspace{-.3cm}
\parbox{6cm}{
\psfig{file=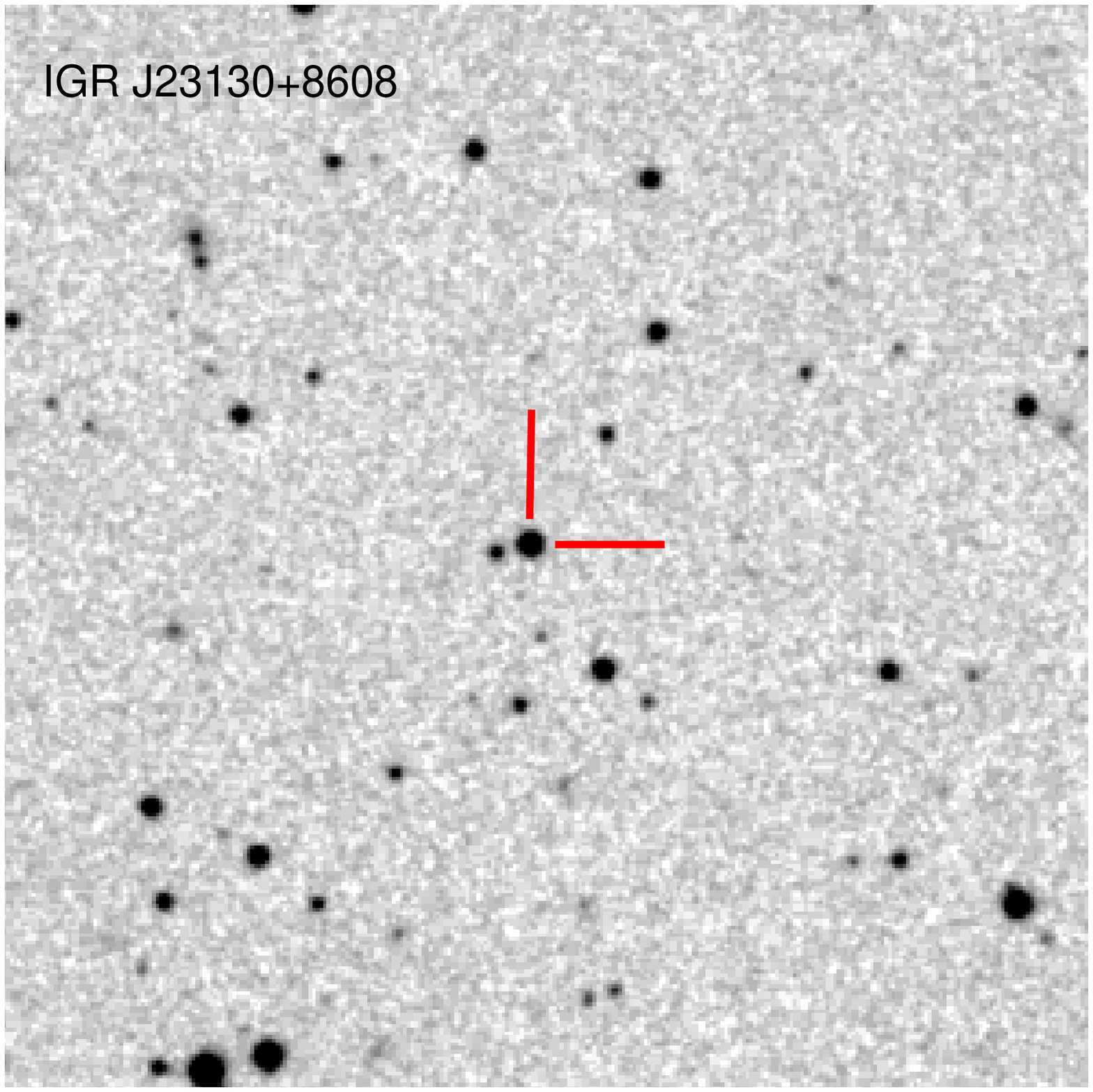,width=5.9cm}
}
\hspace{0.8cm}
\parbox{11cm}{
\vspace{-.8cm}
\hspace{-.3cm}
\caption{As Fig. 1, but for 7 more {\it INTEGRAL} sources of our
sample (see Table 1). The image of the field of IGR J20450+7530 
(centre right panel) has been extracted from the DSS-II-Infrared survey,
whereas the one of IGR J18371+2634 (upper right panel) shows the optical
counterparts of both X--ray objects reported in Landi et al. (2011; 
see also Section 4.3 of the present work).}}
%\end{center}
\end{figure*}

One of the main objectives of the {\it INTEGRAL} mission (Winkler et al. 
2003) is a regular survey of the whole sky in the hard X--ray band. This 
makes use of the unique imaging capability of the IBIS instrument 
(Ubertini et al. 2003) which allows the detection of sources at the 
mCrab\footnote{1 mCrab $\cong$ 2$\times$10$^{-11}$ erg cm$^{-2}$ 
s$^{-1}$, with a slight dependence on the reference energy range.} 
level with a typical localization accuracy of a few arcmin. The latest 
(4$^{\rm th}$ of the series) IBIS catalogue published up to now by Bird et 
al. (2010) contains more than 700 hard X--ray sources detected in the 
20--100 keV band down to an average flux level of about 1 mCrab and with 
positional accuracy better than $\sim$5 arcmin.

However, a substantial fraction ($\sim$30\%) of them had no obvious 
counterpart at other wavelengths and therefore could not yet be associated 
with any known class of high-energy emitting objects. A similar situation 
is present in the 7-year {\it INTEGRAL}/IBIS all-sky survey of Krivonos et 
al. (2010)\footnote{An up-to-date online version of this catalog can be 
found at \\ {\tt http://hea.iki.rssi.ru/integral/survey/catalog.php}}, 
obtained with a parallel (and partially overlapping) {\it INTEGRAL} data 
set and which contains more than 500 sources (in part coincident with 
those in Bird et al. 2010).

Therefore, a multiwavelength observational campaign on these unidentified 
sources is fundamental to pinpoint their nature, and is especially 
important given that they build up to one third of the whole set of IBIS 
detections.

X--ray analysis methods have shown their usefulness in identifying the 
nature of new {\it INTEGRAL} sources, such as for example X--ray timing by 
the detection of pulsations, orbital periods or X--ray bursts (e.g., 
Walter et al. 2006; Sguera et al. 2007; Del Santo et al. 2007; Chenevez et 
al. 2007; Bird et al. 2009; La Parola et al. 2010), or X--ray spectroscopy 
and imaging (see for instance Tomsick et al. 2009, Rodriguez et al. 2010a, 
Malizia et al. 2010, Fiocchi et al. 2010, and references therein). 
Alternatively, and also quite effectively, cross-correlation with soft 
X--ray catalogs and consequently optical spectroscopy on thereby selected 
candidates allows the determination of the nature and the main 
multiwavelength characteristics of unidentified or poorly studied hard 
X--ray objects.

Here, we continue the identification work on {\it 
INTEGRAL} sources we started in 2004 and which permitted us to, so far, 
identify more than 150 sources by means of optical spectroscopy (see Masetti 
et al. 2004, 2006abcd, 2008a, 2009, 2010 [hereafter Papers I-VIII, 
respectively]; Masetti et al. 2007, 2008b; Maiorano et al. 2011). 
We report here on optical spectra of firm or likely counterparts 
of 22 unidentified, unclassified or poorly studied sources belonging to 
one or more IBIS surveys (Bird et al. 2007, 2010; Krivonos et al. 2010); 
we also added to our sample the unidentified source IGR J19173+0747 
reported in Pavan et al. (2011). Optical spectroscopy for these objects was 
acquired using 6 different telescopes and one public spectroscopic 
archive.

The paper is structured as follows. In Sect. 2 we explain the criteria 
used to choose the sample of {\it INTEGRAL} and optical objects considered 
in this work. In Sect. 3 a brief description of the observations is 
presented. Section 4 illustrates and discusses the results together with an 
update of the statistical outline of the identifications of {\it INTEGRAL} 
sources obtained until now. Conclusions are reported in Sect. 5.

As in our previous Papers I-VIII on the subject, the main results of this 
work, along with the information about the {\it INTEGRAL} sources that 
have been identified (by us or by other groups) using optical or 
near-infrared (NIR) observations, are listed in a web page\footnote{{\tt 
http://www.iasfbo.inaf.it/extras/IGR/main.html}} that we maintain as a 
service to the scientific community (Masetti \& Schiavone 2008).

In this paper, unless otherwise stated, errors and limits are reported at 
1$\sigma$ and 3$\sigma$ confidence levels, respectively. We also note that 
this work supersedes the results presented in the preliminary analyses of 
Parisi et al. (2010) and Masetti et al. (2011).

\section{Sample selection}

Using the same approach we applied in our previous Papers I-VIII, we 
considered the IBIS surveys of Bird et al. (2007, 2010) and Krivonos et 
al. (2010), and we selected unidentified or unclassified hard X--ray 
sources that contain, within the IBIS 90\% confidence level error box, a 
single bright soft X--ray object detected either in the {\it ROSAT} 
all-sky surveys (Voges et al. 1999, 2000), or with {\it Swift}/XRT (from 
Page et al. 2007, Lutovinov et al. 2010, Rodriguez et al. 2010ab, Landi et 
al. 2010abc, 2011, Krivonos et al. 2011, as well as from the XRT 
archive\footnote{XRT archival data are freely available at \\ {\tt 
http://www.asdc.asi.it/}}), or in the Slew Survey (Saxton et al. 2008) or 
Serendipitous Source Catalog (Watson et al. 2009) of {\it XMM-Newton}, or 
with pointed {\it Chandra} (Fiocchi et al. 2010) and {\it XMM-Newton} 
(Halpern \& Gotthelf 2010; Farrell et al. 2010) observations. This 
approach was proven by Stephen et al. (2006) to be very effective in 
associating, with a high degree of probability, IBIS sources with a softer 
X--ray counterpart and in turn drastically reducing their positional error 
circles to better than a few arcsec in radius, making the search area 
smaller by a factor of $\sim$10$^4$.

After this first selection, we chose among these objects those that had, 
within their refined 90\% confidence level soft X--ray error 
boxes\footnote{When needed, for the cases in which the soft X--ray 
positional error is given at 1$\sigma$ confidence level (basically, the
{\it ROSAT} and {\it XMM-Newton} Slew Survey sources) we rescaled it
to the corresponding 90\% confidence level assuming a Gaussian probability 
distribution.}, a single possible optical counterpart with magnitude 
$R \la$ 20 in the DSS-II-Red survey\footnote{Available at 
{\tt http://archive.eso.org/dss/dss}}, so that optical spectroscopy 
could be obtained with reasonable signal-to-noise ratio (S/N) with
telescopes having aperture of at least 1.5 metres.
This allowed us to select 13 IBIS sources.

We moreover added the newly-discovered source IGR J19173+0747 (Pavan et al. 
2011) to our list, as these authors also report an arcsecond-sized 
position obtained with {\it Swift} and a possible optical counterpart. 

To increase the size of this sample, we then used the following independent 
approaches. 

First, we cross-correlated the above IBIS 
surveys with radio catalogs such as the NVSS (Condon et al. 1998), SUMSS 
(Mauch et al. 2003), and MGPS (Murphy et al. 2007) surveys when a soft 
X--ray observation of the hard X--ray source field was not available. 
This step provided 1 more {\it INTEGRAL} source (IGR J06073$-$0024) with 
a likely optical counterpart. In a few cases (e.g. IGR J13550$-$7218; Landi 
et al. 2010d), the use of this method on the IBIS sources already selected 
allowed us to further reduce the soft X--ray position uncertainty down 
to less than 2$''$ and thus to pinpoint the actual counterpart with a higher 
degree of certainty.

Next, we relaxed the search described above by considering the IBIS 99\% 
error circles. This allowed us to pinpoint 7 further cases to be followed 
up with optical spectroscopy (see also Rodriguez et al. 2010a, Fiocchi et 
al. 2010, and Landi et al. 2010a, 2011).

We wish to point out that the 8 sources thus selected should be considered 
to have only a tentative, albeit likely, counterpart: this is indicated 
with an asterisk alongside their names in Table 1 and by a question 
mark near the name of the corresponding hard X--ray source in the optical 
spectra of Figs. 4, 5, 7 and 8. This caution should particularly be 
applied to the IBIS source selected using radio surveys (but see Maiorano 
et al. 2011), as no thorough statistical study on the confidence of the 
hard X--ray/radio positional correlation is available up to now for the 
IBIS catalogues. The reader is moreover referred to Paper III for the 
caveats and the shortcomings of choosing, within an IBIS error box, 
``peculiar" sources that are not straightforwardly linked to an 
arcsec-sized soft X--ray position.

Thus, in total we gathered a sample of 22 {\it INTEGRAL} objects with 
possible optical counterparts, which we explored by means of optical 
spectroscopy. Their names and accurate coordinates (to 0$\farcs$2 or 
better; see next section) are reported in Table 1, while their optical 
finding charts are shown in Figs. 1-3, with the corresponding putative 
counterparts indicated with tick marks.

Finally, we stress here that in our final sample there are 5 {\it 
INTEGRAL} sources (IGR J03249+4041, Lutovinov et al. 2010; RX J0525.3+2413, 
Torres et al. 2007; IGR J06253+5334, Halpern 2011; AX J1740.2$-$2903,
Halpern \& Gotthelf 2010; and IGR J19491$-$1035, Krivonos et al. 2011) 
that, although already identified by these authors, have incomplete 
information at longer wavelengths or were independently observed by us 
before their identification was published. Our observations are thus 
presented here to confirm the nature of these objects and to improve 
their classification and the amount of information known about them.

\begin{table*}[th!]
\caption[]{Log of the spectroscopic observations presented in this paper
(see text for details).}
\scriptsize
%\hspace{-.3cm}
%\vspace{-1.3cm}
\begin{center}
\begin{tabular}{llllcccr}
\noalign{\smallskip}
\hline
\hline
\noalign{\smallskip}
\multicolumn{1}{c}{{\it (1)}} & \multicolumn{1}{c}{{\it (2)}} & \multicolumn{1}{c}{{\it (3)}} & \multicolumn{1}{c}{{\it (4)}} & 
{\it (5)} & {\it (6)} & {\it (7)} & \multicolumn{1}{c}{{\it (8)}} \\
\multicolumn{1}{c}{Object} & \multicolumn{1}{c}{RA} & \multicolumn{1}{c}{Dec} & 
\multicolumn{1}{c}{Telescope+instrument} & $\lambda$ range & Disp. & \multicolumn{1}{c}{UT Date \& Time}  & Exposure \\
 & \multicolumn{1}{c}{(J2000)} & \multicolumn{1}{c}{(J2000)} & & (\AA) & (\AA/pix) & 
\multicolumn{1}{c}{at mid-exposure} & time (s)  \\

\noalign{\smallskip}
\hline
\noalign{\smallskip}

IGR J03184$-$0014$^*$     & 03:18:17.54 & $-$00:17:50.2 & GTC+OSIRIS          & 3600-10000 & 5.2 & 05 Sep 2010, 06:25 & 3$\times$300  \\ % R300B,  1.23
IGR J03249+4041:          & & & & & & & \\
~~~~~~~LEDA 4678815       & 03:25:12.23 &   +40:42:02.2 & SPM 2.1m+B\&C Spec. & 3500-7800  & 4.0 & 01 Nov 2010, 11:33 & 2$\times$1800 \\ % 300,    2.5
~~~~~~~LEDA 97012         & 03:25:12.96 &   +40:41:52.8 & SPM 2.1m+B\&C Spec. & 3500-7800  & 4.0 & 01 Nov 2010, 10:28 & 2$\times$1800 \\ % 300,    2.5
IGR J05255$-$0711$^*$     & 05:25:09.77$^\dagger$ & $-$07:07:45.1$^\dagger$ & TNG+DOLoReS         & 3800-8000  & 2.5 & 09 Oct 2010, 03:27 & 3$\times$1200 \\ % LR-B,    2.0
RX J0525.3+2413           & 05:25:22.70 &   +24:13:33.3 & SPM 2.1m+B\&C Spec. & 3500-7800  & 4.0 & 31 Oct 2010, 10:24 & 2$\times$1800 \\ % 300,    2.5
IGR J06073$-$0024$^*$     & 06:06:57.43$^\dagger$ & $-$00:24:57.7$^\dagger$ & TNG+DOLoReS         & 3800-8000  & 2.5 & 01 Dec 2010, 00:19 & 3$\times$1500 \\ % LR-B,    2.0
IGR J06523+5334           & 06:52:31.40 &   +53:31:31.5 & SPM 2.1m+B\&C Spec. & 3500-7800  & 4.0 & 13 Mar 2010, 04:18 & 3$\times$1800 \\ % 300,    2.5
IGR J08262+4051           & 08:26:17.92$^\dagger$ &   +40:47:58.8$^\dagger$ & TNG+DOLoReS         & 3800-8000  & 2.5 & 01 Dec 2010, 02:07 & 3$\times$1800 \\ % LR-B,    2.0
IGR J12319$-$0749         & 12:31:57.71$^\dagger$ & $-$07:47:18.0$^\dagger$ & TNG+DOLoReS         & 3800-8000  & 2.5 & 01 Jan 2011, 05:14 & 840 \\ % LR-B,    1.5
IGR J12562+2554$^*$       & 12:56:10.43$^\dagger$ &   +26:01:03.7$^\dagger$ & GTC+OSIRIS & 3600-10000 & 5.2 & 26 Dec 2010, 05:07 & 2$\times$300 \\ % R300B,    1.23
IGR J13466+1921           & 13:46:28.43 &   +19:22:43.1 & SPM 2.1m+B\&C Spec. & 3500-7800  & 4.0 & 13 Mar 2010, 10:33 & 2$\times$1800 \\ % 300,    2.5
IGR J13550$-$7218         & 13:55:13.53 & $-$72:19:15.0 & CTIO 1.5m+RC Spec.  & 3300-10500 & 5.7 & 09 Feb 2011, 07:23 & 2$\times$1500 \\ % Gr13/I, 1.5
IGR J14385+8553$^*$       & 14:44:27.71 &   +86:00:56.5 & Copernicus+AFOSC    & 3500-7800  & 4.2 & 01 Mar 2011, 23:41 & 2$\times$1200 \\ % Gr4,    2.1
IGR J16388+3557$^*$       & 16:38:21.23$^\dagger$ &   +36:02:26.4$^\dagger$ & TNG+DOLoReS         & 3800-8000  & 2.5 & 1 Aug 2011, 00:09 & 2$\times$1800 \\ % LR-B,    1.5
IGR J16443+0131           & 16:44:37.93 &   +01:29:01.4 & TNG+DOLoReS         & 3800-8000  & 2.5 & 11 Jun 2011, 04:24 & 2$\times$1200 \\ % LR-B,   1.5
IGR J17197$-$3010$^*$     & 17:19:51.83 & $-$30:02:00.3 & SPM 2.1m+B\&C Spec. & 3500-7800  & 4.0 & 16 Jul 2010, 04:37 & 2$\times$1800 \\ % 300,    2.5
AX J1740.2$-$2903         & 17:40:16.11$^\dagger$ & $-$29:03:37.9$^\dagger$ & SPM 2.1m+B\&C Spec. & 3500-7800  & 4.0 & 17 Jul 2010, 06:05 & 2400 \\ % 300,    2.5
1RXS J175252.0$-$053210   & 17:52:52.47 & $-$05:32:06.4 & SPM 2.1m+B\&C Spec. & 3500-7800  & 4.0 & 15 Mar 2010, 11:57 & 2$\times$1800 \\ % 300,    2.5
IGR J18371+2634$^*$       & 18:37:28.88 &   +26:32:29.0 & Cassini+BFOSC       & 3500-8700  & 4.0 & 07 May 2011, 23:52 & 2$\times$1200 \\ % Gr4,    2.0
%                          & 18:37:28.35 &   +26:29:42.7 & Cassini+BFOSC       & 3500-8700  & 4.0 & 07 May 2011, 23:52 & 2$\times$1200 \\ % Gr4,    2.0
IGR J19173+0747           & 19:17:20.78 &   +07:47:50.7 & SPM 2.1m+B\&C Spec. & 3500-7800  & 4.0 & 14 Jul 2010, 08:22 & 2$\times$1800 \\ % 300,    2.5
IGR J19491$-$1035         & 19:49:09.27 & $-$10:34:25.0 & AAT+6dF             & 3900-7600  & 1.6 & 15 Jun 2004, 12:10 &      1200+600 \\ % 580V+425R, 6.7 fibre
IGR J20450+7530           & 20:44:34.41 &   +75:31:58.9 & Cassini+BFOSC       & 3500-8700  & 4.0 & 03 Aug 2010, 00:09 &          1800 \\ % Gr4,    2.0
IGR J23130+8608           & 23:08:59.80 &   +86:05:52.6 & TNG+DOLoReS         & 3800-8000  & 2.5 & 01 Sep 2010, 23:05 & 3$\times$1500 \\ % LR-B,    1.5

\noalign{\smallskip}
\hline
\noalign{\smallskip}
\multicolumn{8}{l}{Note: if not indicated otherwise, source coordinates were extracted from the 2MASS 
catalog and have an accuracy better than 0$\farcs$1.}\\
\multicolumn{8}{l}{$^*$: tentative association (see text).}\\
\multicolumn{8}{l}{$^\dagger$: coordinates extracted from the USNO catalogs, having 
an accuracy of about 0$\farcs$2 (Deutsch 1999; Assafin et al. 2001; Monet et al. 2003).}\\
\noalign{\smallskip}
\hline
\hline
\noalign{\smallskip}
\end{tabular}
\end{center}
\end{table*}

\section{Optical spectroscopy}

Analogously to Papers VI-VIII, almost all of the data presented in this work 
were collected in the course of a campaign that lasted more than one year 
(between March 2010 and August 2011) and that involved observations at the 
following telescopes:

\begin{itemize}
\item the 1.5m at the Cerro Tololo Interamerican Observatory (CTIO), Chile;
\item the 1.52m ``Cassini'' telescope of the Astronomical Observatory of 
Bologna, in Loiano, Italy; 
\item the 1.82m ``Copernicus'' telescope of the Astronomical Observatory of
Asiago, Italy;
\item the 2.1m telescope of the Observatorio Astron\'omico Nacional in San 
Pedro Martir (SPM), M\'exico;
\item the 3.58m ``Telescopio Nazionale Galileo" (TNG) at the Roque de 
Los Muchachos Observatory in La Palma, Spain; 
\item the 10.4m ``Gran Telescopio Canarias'' (GTC), again at the Roque de 
Los Muchachos Observatory in La Palma, Spain;
\end{itemize}

The spectroscopic data acquired at these telescopes have been optimally 
extracted (Horne 1986) and reduced following standard procedures using 
IRAF\footnote{IRAF is the Image Reduction and Analysis Facility made 
available to the astronomical community by the National Optical Astronomy 
Observatories, which are operated by AURA, Inc., under contract with the 
U.S. National Science Foundation. It is available at {\tt 
http://iraf.noao.edu/}}.  Calibration frames (flat fields and bias) were 
taken on the day preceeding or following the observing night. The 
wavelength calibration was performed using lamp data acquired soon after 
each on-target spectroscopic acquisiton; the uncertainty in this 
calibration was $\sim$0.5~\AA~in all cases according to our checks made 
using the positions of background night sky lines. Flux calibration was 
performed using catalogued spectrophotometric standards.

The spectrum of the optical counterpart of IGR J19491$-$1035 was
retrieved from the Six-degree Field Galaxy Survey\footnote{{\tt 
http://www.aao.gov.au/local/www/6df/}} (6dFGS) archive (Jones et al. 
2004), collected using the 3.9m Anglo-Australian Telescope of the
Anglo-Australian Observatory in Siding Spring (Australia). Since the 
6dFGS archive provides spectra that are not flux-calibrated, we used 
the optical photometric information in Jones et al. (2005) to calibrate 
the spectrum of this object.

In Table 1, we provide a detailed log of all the observations. Column 1 
indicates the names of the observed {\it INTEGRAL} sources. In Cols. 2 
and 3 we list the possible optical counterpart coordinates, extracted 
from the 2MASS (with an accuracy of $\leq$0$\farcs$1, according to 
Skrutskie et al. 2006) or USNO catalogs (with uncertainties of 
about 0$\farcs$2: Deutsch 1999; Assafin et al. 2001; Monet et al. 2003).
The telescope and instrument used for the observations are reported in 
Col. 4, while characteristics of each spectrograph are presented in 
Cols. 5 and 6. Column 7 reports the observation date and the UT time at 
mid-exposure, and Col. 8 provides the exposure times and the number of 
observations for each source.

For the source naming in Table 1, we adopted the names as they are reported
in the relevant surveys (Bird et al. 2007, 2010; Krivonos et al. 2010) or 
papers (Pavan et al. 2011), and the ``IGR" alias when available.

\section{Results}

In the following, we give a description of the adopted identification and 
classification criteria for the optical spectra of the selected sources. 
They are basically the same as in our previous papers (I-VIII); however, 
for the sake of clarity we briefly list them again. The optical magnitudes 
quoted below, if not stated otherwise, are extracted from the USNO-A2.0 
catalog\footnote{available at: \\ {\tt 
http://archive.eso.org/skycat/servers/usnoa}}.

To evaluate the reddening along the line of sight for the Galactic 
sources in our sample, when possible and applicable, we considered an 
intrinsic H$_\alpha$/H$_\beta$ line ratio of 2.86 (Osterbrock 1989) and 
inferred the corresponding color excess by comparing the intrinsic line 
ratio with the measured one by applying the Galactic extinction law of 
Cardelli et al. (1989).

To determine the distances of the compact Galactic X--ray sources of our 
sample, for Cataclysmic Variables (CVs) we assumed an absolute magnitude 
M$_V \sim$ +9 and an intrinsic color index $(V-R)_0 \sim$ 0 mag (Warner 1995), 
whereas for the single high-mass X--ray binary (HMXB) and symbiotic star of 
our sample we used the intrinsic stellar color indices and absolute magnitudes 
reported in Lang (1992), Wegner (1994) and Ducati et al. (2001). 
For the active stars identified in our selection of sources we assumed 
similarity with object II Peg, as suggested in Rodriguez et al. (2010a).
Although these methods basically provide an order-of-magnitude value for the 
distance of Galactic sources, our past experience (Papers I-VIII) tells us 
that these estimates are in general correct to within 50\% of the refined 
value subsequently determined with more precise approaches.

For the classification of active galactic nuclei (AGNs), we used 
the criteria of Veilleux \& Osterbrock (1987) and the line ratio 
diagnostics of both Ho et al. (1993, 1997) and Kauffmann et al. (2003) 
and, for assigning the subclasses of Seyfert 1 galaxies, we used the 
H$_\beta$/[O {\sc iii}]$\lambda$5007 line flux ratio criterion described 
in Winkler (1992).

The AGN spectra shown here were not corrected for starlight contamination 
(see, e.g., Ho et al. 1993, 1997) because of the limited S/N and spectral 
resolution. We do not expect this to affect any of our main results and 
conclusions.

In the following, we consider a cosmology with $H_{\rm 0}$ = 65 km 
s$^{-1}$ Mpc$^{-1}$, $\Omega_{\Lambda}$ = 0.7, and $\Omega_{\rm m}$ = 0.3; 
the luminosity distances of the extragalactic objects reported in this 
paper were computed for these parameters using the Cosmology Calculator of 
Wright (2006). When not explicitly stated otherwise, for our X--ray flux 
estimates we assume a Crab-like spectrum except for the {\it XMM-Newton} 
sources, for which we considered the fluxes reported in Saxton et al. 
(2008) or in Watson et al. (2009). The X--ray luminosities reported 
in Tables 2, 3, 4, 6, 7 and 8 are associated with a letter indicating the 
satellite and/or the instrument with which the measurement of the 
corresponding X--ray flux was obtained, namely {\it ASCA} ({\it A}), 
{\it Swift}/BAT ({\it B}), {\it Chandra} ({\it C}), {\it INTEGRAL} 
({\it I}), {\it XMM-Newton} ({\it N}), {\it ROSAT} ({\it R}), and 
{\it Swift}/XRT ({\it X}).

Hereafter, we present the object identifications by dividing them into three 
broad classes (AGNs, accreting binaries, and other Galactic objects).

\subsection{AGNs}

\begin{table*}%[th!]
\caption[]{Synoptic table containing the main results for the 8
low-$z$ broad emission-line AGNs (Fig. 4) identified or observed in the 
present sample of {\it INTEGRAL} sources.}
\scriptsize
\begin{center}
%\hspace{-1cm}
\begin{tabular}{lcccccrcr}
\noalign{\smallskip}
\hline
\hline
\noalign{\smallskip}
\multicolumn{1}{c}{Object} & $F_{\rm H_\alpha}$ & $F_{\rm H_\beta}$ &
$F_{\rm [OIII]}$ & Class & $z$ &
\multicolumn{1}{c}{$D_L$ (Mpc)} & $E(B-V)_{\rm Gal}$ & \multicolumn{1}{c}{$L_{\rm X}$} \\
%\cline{8-9}
\noalign{\smallskip}
\hline
\noalign{\smallskip}

IGR J03184$-$0014 & * & $<$0.05 & 0.11$\pm$0.02 & Sy1.9 & 0.330 & 1867.2 & 0.063 & 6.3 (2--10; {\it X}) \\
 & * & [$<$0.07] & [0.14$\pm$0.02] & & & & & 1200 (20--40; {\it I}) \\
 & & & & & & & & $<$880 (40--100; {\it I}) \\

 & & & & & & & & \\

IGR J06523+5334 & --- & 10.0$\pm$1.0 & 3.6$\pm$0.3 & Sy1.2 & 0.301 & 1678.6 & 0.062 & 5.4 (0.1--2.4; {\it R}) \\
 & --- & [11.8$\pm$1.2] & [4.2$\pm$0.4] & & & & & 5.7 (2--10; {\it X}) \\
 & & & & & & & & 740 (20--40; {\it I}) \\
 & & & & & & & & $<$870 (40--100; {\it I}) \\

 & & & & & & & & \\

IGR J13466+1921 & * & 39$\pm$4 & 14.6$\pm$1.0 & Sy1.2 & 0.085 & 417.1 & 0.024 & 0.73 (0.1--2.4; {\it R}) \\
 & * & [42$\pm$4] & [15.5$\pm$1.1] & & & & & 33 (17--60; {\it I}) \\
 & & & & & & & & 23 (14--195; {\it B}) \\

 & & & & & & & & \\

IGR J14385+8553 & * & 11.5$\pm$1.7 & 7.4$\pm$0.7 & Sy1.5 & 0.081 & 396.4 & 0.191 & 6.9 (0.2--12; {\it N}) \\
 & * & [20$\pm$3] & [12.4$\pm$1.2] & & & & & $<$10 (20--40; {\it I}) \\
 & & & & & & & & $<$21 (40--100; {\it I}) \\

 & & & & & & & & \\

IGR J16443+0131 & --- & 1.2$\pm$0.2 & 1.60$\pm$0.16 & Sy1.5 & 0.342 & 1946.5 & 0.071 & 12 (2--10; {\it X}) \\
 & --- & [1.8$\pm$0.3] & [1.9$\pm$0.2] & & & & & 340 (20--40; {\it I}) \\
 & & & & & & & & $<$430 (40--100; {\it I}) \\

 & & & & & & & & \\

1RXS J175252.0$-$053210 & * & 20$\pm$2 & 6.0$\pm$1.0 & Sy1.2 & 0.136 & 690.2 & 0.995 & 2.7 (0.1--2.4; {\it R}) \\
 & * & [370$\pm$40] & [100$\pm$20] & & & & & 57 (20--100; {\it I}) \\

 & & & & & & & & \\

IGR J19491$-$1035 & * & 220$\pm$20 & 510$\pm$50 & Sy1.2 & 0.0240 & 112.7 & 0.238 & 1.2 (2--10; {\it X}) \\
 & * & [480$\pm$50] & [1000$\pm$100] & & & & & 1.7 (17--60; {\it I}) \\

 & & & & & & & & \\

IGR J20450+7530 & * & 11$\pm$2 & $<$3 & Sy1 & 0.095 & 469.3 & 0.382 & 0.84 (2--10; {\it X}) \\
 & * & [35$\pm$8] & [$<$10] & & & & & 53 (20--40; {\it I}) \\
 & & & & & & & & $<$42 (40--100; {\it I}) \\

\noalign{\smallskip} 
\hline
\noalign{\smallskip} 
\multicolumn{9}{l}{Note: emission-line fluxes are reported both as 
observed and (between square brackets) corrected for the intervening Galactic} \\ 
\multicolumn{9}{l}{absorption $E(B-V)_{\rm Gal}$ along the object line of sight 
(from Schlegel et al. 1998). Line fluxes are in units of 10$^{-15}$ erg cm$^{-2}$ s$^{-1}$,} \\
\multicolumn{9}{l}{whereas X--ray luminosities are in units of 10$^{43}$ erg s$^{-1}$ 
and the reference band (between round brackets) is expressed in keV.} \\ 
\multicolumn{9}{l}{In the last column, the upper-case letter indicates the satellite 
and/or the instrument with which the} \\
\multicolumn{9}{l}{corresponding X--ray flux measurement was obtained (see text).} \\
\multicolumn{9}{l}{The typical error in the redshift measurement is $\pm$0.001 
but for the 6dFGS spectrum, for which an uncertainty} \\
\multicolumn{9}{l}{of $\pm$0.0003 can be assumed.} \\
\multicolumn{9}{l}{$^*$: heavily blended with [N {\sc ii}] lines} \\
\noalign{\smallskip} 
\hline
\hline
\end{tabular} 
\end{center} 
\end{table*}

\begin{table*}%[th!]
\caption[]{Synoptic table containing the main results for the 6 
high-redshift broad-line QSOs ($z >$ 0.5; Fig. 5) identified in the 
present sample of {\it INTEGRAL} sources.}
\scriptsize
\begin{center}
%\hspace{-1cm}
\begin{tabular}{lcccccrcr}
\noalign{\smallskip}
\hline
\hline
\noalign{\smallskip}
\multicolumn{1}{c}{Object} & $F_{\rm H_\beta}$ & $F_{\rm MgII}$ &
$F_{\rm C III]}$ & $F_{\rm C IV}$ & $z$ &
\multicolumn{1}{c}{$D_L$ (Mpc)} & $E(B-V)_{\rm Gal}$ & \multicolumn{1}{c}{$L_{\rm X}$} \\
%\cline{8-9}
\noalign{\smallskip}
\hline
\noalign{\smallskip}

IGR J05255$-$0711 & 4.3$\pm$0.4 & 5.5$\pm$0.5 & --- & --- & 0.634 & 4065.0 & 0.113 & 32 (2--10; {\it X}) \\
 & 4.9$\pm$0.5 & 7.4$\pm$0.7 & --- & --- & & & & 4000 (20--40; {\it I}) \\
 & & & & & & & & $<$4700 (40--100; {\it I}) \\

 & & & & & & & & \\

IGR J06073$-$0024 & --- & 2.2$\pm$0.2 & 1.9$\pm$0.4 & --- & 1.028 & 7362.5 & 0.438 & 7800 (20--40; {\it I}) \\
 & --- & [7.2$\pm$0.7] & [12$\pm$3] & --- & & & & $<$14000 (40--100; {\it I}) \\
 & & & & & & & & \\

IGR J08262+4051 & --- & 1.65$\pm$0.17 & 2.4$\pm$0.2 & --- & 1.038 & 7451.1 & 0.042 & 6600 (20--40; {\it I}) \\
 & --- & [1.90$\pm$0.19] & [2.6$\pm$0.3] & & & & & $<$11000 (40--100; {\it I}) \\

 & & & & & & & & \\

IGR J12319$-$0749 & --- & --- & --- & 0.9$\pm$0.3 & 3.12 & 28692.5 & 0.023 & 5300 (0.1--2.4; {\it R}) \\
 & --- & --- & --- & [1.0$\pm$0.3] & & & & 44000 (20--40; {\it I}) \\
 & & & & & & & & $<$46000 (40--100; {\it I}) \\

 & & & & & & & & \\

IGR J12562+2554 & --- & 6.1$\pm$0.3 & 3.0$\pm$0.4 & --- & 1.199 & 8906.3 & 0.010 & $\approx$100 (0.1--2; {\it R}) \\
 & --- & [6.3$\pm$0.3] & [3.2$\pm$0.4] & & & & & 250 (0.2--12; {\it N}) \\
 & & & & & & & & 240 (0.3--10; {\it C}) \\
 & & & & & & & & 12000 (20--100; {\it I}) \\

 & & & & & & & & \\

IGR J16388+3557 & --- & 4.5$\pm$0.5 & --- & --- & 0.675 & 4387.7 & 0.021 & 5100 (20--100; {\it I}) \\
 & --- & [5.0$\pm$0.5] & --- & --- & & & & \\

\noalign{\smallskip} 
\hline
\noalign{\smallskip} 
\multicolumn{9}{l}{Note: emission-line fluxes are reported both as 
observed and (between square brackets) corrected for the intervening Galactic} \\ 
\multicolumn{9}{l}{absorption $E(B-V)_{\rm Gal}$ along the object line of sight 
(from Schlegel et al. 1998). Line fluxes are in units of 10$^{-15}$ erg cm$^{-2}$ s$^{-1}$,} \\
\multicolumn{9}{l}{whereas X--ray luminosities are in units of 10$^{43}$ erg s$^{-1}$ 
and the reference band (between round brackets) is expressed in keV.} \\
\multicolumn{9}{l}{In the last column, the upper-case letter indicates the satellite and/or the 
instrument with which the} \\
\multicolumn{9}{l}{corresponding X--ray flux measurement was obtained (see text).} \\
 \multicolumn{9}{l}{The typical error in the redshift measurement is $\pm$0.001 
but for the spectrum of IGR J12319$-$0749, for which an uncertainty} \\
\multicolumn{9}{l}{of $\pm$0.01 can be assumed.} \\
\noalign{\smallskip} 
\hline
\hline
\end{tabular} 
\end{center} 
\end{table*}

\begin{table*}%[t!]
\caption[]{Synoptic table containing the main results for the 3
narrow emission-line AGNs (Fig. 6) identified or observed in the present 
sample of {\it INTEGRAL} sources.}
\scriptsize
%\vspace{-.3cm}
\begin{center}
%\hspace{-.5cm}
\begin{tabular}{lcccccrccr}
\noalign{\smallskip}
\hline
\hline
\noalign{\smallskip}
\multicolumn{1}{c}{Object} & $F_{\rm H_\alpha}$ & $F_{\rm H_\beta}$ & $F_{\rm [OIII]}$ & Class & $z$ &
\multicolumn{1}{c}{$D_L$} & \multicolumn{2}{c}{$E(B-V)$} &
\multicolumn{1}{c}{$L_{\rm X}$} \\
\cline{8-9}
\noalign{\smallskip}
 & & & & & & (Mpc) & Gal. & AGN & \\
\noalign{\smallskip}
\hline
\noalign{\smallskip}

IGR J03249+4041 &  &  &  & & & & & & 14 (14--150; {\it B}) \\
 & & & & & & & & & 14 (14--195; {\it B}) \\
 & & & & & & & & & 8.5 (17--60; {\it I}) \\

(LEDA 4678815) & 30.4$\pm$1.5 & 3.9$\pm$0.6 & 27.2$\pm$1.4 & Sy2 & 0.049 & 234.4 & 0.206 & 0.77 & 2.8 (0.2--12; {\it N}) \\
               & [48$\pm$3] & [7.8$\pm$1.2] & [51$\pm$3] & & & & & & \\

(LEDA 97012)   & 29$\pm$3 & 1.3$\pm$0.3 & 17.3$\pm$1.2 & Sy2 & 0.049 & 234.4 & 0.206 & 1.98 & 0.059 (0.2--12; {\it N}) \\
               & [46$\pm$5] & [2.3$\pm$0.6] & [33$\pm$2] & & & & & & \\

 & & & & & & & & & \\

IGR J13550$-$7218 & 26$\pm$2 & 8.7$\pm$0.9 & 77$\pm$2 & Sy2 & 0.071 & 345.0 & 0.360 & 0 & 3.0 (0.5--10; {\it X}) \\
 & [56$\pm$5] & [25$\pm$4] & [218$\pm$7] & & & & & & 21 (20--100; {\it I}) \\

\noalign{\smallskip} 
\hline
\noalign{\smallskip} 
\multicolumn{10}{l}{Note: emission-line fluxes are reported both as 
observed and (between square brackets) corrected for the intervening Galactic} \\ 
\multicolumn{10}{l}{absorption $E(B-V)_{\rm Gal}$ along the object line of sight 
(from Schlegel et al. 1998). Line fluxes are in units of 10$^{-15}$ erg cm$^{-2}$ s$^{-1}$,} \\
\multicolumn{10}{l}{whereas X--ray luminosities are in units of 10$^{43}$ erg s$^{-1}$ 
and the reference band (between round brackets) is expressed in keV.} \\ 
\multicolumn{10}{l}{In the last column, the upper-case letter indicates the satellite and/or the 
instrument with which the} \\
\multicolumn{10}{l}{corresponding X--ray flux measurement was obtained (see text).} \\
\noalign{\smallskip} 
\hline
\hline
\end{tabular}
\end{center}
\end{table*}

\begin{figure*}%[th!]
%\begin{center}
%\hspace{.1cm}
%\hspace{-.8cm}
%\centering{
\mbox{\psfig{file=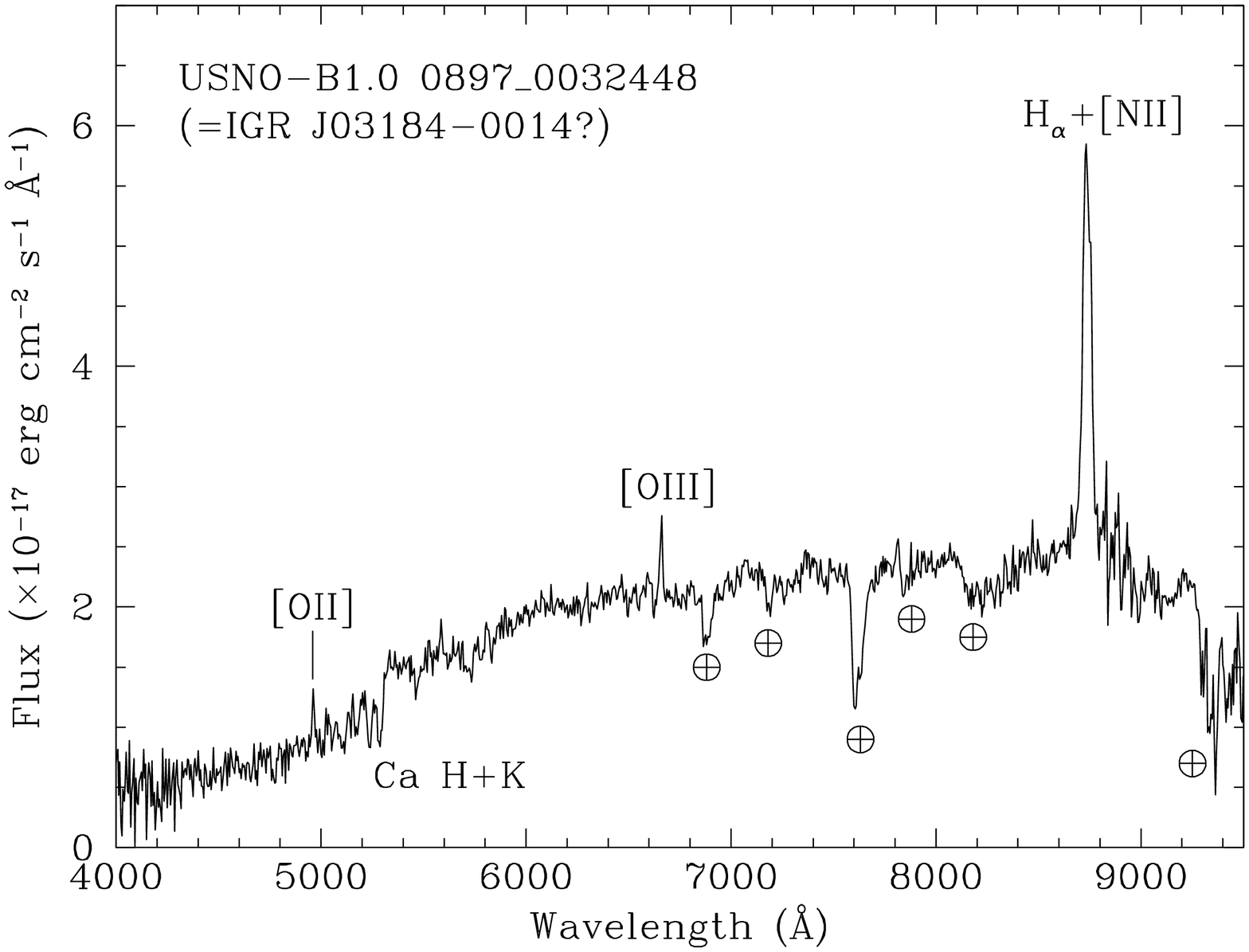,width=9cm,angle=270}}%}
%\hspace{-1.0cm}
%\centering{
\mbox{\psfig{file=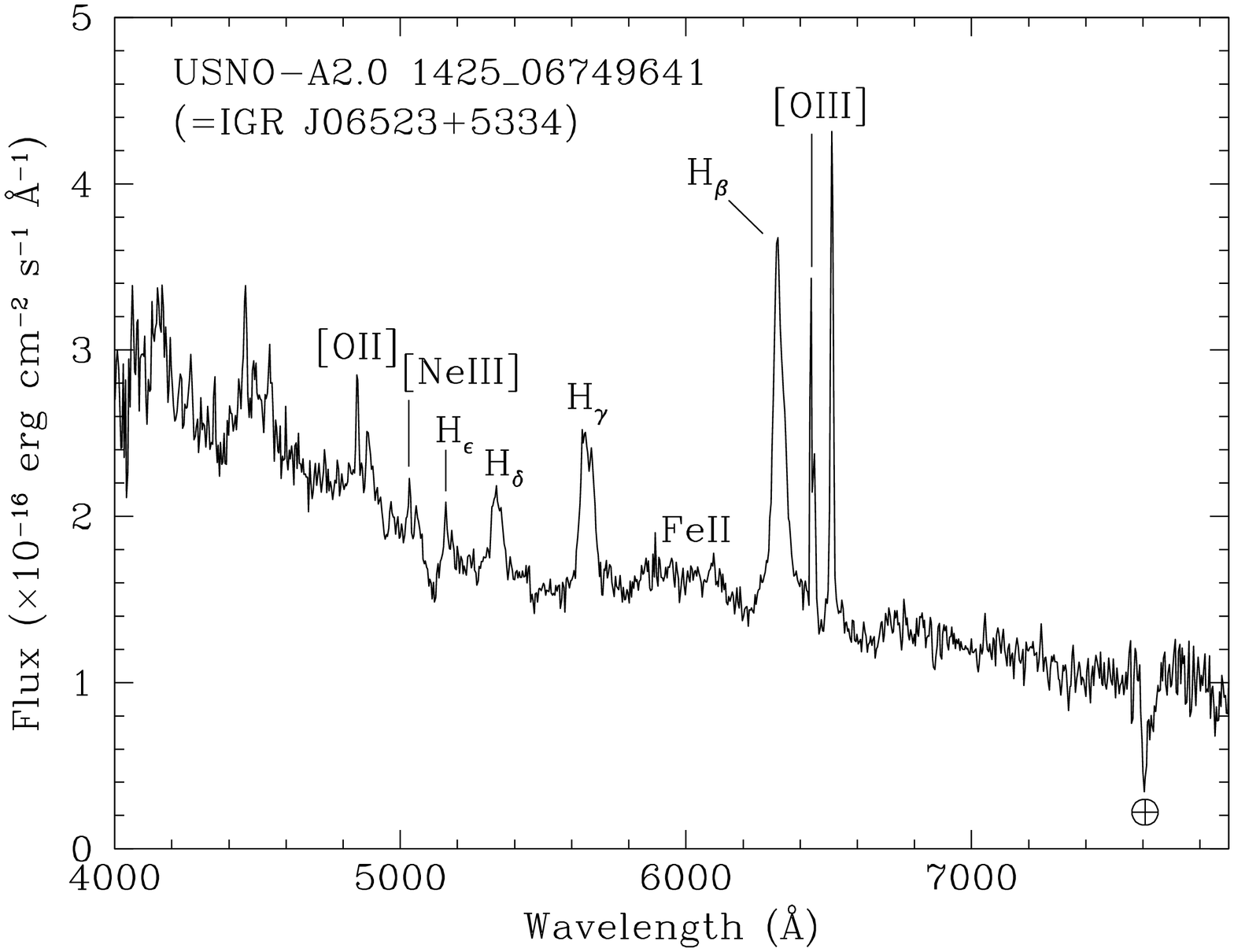,width=9cm,angle=270}}%}

\vspace{-.9cm}
%\hspace{-.8cm}
%\centering{
%\mbox{\hspace{8.9cm}}
\mbox{\psfig{file=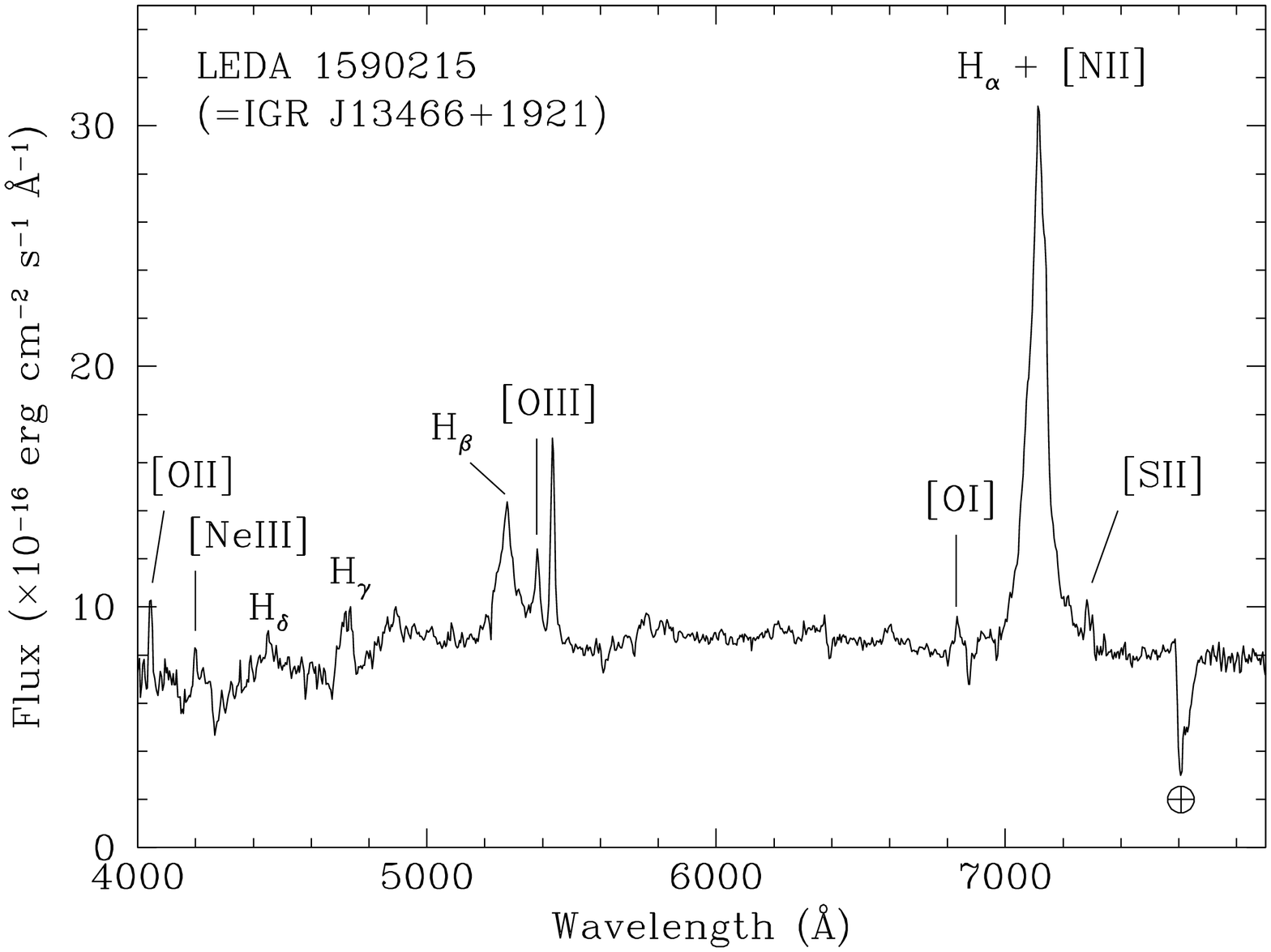,width=9cm,angle=270}}%}
%\hspace{-1cm}
%\centering{
\mbox{\psfig{file=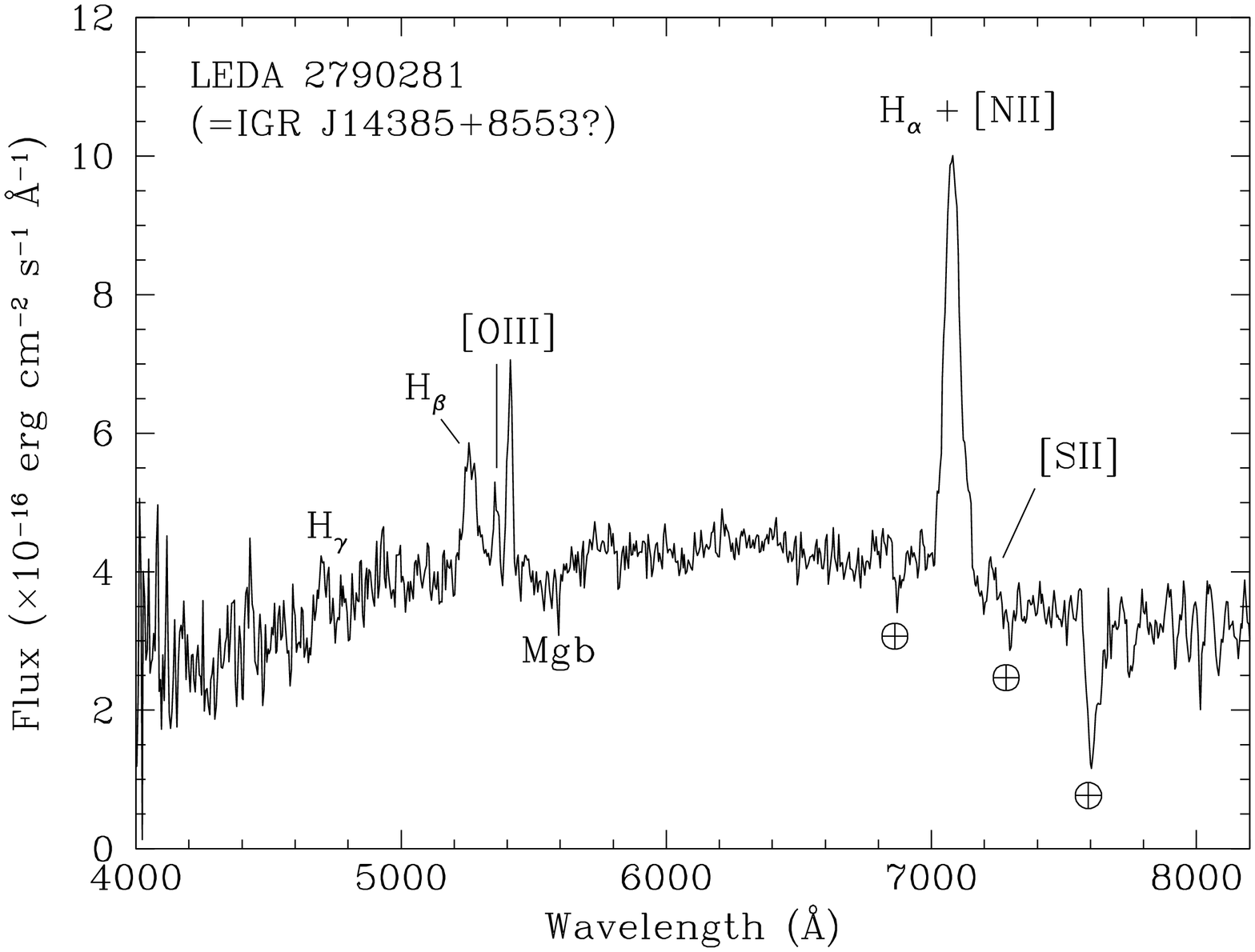,width=9cm,angle=270}}%}

\vspace{-.9cm}
%\hspace{-.8cm}
%\centering{
%\hspace{10cm}
\mbox{\psfig{file=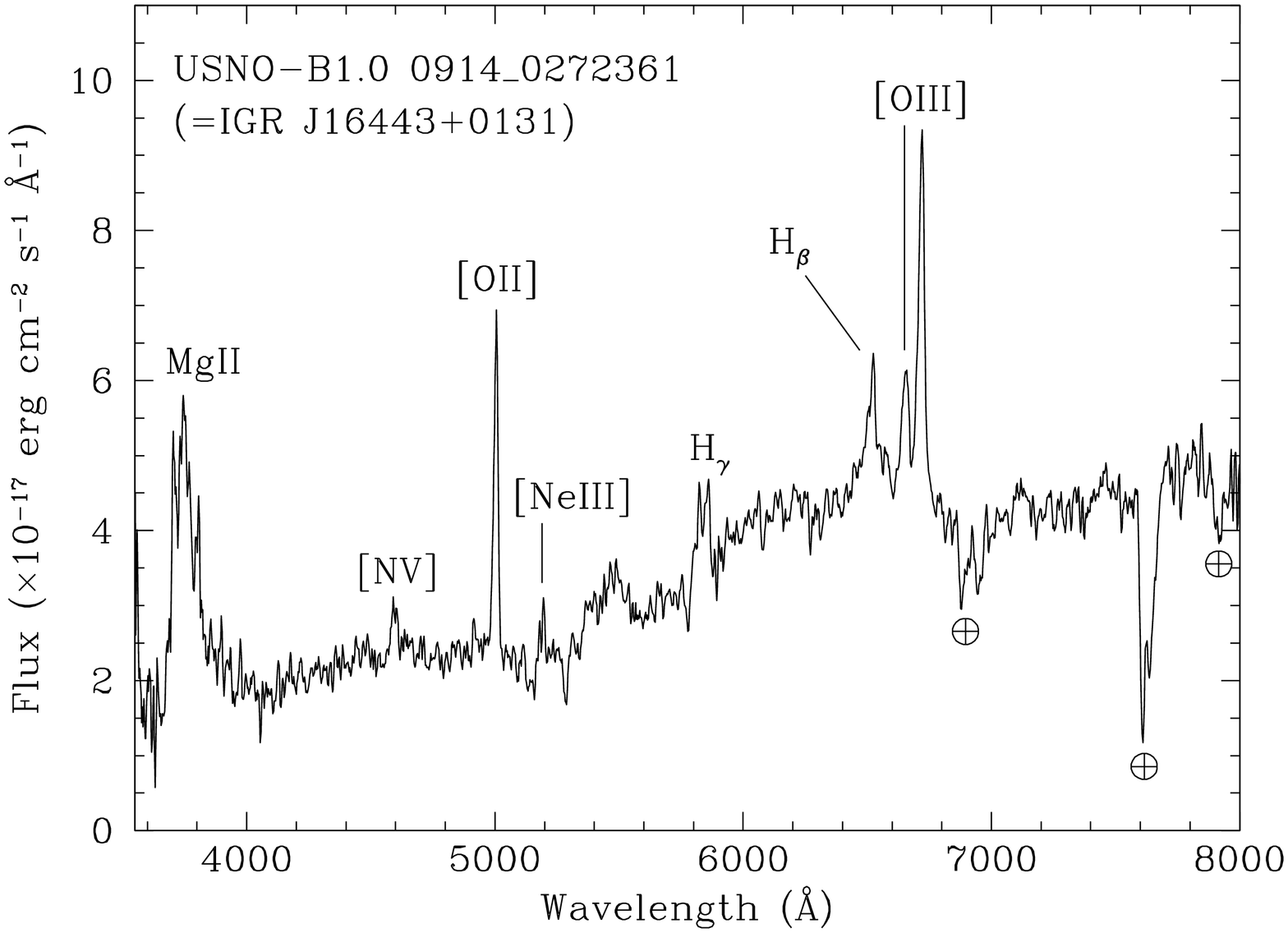,width=9cm,angle=270}}%}
%\hspace{-1cm}
%\centering{
\mbox{\psfig{file=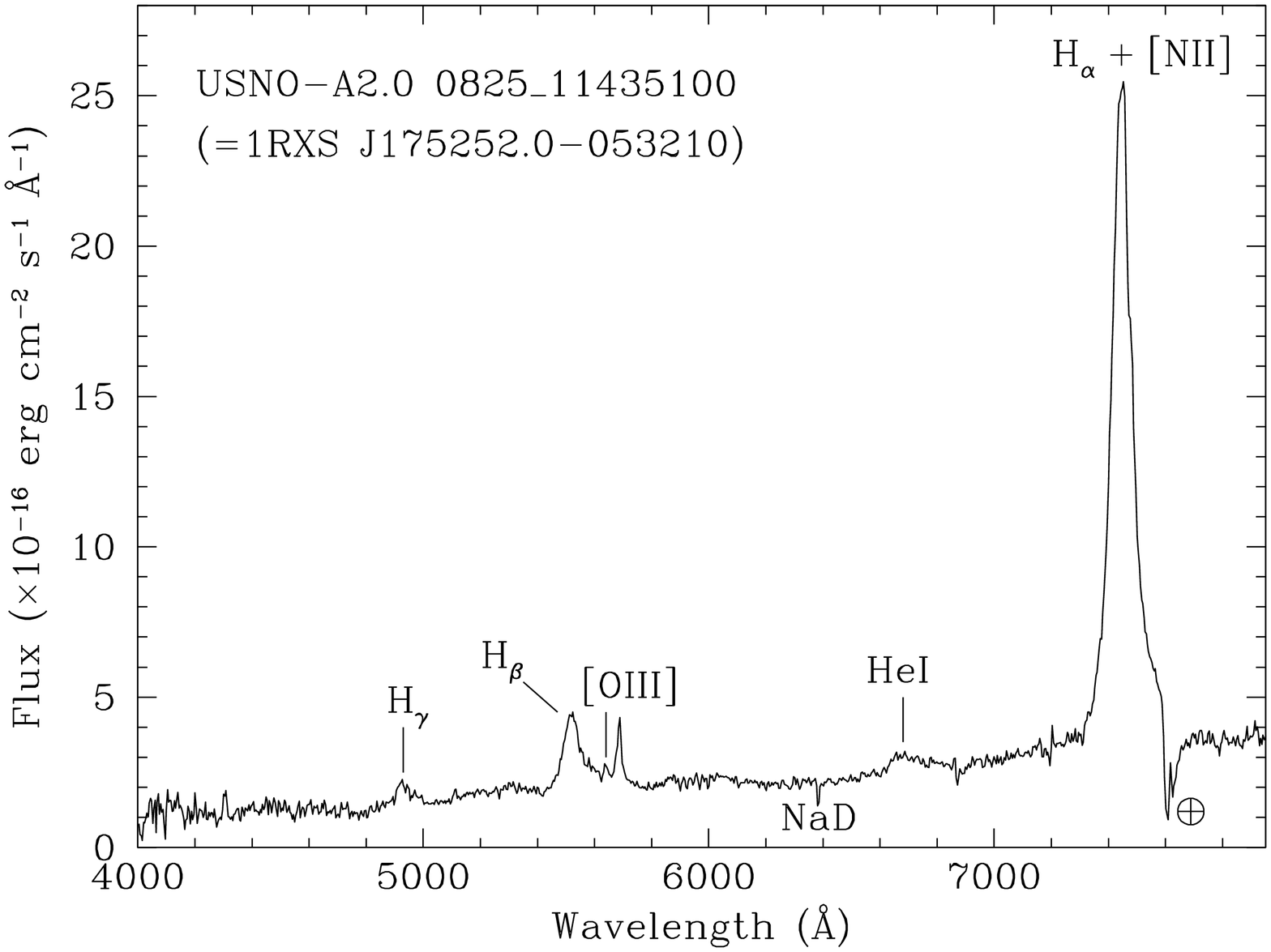,width=9cm,angle=270}}%}
%\begin{center}

\vspace{-.9cm}
%\hspace{-.8cm}
%\centering{
%\hspace{10cm}
\mbox{\psfig{file=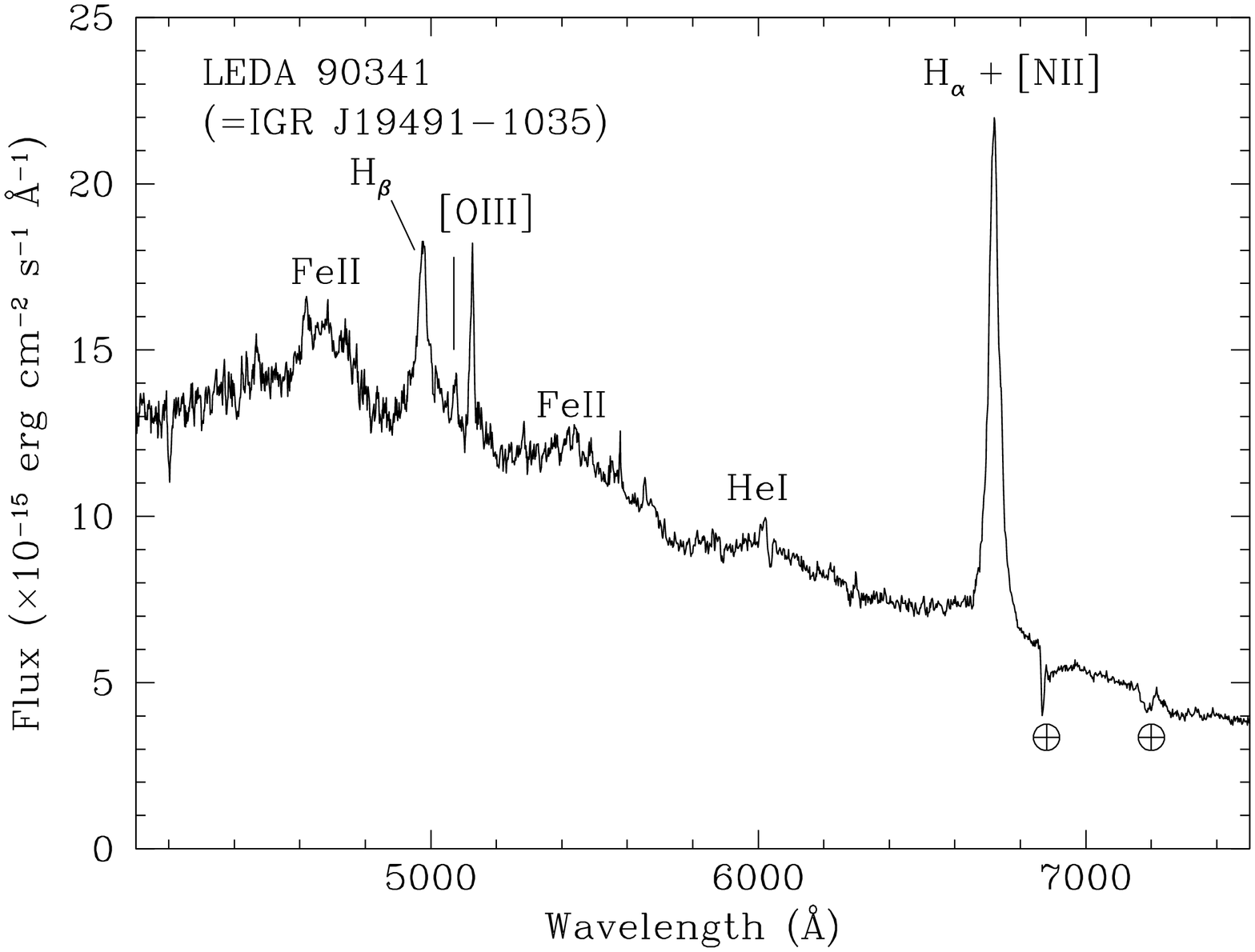,width=9cm,angle=270}}%}
%\hspace{-1cm}
%\centering{
\mbox{\psfig{file=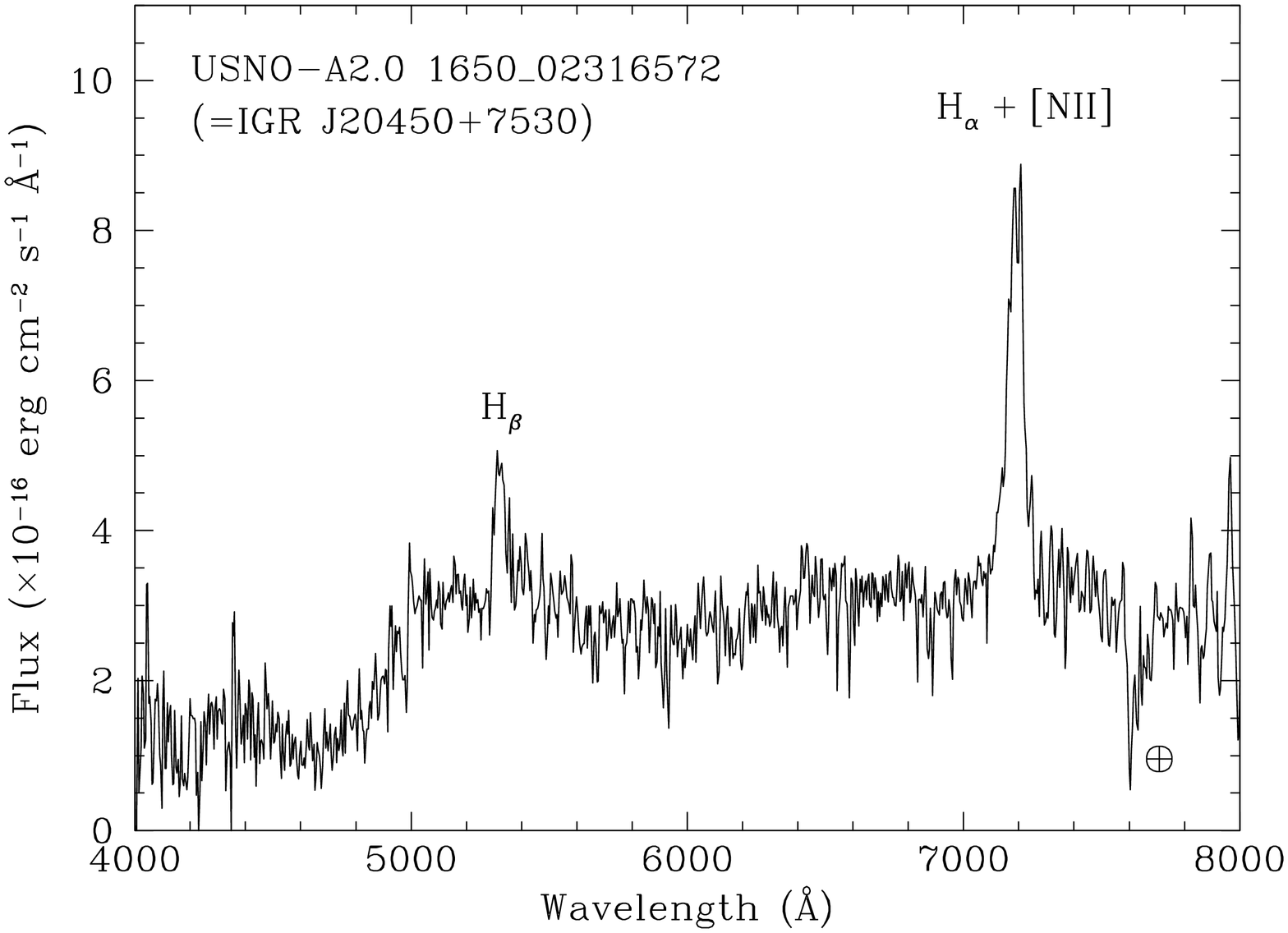,width=9cm,angle=270}}%}
%\begin{center}

%\vspace{1.5cm}
\caption{Spectra (not corrected for the intervening Galactic absorption) 
of the optical counterparts of the 8 low-redshift, broad emission-line 
AGNs belonging to the sample of {\it INTEGRAL} sources presented in 
this paper. For each spectrum, the main spectral features are labeled. 
The symbol $\oplus$ indicates atmospheric telluric absorption bands.
The TNG spectrum has been smoothed using a Gaussian filter with
$\sigma$ = 3 \AA.}
%\end{center}
\end{figure*}

\begin{figure*}%[th!]
%\begin{center}
%\hspace{.1cm}
%\hspace{-.8cm}
%\centering{
\mbox{\psfig{file=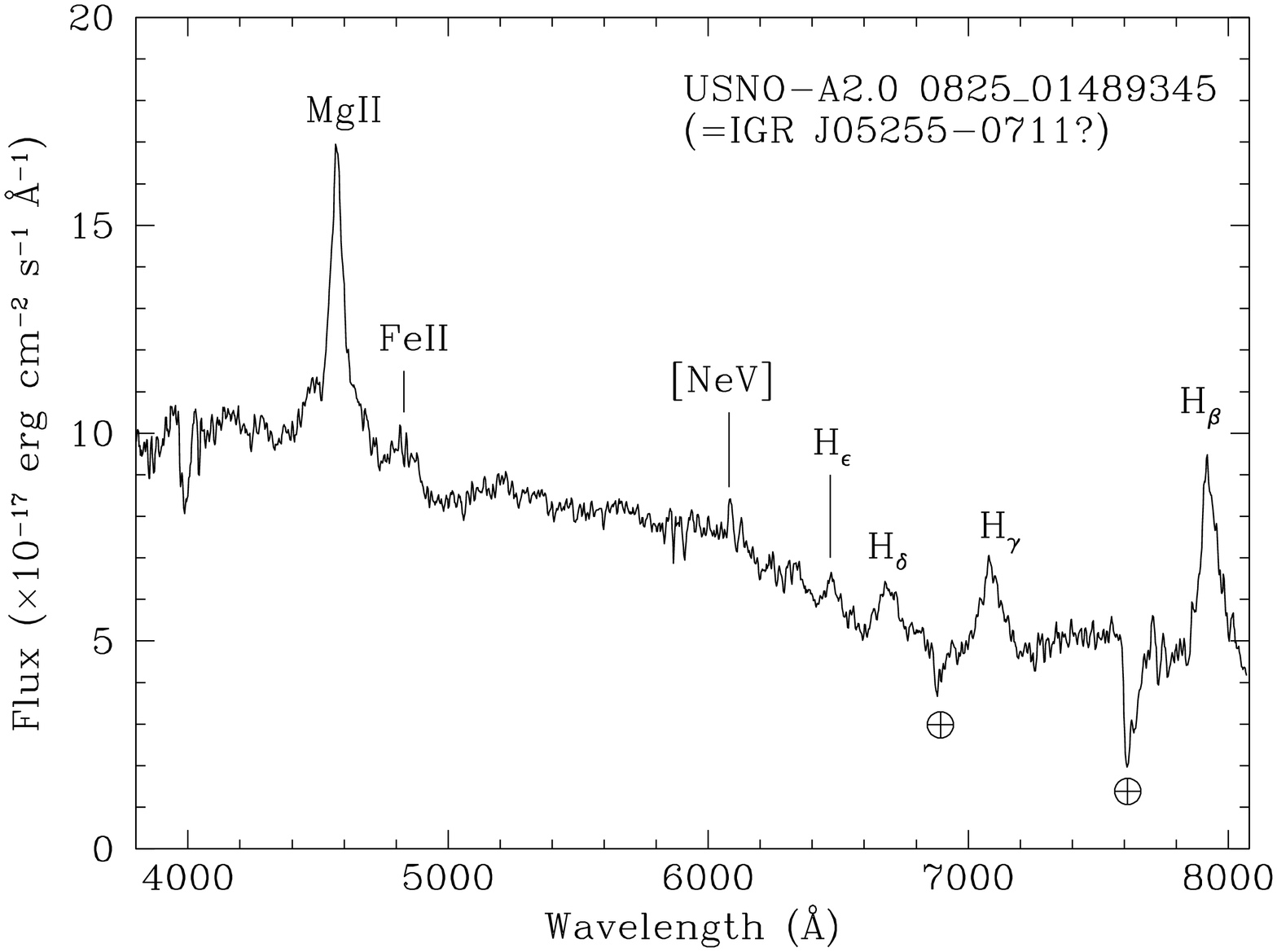,width=9cm,angle=270}}%}
%\hspace{-1.0cm}
%\centering{
\mbox{\psfig{file=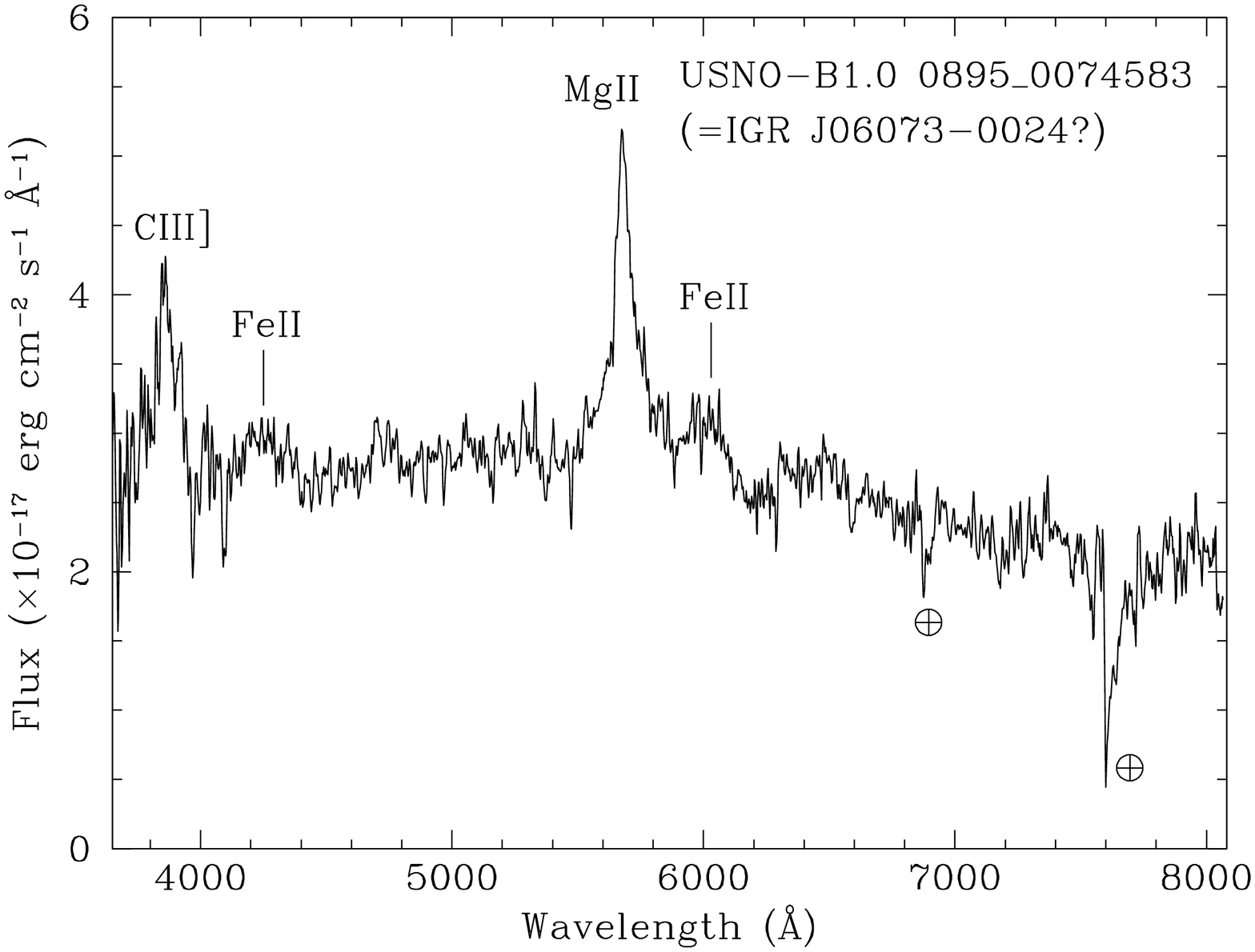,width=9cm,angle=270}}%}

\vspace{-.9cm}
%\hspace{-.8cm}
%\centering{
%\mbox{\hspace{8.9cm}}
\mbox{\psfig{file=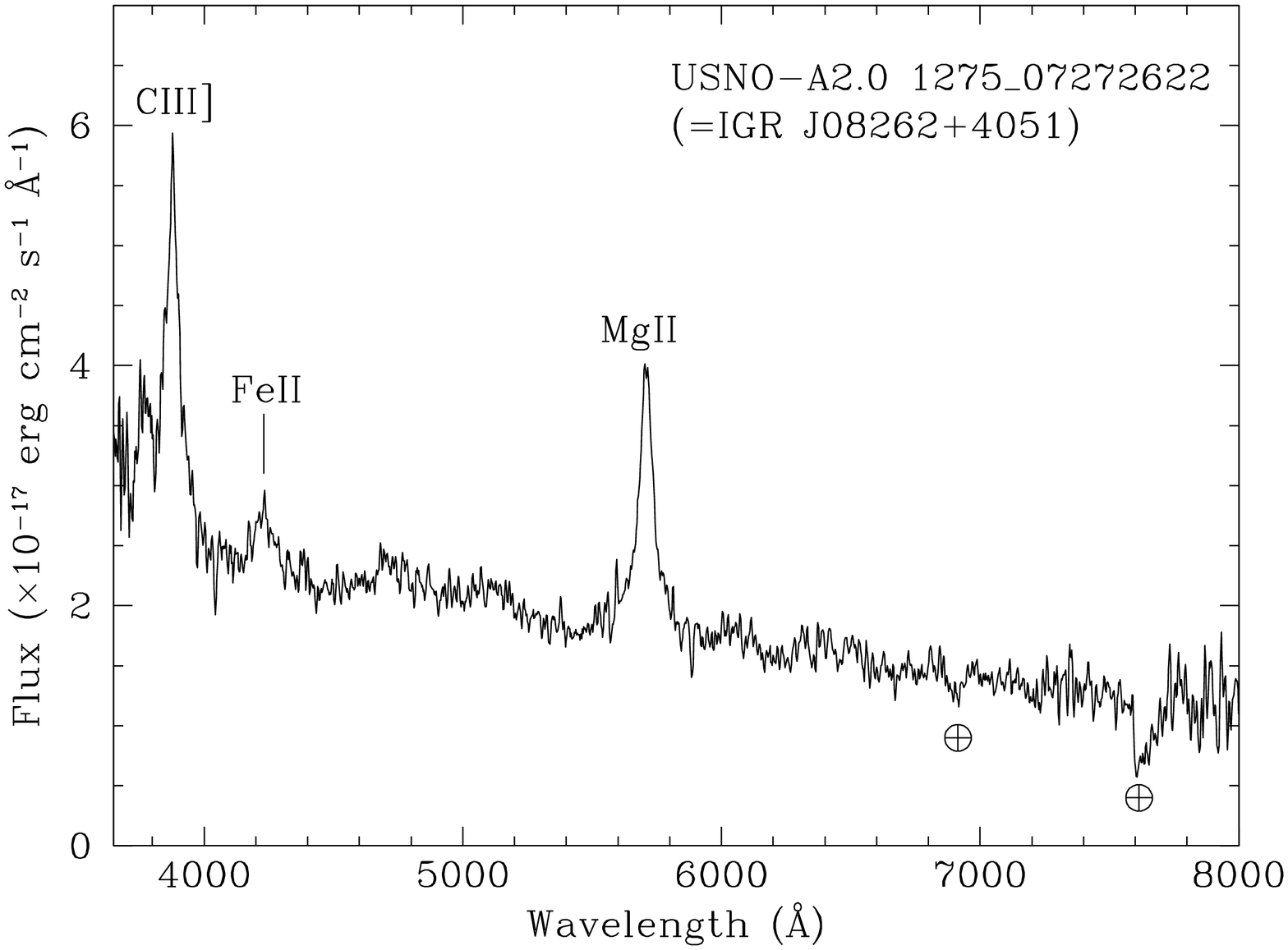,width=9cm,angle=270}}%}
%\hspace{-1cm}
%\centering{
\mbox{\psfig{file=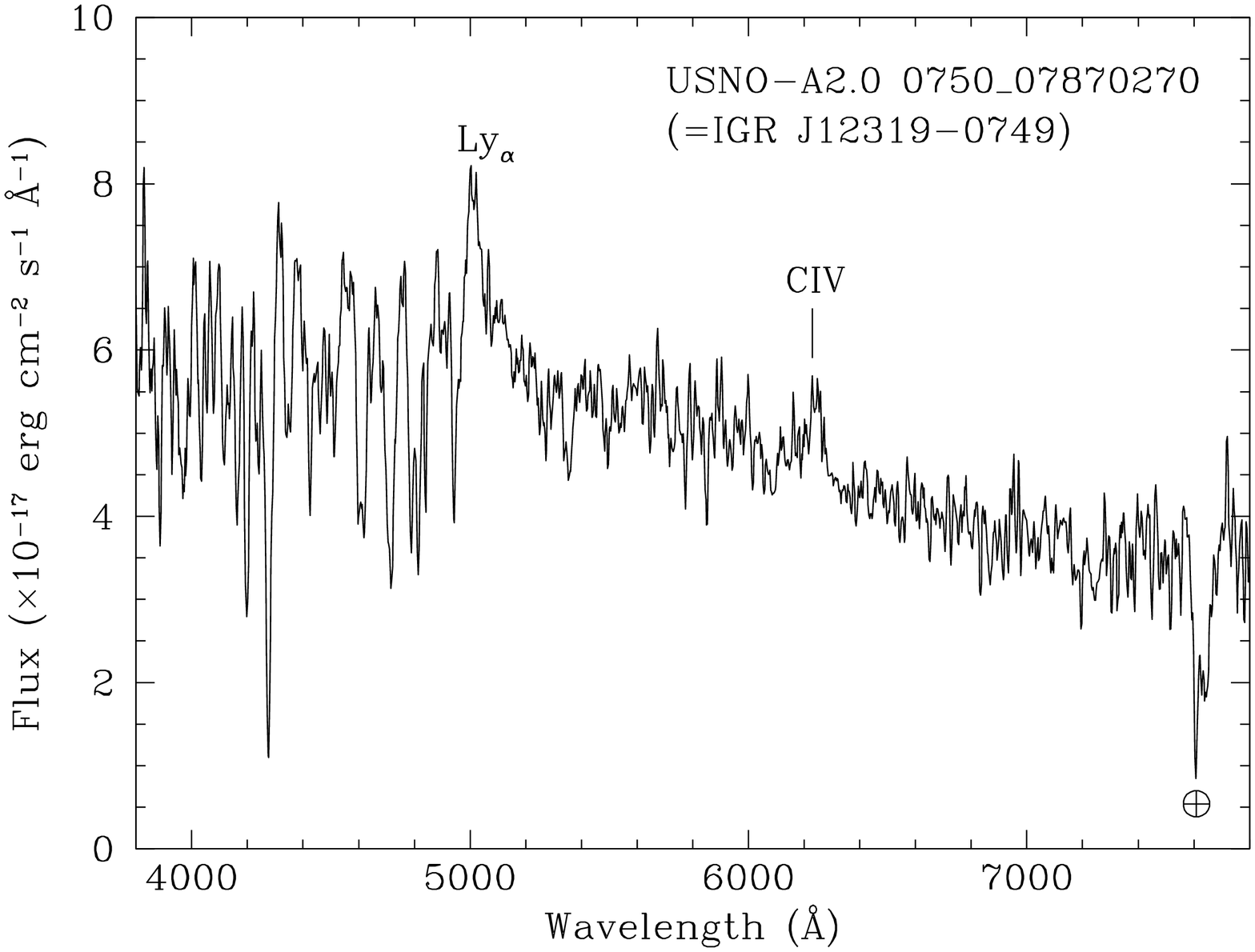,width=9cm,angle=270}}%}

\vspace{-.9cm}
%\hspace{-.8cm}
%\centering{
%\hspace{10cm}
\mbox{\psfig{file=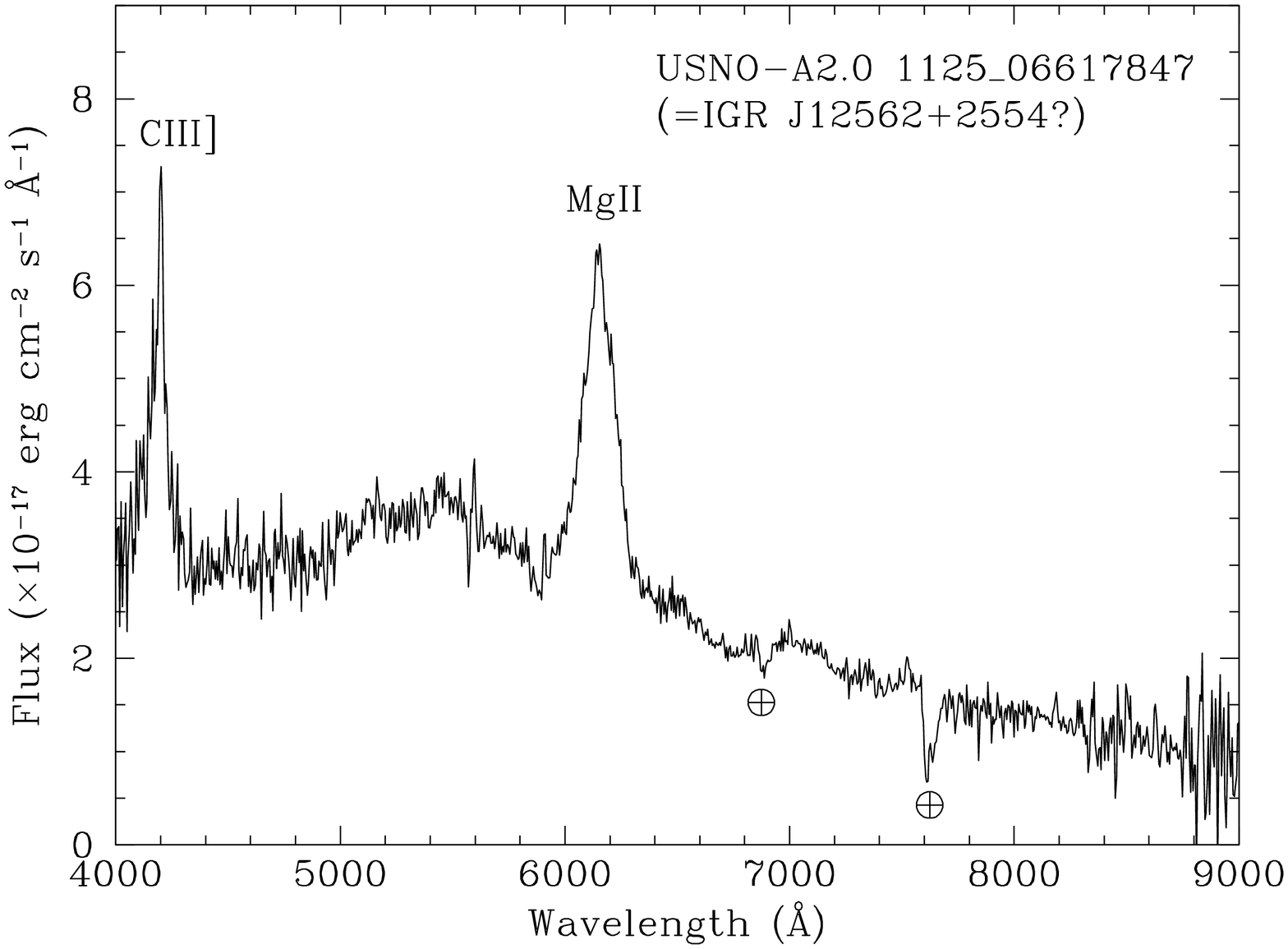,width=9cm,angle=270}}%}
%\hspace{-1cm}
%\centering{
\mbox{\psfig{file=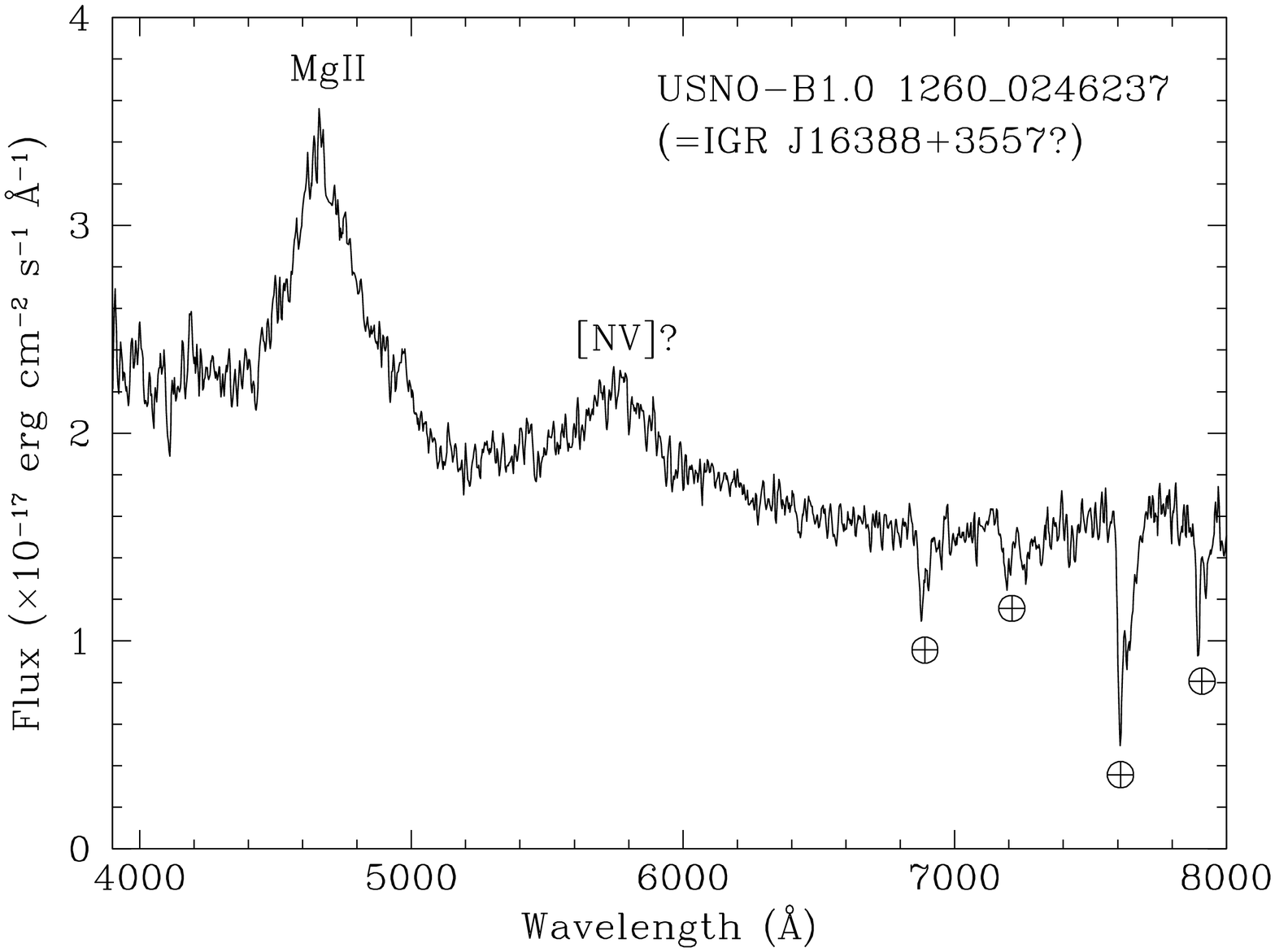,width=9cm,angle=270}}%}
%\begin{center}

%\vspace{1.5cm}
\caption{Spectra (not corrected for the intervening Galactic absorption) 
of the optical counterparts of the 6 high-redshift QSOs belonging to the 
sample of {\it INTEGRAL} sources presented in this paper.
For each spectrum, the main spectral features are labeled. The 
symbol $\oplus$ indicates atmospheric telluric absorption bands.
The TNG spectra have been smoothed using a Gaussian filter with 
$\sigma$ = 3 \AA.}
%\end{center}
\end{figure*}

\begin{figure*}%[th!]
%\begin{center}
%\hspace{.1cm}
%\hspace{-.8cm}
%\centering{
\mbox{\psfig{file=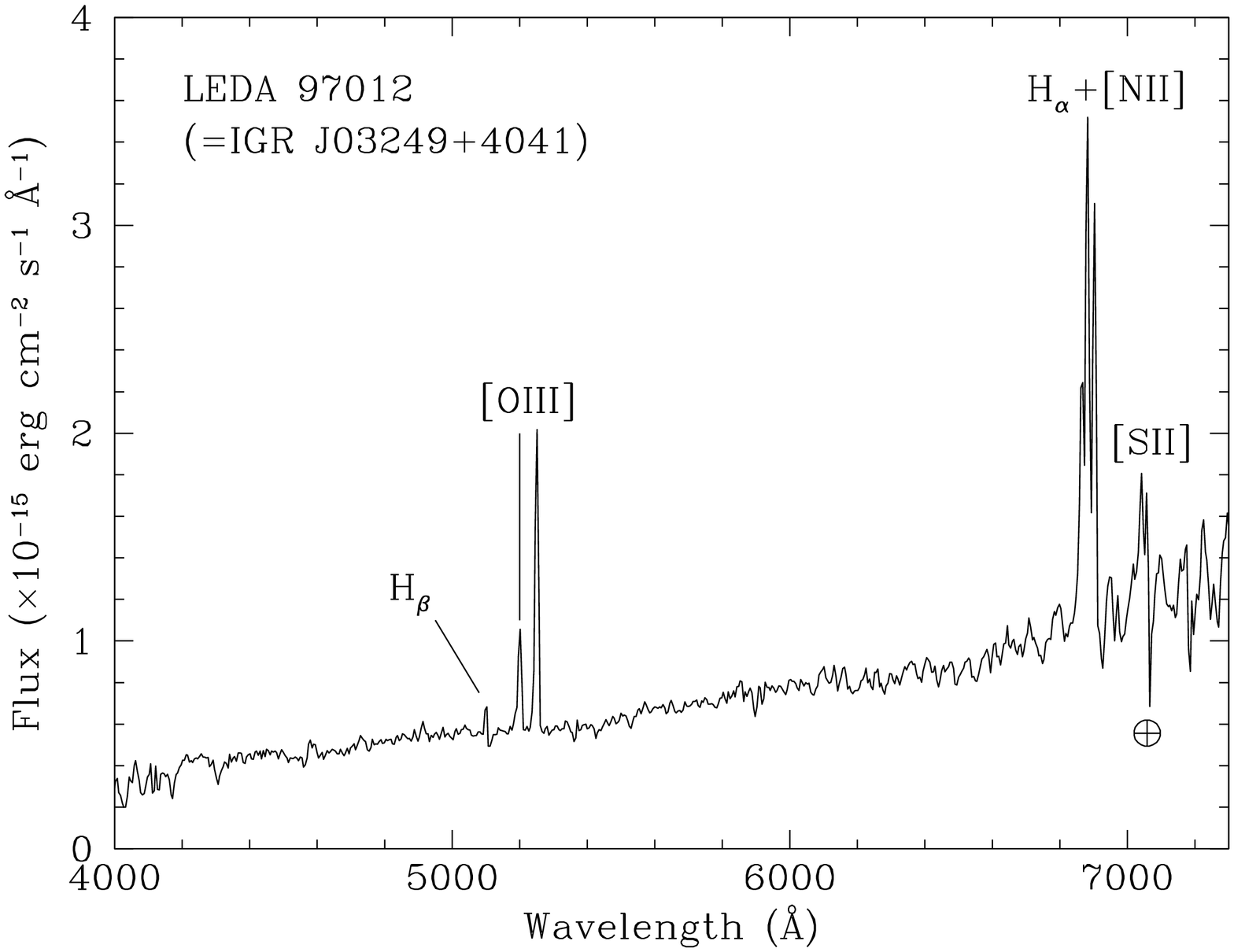,width=9cm,angle=270}}%}
%\hspace{-1.0cm}
%\centering{
\mbox{\psfig{file=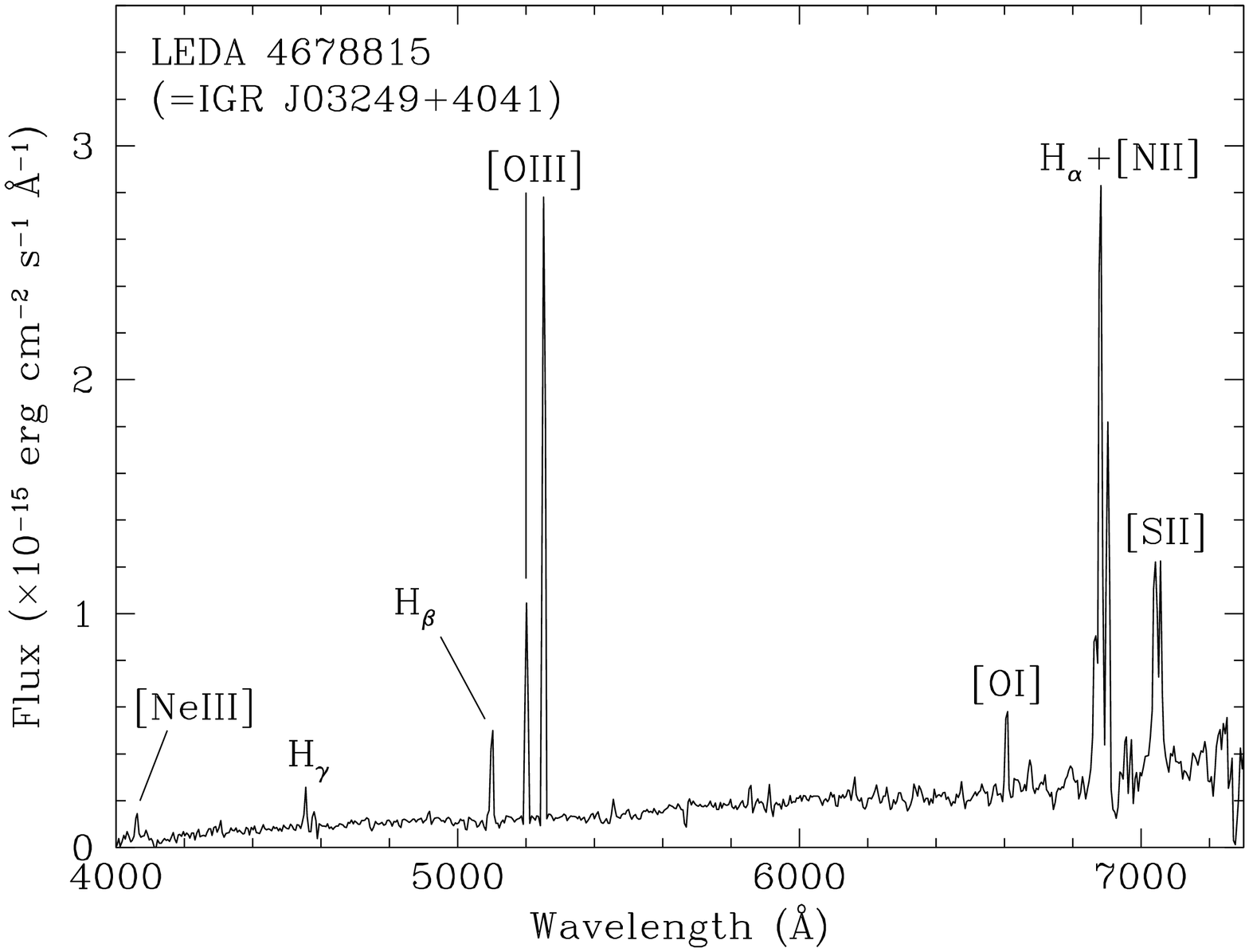,width=9cm,angle=270}}%}

\vspace{-.9cm}
\parbox{9.5cm}{
\psfig{file=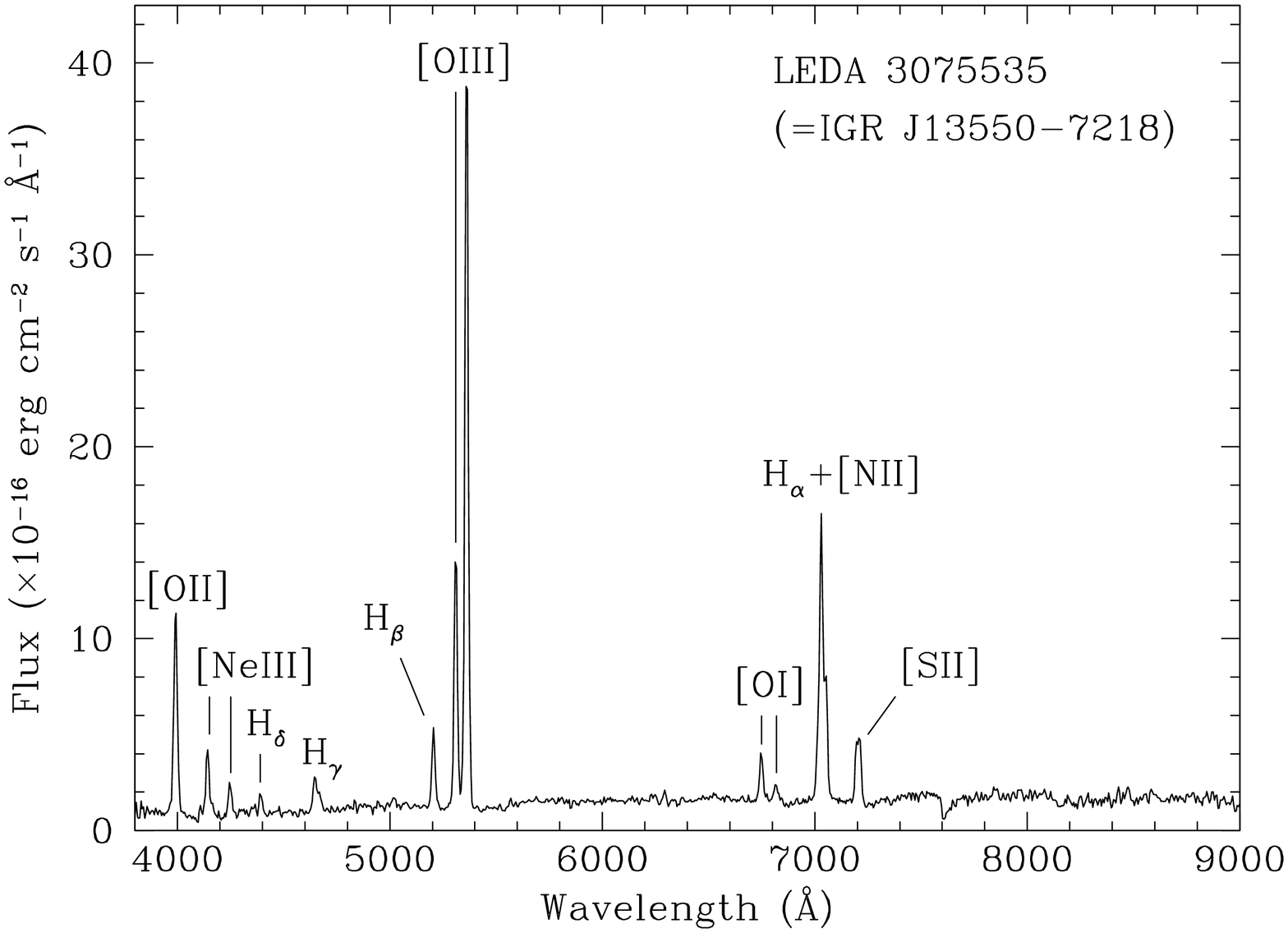,width=9cm,angle=270}
}
\hspace{0.8cm}
\parbox{8cm}{
\vspace{-.5cm}
%\hspace{-.3cm}
\caption{Spectra (not corrected for the intervening Galactic absorption) 
of the optical counterparts of the 3 narrow emission-line AGNs 
belonging to the sample of {\it INTEGRAL} sources presented in this paper.
For each spectrum, the main spectral features are labeled. The 
symbol $\oplus$ indicates atmospheric telluric absorption bands.}}
%\end{center}
\end{figure*}

We found that 16 objects in our sample have optical spectra that allow 
us to classify them as AGNs (see Figs. 4-6): indeed, all of them exhibit
strong, redshifted broad and/or narrow emission lines typical of nuclear 
galactic activity: 14 of them can be classified as Type 1 (broad-line) and 
2 as Type 2 (narrow-line) AGNs; in detail, see Table 2 for the breakdown of 
low-redshift Type 1 AGNs in terms of subclasses. 

Moreover, we stress that 6 of the broad-line AGN identified here lie at 
high redshift ($z >$ 0.5, and in four cases $z >$ 1; see Table 3). In 
particular, our data allow us to state that IGR J12319$-$0749, at redshift 
$z$ = 3.12, is the second most distant and persistently emitting hard X--ray
object detected by {\it INTEGRAL} up to now, after the blazar IGR J22517+2218, 
which lies at redshift $z$ = 3.668 (Bassani et al. 2007 and references 
therein).

The main observed and inferred parameters for each of these two broad 
classes of AGNs are reported in Tables 2-4. In these tables, X--ray 
luminosities were computed from the fluxes reported in Voges et al. 
(1999, 2000), {\it ROSAT} Team (2000), Bird et al. (2007, 2010), Saxton et al. 
(2008), Watson et al. (2009), Cusumano et al. (2010), Krivonos et al. (2010, 
2011), Fiocchi et al. (2010), Landi et al. (2010abc, 2011), Rodriguez et 
al. (2010ab), and Tueller et al. (2010).

For the large majority of the AGNs in our sample (that is, 13 out of 16), 
the redshift value was determined in this work for the first time. The 
redshifts of the remaining 3 cases are consistent with those reported in 
the literature (Lutovinov et al. 2010; Halpern 2011; Krivonos et al. 
2011). We also give here a more accurate classification of the sources IGR 
J06523+5334 and IGR J19491$-$1035, independently identified by Halpern 
(2011) and Krivonos et al. (2011), respectively: both are Seyfert 1.2 
galaxies.

When we examined the optical and X--ray properties
of the AGN sources of our sample in detail, we found the following noteworthy 
issues for some selected cases. 

Concerning narrow emission line galaxies, we confirm the double Seyfert 2 
nature of the interacting galaxy pair LEDA 97012 and LEDA 4678815, 
associated with the {\it INTEGRAL} source IGR J03249+4041, as first 
proposed by Lutovinov et al. (2010). 

Moreover, we found that LEDA 3075535 (the counterpart of IGR 
J13550$-$7218) does not appear to show any substantial reddening local to 
the AGN (see Table 4). This suggests that this source may be a ``naked" 
Seyfert 2 galaxy (e.g., Panessa \& Bassani 2002, Bianchi et al. 2008), 
i.e. an AGN that lacks the broad-line region (BLR): these objects, despite 
first appearances, are not rare among the sources detected with {\it 
INTEGRAL} (see for instance Paper VI). It may be noted that Rodriguez et 
al. (2010b) found a relatively high column density ($N_{\rm H} \sim$ 
2$\times$10$^{23}$ cm$^{-2}$) in the X--ray spectrum of IGR J13550$-$7218: 
this differs from our optical result, although an X--ray spectrum with 
higher S/N is mandatory for a definite comparison between the two 
estimates.

Using the diagnostic $T$ of Bassani et al. (1999), that is, the 
ratio of the measured 2--10 keV X--ray flux to the unabsorbed flux of the 
[O {\sc iii}]$\lambda$5007 forbidden emission line, we can determine
the Compton nature of the Seyfert 2 AGNs in our sample. 
After correcting the [O {\sc iii}]$\lambda$5007 emission line flux 
of LEDA 97012, LEDA 4678815 and LEDA 3075535 for the absorption local to 
the corresponding AGN (see Table 4), we found that the parameter $T$
has values 0.005, 7.0 and 9.6, respectively, indicating that LEDA 97012 
is likely a Compton thick source, whereas the other two AGNs are in the 
Compton thin regime.
We remark that the figures above should be considered as upper limits 
to the actual values of $T$ because the X--ray fluxes that we used refer 
to bands which are wider than the ones for which this method is intended to be 
applied (see Table 4).

These results can be checked following an independent method, that is,
the diagnostic of Malizia et al. (2007), which uses the ratio of the flux 
measurement in the 2--10 keV band to that in the 20--100 keV band. 
For this parameter we found values of 0.007, 0.33, and 0.14, respectively, 
for the three sources considered above: comparing these numbers with those 
of the sample of Malizia et al. (2007, their Fig. 5), we again found that 
the one corresponding to LEDA 97012 falls in the locus in which possible 
Compton thick AGNs are segregated. This result therefore independently 
confirms those obtained with the method of Bassani et al. (1999).
Once again, we caution the reader that, despite the definition 
of the diagnostic of Malizia et al. (2007), the X--ray fluxes available
allow us to actually compute strict upper limits to the above ratio.
We also note that, for galaxies LEDA 97012 and LEDA 4678815, we used in 
the above computations their combined hard X--ray flux detected by 
{\it INTEGRAL} as IGR J03249+4041.

With regards to the galaxy LEDA 97012, it should be noted that Malizia et al. (in 
preparation) found an upper limit for its local hydrogen column density 
($<$1.5$\times$10$^{21}$ cm$^{-2}$) which argues against a Compton Thick
interpretation for this AGN. However, the difference with our results can 
be (at least partially) explained by the relatively low S/N of the 
spectra, and by the source confusion produced by the vicinity to LEDA 
4678815, the other X--ray emitting AGN in this galaxy pair. Thus, a 
clearer analysis on this issue may only be obtained using X--ray 
satellites affording very high angular resolution such as {\it Chandra}.

Regarding the broad-line AGNs of our sample, we point out that according to
Landi et al. (2010a), of the soft X--ray sources which are found within the 
99\% error circle of IGR J05255$-$0711, only one (source \#2; see also Table 
1) has detectable emission above 3 keV, and this object is indeed the one for 
which we report the optical spectrum in Fig. 5, upper left panel. Actually, 
we also acquired spectroscopy for the source labeled \#1 by Landi et al. (2010a) 
with the 1.5m CTIO telescope on 17 October 2010: the data show 
that it is a Galactic star with no peculiarities. Therefore we will not 
discuss this object further.

Finally, we applied the prescriptions of Wu et al. (2004) and Kaspi et al. 
(2000), which use the width and the strength of the broad component of the 
H$_\beta$ emission as a probe of the orbital velocity and the size of the 
BLR (this procedure could not be applied to IGR J03184$-$0014 as no 
broad H$_\beta$ emission component was detected in its spectrum). In the 
cases in which H$_\beta$ falls outside the optical range covered by our 
spectroscopy, we apply the formulae of McLure \& Jarvis (2002) or Vestergaard 
(2002) which use the information conveyed by the Mg {\sc ii} or C {\sc iv} 
broad emissions, respectively.
With these approaches we thus calculated an estimate of the mass of the 
central black hole in 13 of the 14 broad-line AGNs of our sample.

The corresponding black hole masses for these 13 cases are reported in 
Table 5. Here we assumed a null local absorption for all Type 1 AGNs. The 
main sources of error in these mass estimates generally come from the 
determination of the emission flux of the reference emission lines, which 
spans from 5\% to 30\% in our sample (see Tables 2 and 3), and from the 
scatter in the $R_{\rm BLR} - L_{\rm H_\beta}$ scaling relation, which 
introduces typical uncertainties of 0.4--0.5 dex (i.e., logarithmic 
decimals) in the black hole mass estimate (Vestergaard 2004). As a 
whole, we expect the typical error to be about 50\% of the value.

In Table 5 we also report the apparent Eddington ratios for the listed 
AGNs: these were obtained using the observed X--ray fluxes and/or upper 
limits in the 20--100 keV band. We considered this spectral range because, 
looking at Tables 2 and 3, all objects for which we could provide a black 
hole mass estimate have a flux measurement in this band and apparently 
they emit the bulk of their X--ray luminosity in this range. When needed, 
we rescaled the observed luminosities to the above spectral range assuming 
a photon index $\Gamma$ = 1.8.

Although an in-depth study of these AGNs is beyond the scope of this 
paper, and more focused observations on these sources are needed to 
determine their physical properties such as broadband spectral energy 
distribution, Lorentz factor and true luminosity, one can see from Table 5 
that the Eddington ratios can be around unity (or even more) for a number 
of these objects, especially for those at high redshift. This hints at an 
extreme blazar nature for them, as already found for various high-$z$ 
objects detected or identified in hard X--ray surveys (see e.g. Bassani et 
al. 2007; Sambruna et al. 2007; Masetti et al. 2008b; Lanzuisi et al. 
2011).

The properties of the high-redshift QSOs reported in Table 5 thus 
support the suggestion by Ghisellini et al. (2010, 2011) that the most 
powerful blazars have a spectral energy distribution with a high energy 
peak at MeV (or even lower) energies. This implies that the most extreme 
blazars can be found more efficiently in hard X--rays, rather than in the 
$\gamma$--ray band above the GeV threshold.

\begin{table}%[h!]
\caption{BLR gas velocities (in km s$^{-1}$), central black 
hole masses (in units of 10$^7$ $M_\odot$) and apparent
Eddington ratios for 13 broad line AGNs belonging to the sample 
presented in this paper.}
\begin{center}
%\vspace{-.3cm}
\begin{tabular}{lrcc}
\noalign{\smallskip}
\hline
\hline
\noalign{\smallskip}
\multicolumn{1}{c}{Object} & \multicolumn{1}{c}{$v_{\rm BLR}$} & $M_{\rm BH}$ & $L_X/L_{\rm Edd}$\\
\noalign{\smallskip}
\hline
\noalign{\smallskip}

IGR J05255$-$0711       &  3000 &  15 &      $<$4.6 \\
IGR J06523+5334         &  2000 & 3.6 &      $<$3.5 \\
IGR J13466+1921         &  3800 & 4.6 &  $\sim$0.05 \\
IGR J14385+8553         &  3000 & 1.7 &     $<$0.15 \\
IGR J16443+0131         &  3700 & 4.3 &      $<$1.4 \\
1RXS J175252.0$-$053210 &  3900 & 40  &       0.011 \\
IGR J19491$-$1035       &  3400 & 3.2 & $\sim$0.005 \\
IGR J20450+7530         &  2700 & 2.4 &     $<$0.32 \\

\noalign{\smallskip}
\hline
\noalign{\smallskip}

IGR J06073$-$0024       &  5300 &  50 & $<$3.5 \\
IGR J08262+4051         &  4200 &  15 & $<$9.3 \\
IGR J12562+2554         &  8400 &  90 &    1.1 \\
IGR J16388+3557         & 23000 & 270 &   0.15 \\

\noalign{\smallskip}
\hline
\noalign{\smallskip}

IGR J12319$-$0749       &  5600 & 280 & $<$2.5 \\

\noalign{\smallskip}
\hline
\noalign{\smallskip}
\multicolumn{4}{l}{Note: the final uncertainties on the black hole mass} \\
\multicolumn{4}{l}{estimates are about 50\% of their values. The} \\
\multicolumn{4}{l}{velocities were determined using H$_\beta$, Mg {\sc ii},} \\
\multicolumn{4}{l}{or C {\sc iv} emissions (upper, central and lower part} \\
\multicolumn{4}{l}{of the table, respectively), whereas the apparent} \\
\multicolumn{4}{l}{Eddington ratios were computed using the (observed} \\
\multicolumn{4}{l}{or rescaled, see text) 20--100 keV luminosities.} \\
\noalign{\smallskip}
\hline
\hline
\noalign{\smallskip}
\end{tabular}
\end{center}
\end{table}

\subsection{Accreting binaries}

\begin{figure*}%[th!]
%\begin{center}
%\hspace{.1cm}
%\hspace{-.8cm}
%\centering{
\mbox{\psfig{file=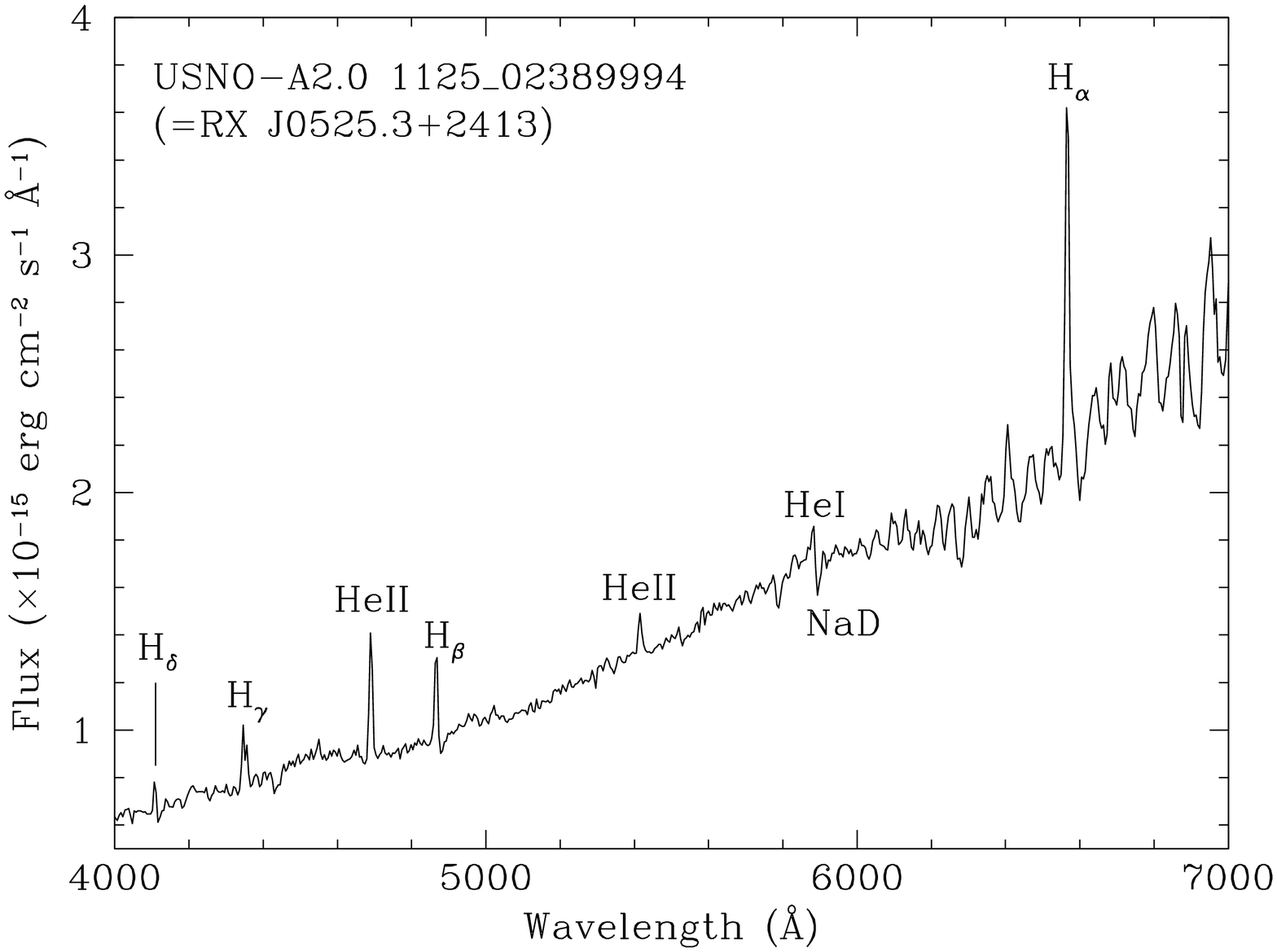,width=9cm,angle=270}}%}
%\hspace{-1.0cm}
%\centering{
\mbox{\psfig{file=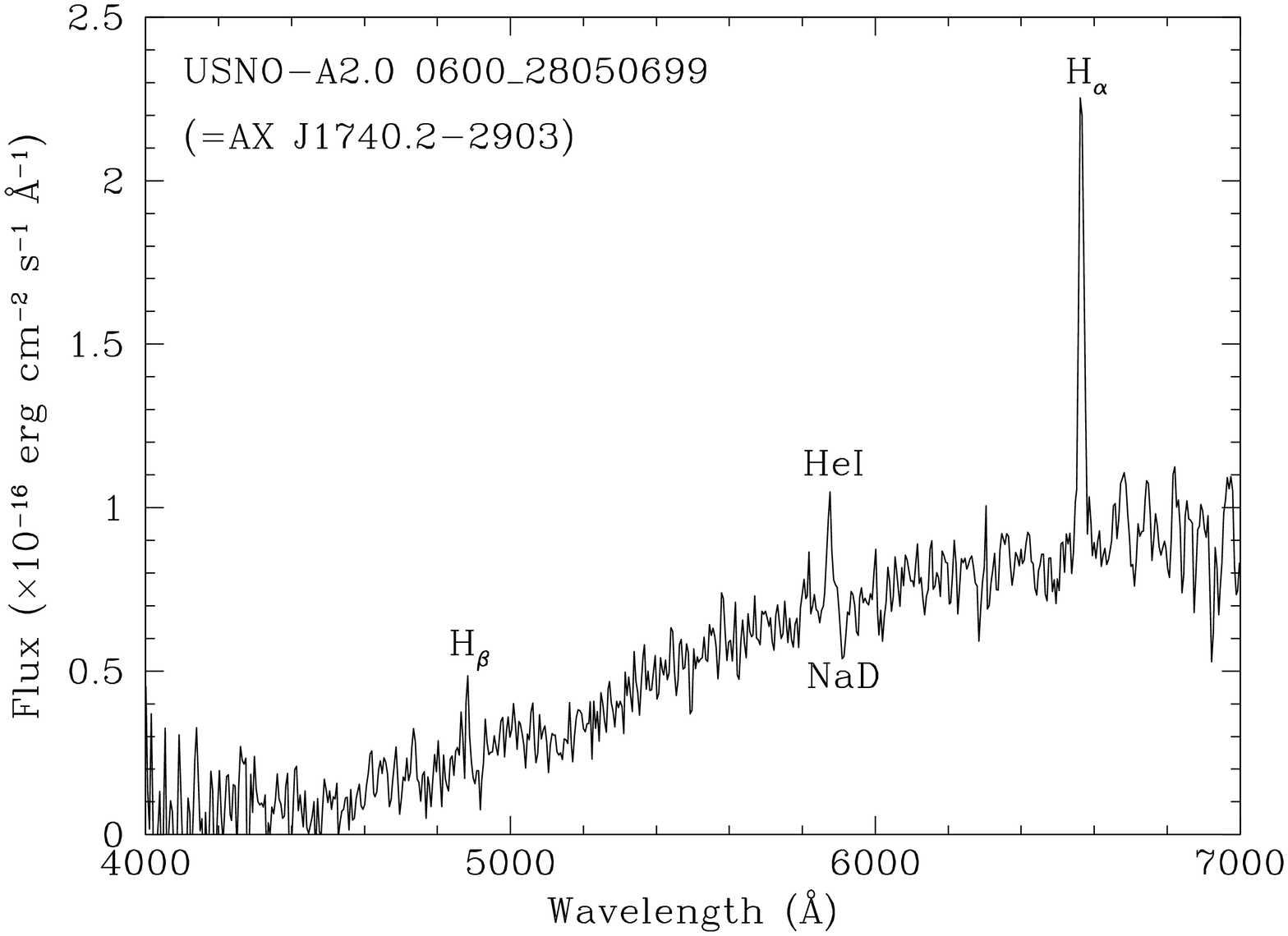,width=9cm,angle=270}}%}

\vspace{-.9cm}
%\hspace{-.8cm}
%\centering{
%\mbox{\hspace{8.9cm}}
\mbox{\psfig{file=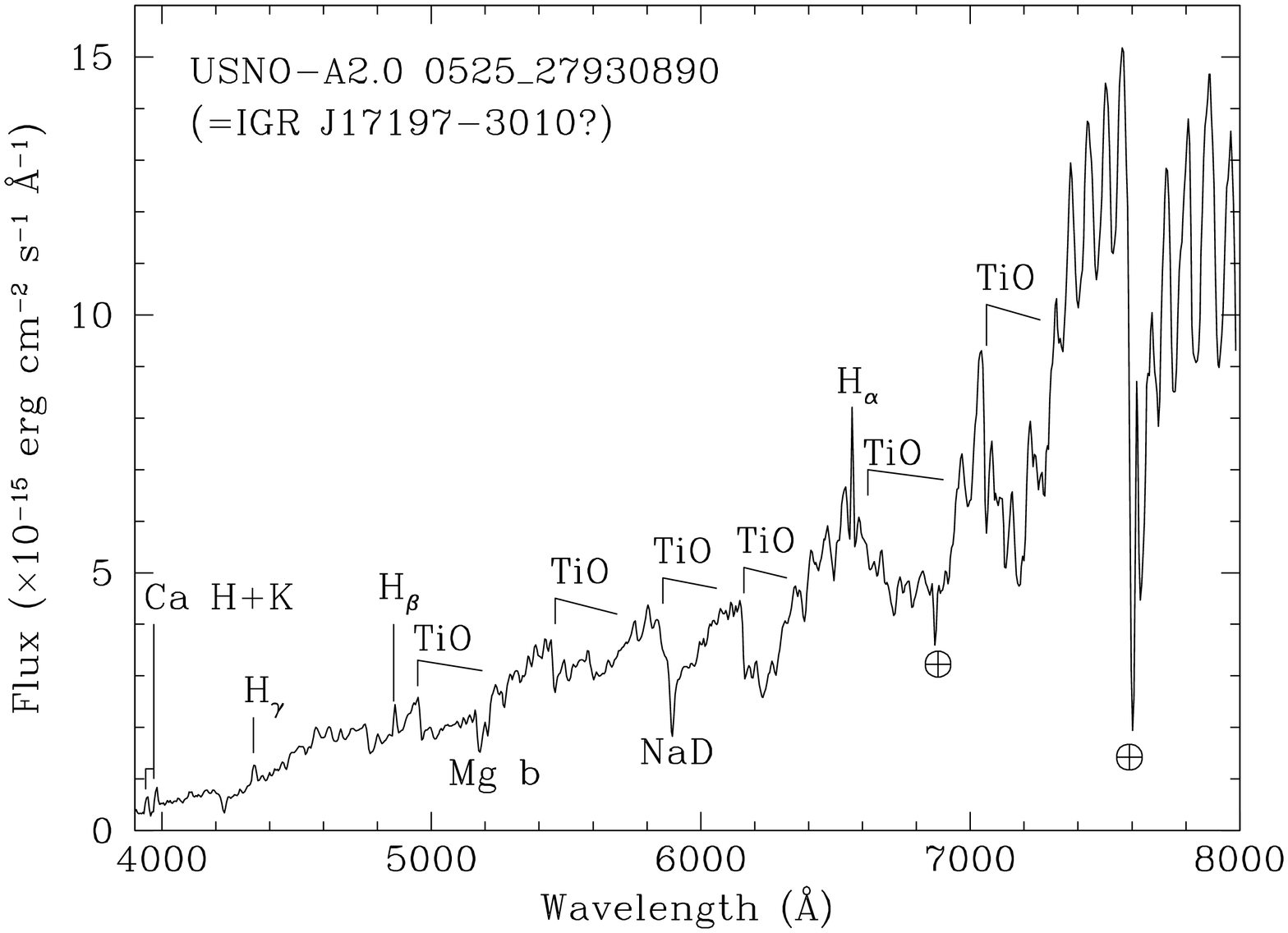,width=9cm,angle=270}}%}
%\hspace{-1cm}
%\centering{
\mbox{\psfig{file=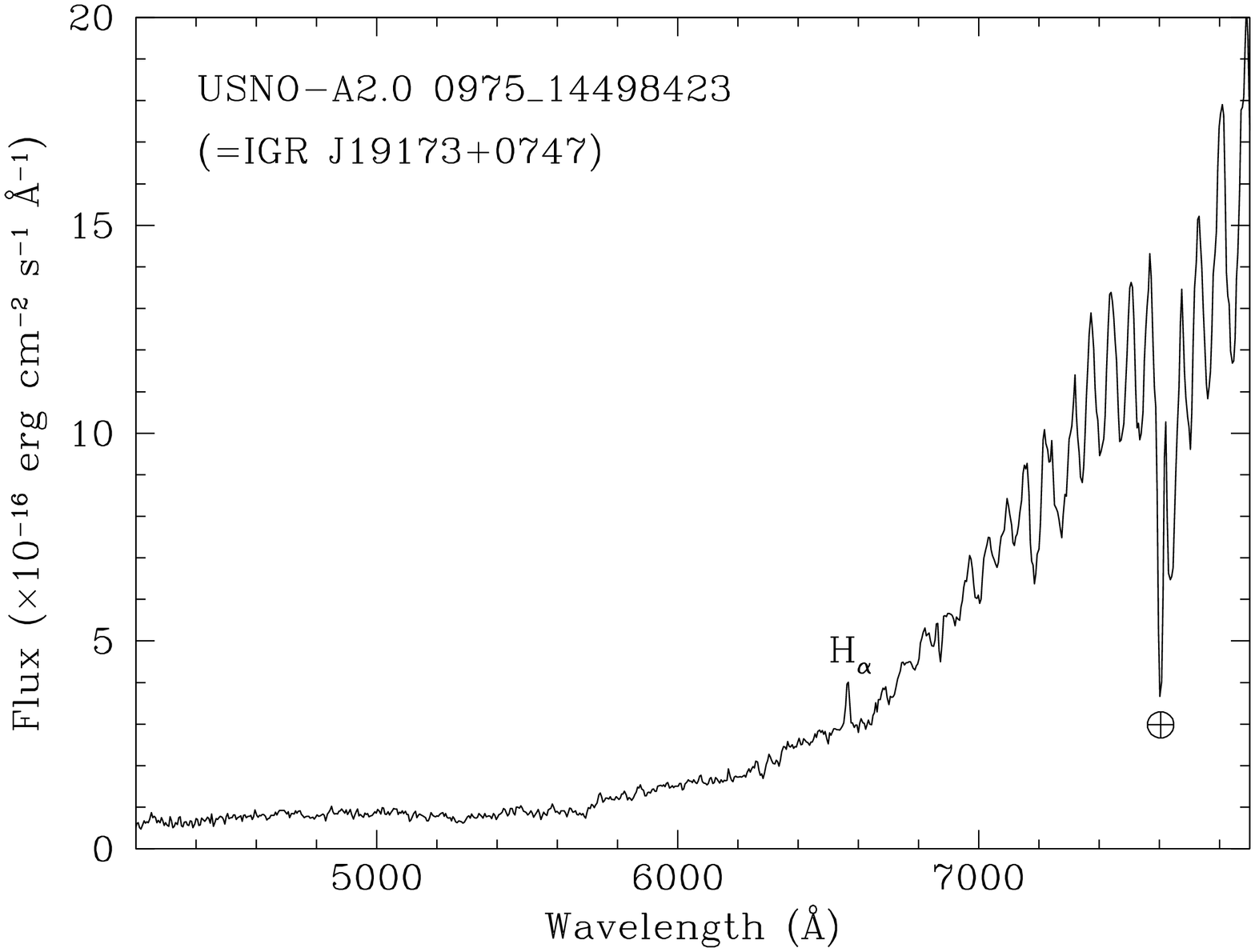,width=9cm,angle=270}}%}

%\vspace{-.5cm}
\caption{Spectra (not corrected for the intervening Galactic absorption) 
of the optical counterparts of the 4 Galactic accreting binaries belonging 
to the sample of {\it INTEGRAL} sources presented in this paper. For each 
spectrum, the main spectral features are labeled. The symbol $\oplus$ 
indicates atmospheric telluric absorption bands.}
%\end{center}
\end{figure*}

Within our sample we identify 4 objects as Galactic accreting binaries.
These are divided into the following subgroups: 2 dwarf nova CVs (RX 
J0525.3+2413 and AX J1740.2$-$2903), one symbiotic binary (IGR J17197$-$3010)
and one HMXB (IGR J19173+0747).

\subsubsection{CVs}

The spectra of the objects identified as CVs (Fig. 7, upper panels) show 
characteristics typical of this class, such as Balmer (up to H$_\delta$ in 
the case of RX J0525.3+2413) and helium lines in emission. All of these 
features lie at redshift $z$ = 0, indicating that these sources are located 
within our Galaxy. Our results confirm the CV nature of sources RX 
J0525.3+2413 and AX J1740.2$-$2903 as first proposed by Torres et al. (2007) 
and Halpern \& Gotthelf (2010), respectively.

The main spectral diagnostic lines of these objects, as well as the main 
astrophysical parameters which can be inferred from the available optical 
and X--ray observational data, are given in Table 6. The X--ray 
luminosities listed in this table for the various objects were computed 
using the fluxes reported in Voges et al. (1999), Sidoli et al. (2001),
Sakano et al. (2002), Saxton et al. (2008), Watson et al. (2009), Krivonos 
et al. (2010), Cusumano et al. (2010), Tueller et al. (2010), Page et al. 
(2007), Farrell et al. (2010) and Malizia et al. (2010).

In particular, source RX J0525.3+2413 shows an He {\sc ii} $\lambda$4686 / 
H$_\beta$ equivalent width (EW) ratio which is larger than 1. As already 
suggested by Torres et al. (2007), this supports the fact that this is 
likely a system hosting a strongly magnetized white dwarf (WD) and 
therefore possibly belonging to the ``polar" subclass of CVs (see e.g. 
Cropper 1990). This classification however needs confirmation through the 
measurement of both the orbital period and the WD spin period, as optical 
spectroscopy is sometimes insufficient to firmly establish the magnetic 
nature of CVs (see e.g. Pretorius 2009 and de Martino et al. 2010).

We also note that the spectral continua of these two CVs appear 
substantially reddened. This may be explained by one (or more) of these 
possibilities: (i) they suffer from local absorption around the source; 
and/or (ii) they are substantially more distant from Earth than derived 
from their dereddened optical magnitudes because the actual magnitude of 
the sources is fainter than the one reported in the USNO-A2.0 catalog. The 
latter issue may be of some importance in the case of AX J1740.2$-$2903, 
for which Farrell et al. (2010) found 4 near-infrared objects within the 
{\it XMM-Newton} error circle: given the relatively low spatial resolution 
of the DSS images, some of them may possibly constitute a blend of sources 
in the USNO-A2.0 catalog, thus giving a counterpart magnitude brighter 
than the actual one for this object. Moreover, the USNO-A2.0 data can 
sometimes present systematic uncertainties of a few tenths of magnitude 
(see Masetti et al. 2003).

It may however be stressed that, for both CVs, the $V$-band absorption 
inferred from the optical spectra is substantially lower than the Galactic 
one along their direction (that is, $\sim$2.7 and $\sim$15 mag, 
respectively: see Schlegel et al. 1998), which may suggest that they do 
not in fact lie at the other side of the Galaxy. As a confirmation of the 
reliability of this result of ours (at least for AX J1740.2$-$2903), we see 
that if we apply the empirical formula of Predehl \& Schmitt (1995) to our 
estimate of the reddening along the source line of sight, we obtain a 
hydrogen column density $N_{\rm H}$ = 6.2$\times$10$^{21}$ cm$^{-2}$, 
which is a value broadly consistent with those obtained by Farrell et al. 
(2010) from the analysis of {\it XMM-Newton} X--ray data.

Again, concerning AX J1740.2$-$2903, we stress that its optical 
spectrum (upper right panel of Fig. 7) allows us to exclude the 
possibility that it is a Symbiotic X--ray binary as proposed by Farrell et 
al. (2010), because in this case the continuum typical of a late-type giant 
star with no emission lines would have clearly been detected (see e.g. 
Masetti et al. 2006e, 2007). We also take this opportunity to remark that 
the arcsec-sized X--ray position of this source reported by Malizia et al. 
(2010) suffers from a typo, and the correct one is reported in Halpern \& 
Gotthelf (2010) and Farrell et al. (2010).

\begin{table*}%[th!]
\caption[]{Synoptic table containing the main results concerning the 2 
CVs (in the upper part) and of the symbiotic star (in the lower part) 
identified in the present sample of {\it INTEGRAL} sources (see Fig. 7).}
\scriptsize
\vspace{-.3cm}
\begin{center}
\begin{tabular}{lcccccccccr}
\noalign{\smallskip}
\hline
\hline
\noalign{\smallskip}
\multicolumn{1}{c}{Object} & \multicolumn{2}{c}{H$_\alpha$} & 
\multicolumn{2}{c}{H$_\beta$} & \multicolumn{2}{c}{He {\sc ii} $\lambda$4686} & 
$R$ & $A_V$ & $d$ & \multicolumn{1}{c}{$L_{\rm X}$} \\
\cline{2-7}
\noalign{\smallskip} 
 & EW & Flux & EW & Flux & EW & Flux & mag & (mag) & (pc) & \\

\noalign{\smallskip}
\hline
\noalign{\smallskip}

RX J0525.3+2413 & 8.0$\pm$0.9 & 18$\pm$2 & 4.0$\pm$0.4 & 3.8$\pm$0.4 & 6.1$\pm$0.6 & 5.4$\pm$0.5 & 
 15.6 & 1.6 & $\sim$110 & 0.10 (0.1--2.4; {\it R}) \\
 & & & & & & & & & & 1.1 (0.2--12; {\it N}) \\
 & & & & & & & & & & 1.5 (0.3--10; {\it X}) \\
 & & & & & & & & & & 1.7 (17--60; {\it I}) \\
 & & & & & & & & & & 3.0 (14--150; {\it B}) \\
 & & & & & & & & & & 2.5 (14--195; {\it B}) \\

& & & & & & & & & & \\ 

AX J1740.2$-$2903 & 31$\pm$3 & 2.6$\pm$0.3 & 15$\pm$5 & 0.3$\pm$0.1 & $<$40 & $<$0.3 & 
 17.4 & 3.46 & $\sim$130 & 0.67 (0.1--2.4; {\it R}) \\
 & & & & & & & & & & 0.80 (0.2--10; {\it N}) \\
 & & & & & & & & & & 0.99 (0.7--10; {\it A}) \\
 & & & & & & & & & & 0.77 (0.2--12; {\it N}) \\
 & & & & & & & & & & 0.65 (2--10; {\it X}) \\
 & & & & & & & & & & 1.4  (20--100; {\it I}) \\

\noalign{\smallskip}
\hline
\noalign{\smallskip}

IGR J17197$-$3010 & 4.5$\pm$0.5 & 27$\pm$3 & 3.8$\pm$0.6 & 7.6$\pm$1.1 & $<$1.8 & $<$3.0 & 
 14.8 & 0.726 & $\la$16600 & $\la$16000 (17--60; {\it I}) \\

\noalign{\smallskip} 
\hline
\noalign{\smallskip} 
\multicolumn{11}{l}{Note: EWs are expressed in \AA, line fluxes are
in units of 10$^{-15}$ erg cm$^{-2}$ s$^{-1}$, whereas X--ray luminosities
are in units of} \\
\multicolumn{11}{l}{10$^{31}$ erg s$^{-1}$ and the reference band (between 
round brackets) is expressed in keV.} \\
\multicolumn{11}{l}{In the last column, the upper-case letter indicates the satellite and/or the 
instrument with which the} \\
\multicolumn{11}{l}{corresponding X--ray flux measurement was obtained (see text).} \\
\noalign{\smallskip} 
\hline
\hline
\noalign{\smallskip} 
\end{tabular} 
\end{center}
\end{table*}

\subsubsection{Symbiotic stars}

As remarked previously, source IGR J17197$-$3010 is identified as a symbiotic star 
given its optical spectral continuum, which shows the typical features of a 
red giant star (namely, broad molecular bands) with superimposed $H_\alpha$ 
and H$_\beta$, H$_\gamma$ and Ca {\sc ii} H+K emissions, again at $z$ = 0 
(Fig. 7, lower left panel).

Using the Bruzual-Persson-Gunn-Stryker (Gunn \& Stryker 1983) and 
Jacoby-Hunter-Christian (Jacoby et al. 1984) spectroscopy atlases, we 
constrain the spectral type of the optical counterpart of this source as 
M1-2 III. We also note a possible excess on the blue side of the optical 
continuum, which is a common feature in symbiotic stars. From this spectral 
information, and assuming colours and absolute $V$ magnitude of a M2 III 
star and considering the measured Balmer line ratio, we obtain a distance 
of $\sim$16.6 kpc for the source, which would place it at the other side 
of the Galaxy.

This suggests that more absorption should occur along the line of sight 
and that this number should rather be used as an (admittedly loose) upper 
limit for the distance to this object. Indeed, if one assumes the total 
Galactic colour excess along the source line of sight, $E(B-V)_{\rm Gal}$ 
= 1.051 mag (Schlegel et al. 1998), we obtain a distance of $\sim$6.3
kpc to the source.

\subsubsection{HMXBs}

We classified the object IGR J19173+0747 within our sample as an X--ray binary 
given the overall optical spectral shape and characteristics of an early-type 
star, which is typical of this class of objects (see Papers II-VIII).
The lower right panel of Fig. 7 indeed shows that the object displays the 
H$_\alpha$ line in emission at redshift 0, superimposed on an intrinsically 
blue continuum modified by intervening reddening: the latter is indicative of 
interstellar dust along its line of sight. This is quite common in 
X--ray binaries detected with {\it INTEGRAL} (e.g., Papers III-VIII) and 
indicates that the object lies relatively far from the Earth. From the optical
magnitudes of IGR J19173+0747 we however find a $V$-band reddening about 2 
mag lower than the Galactic one along its line of sight ($A_V \sim$ 6.1; 
Schlegel et al. 1998).

Table 7 collects the relevant optical spectral information about this 
source, along with the main parameters inferred from the available X--ray 
and optical data. X--ray luminosities in Table 7 were calculated using the 
fluxes in Voges et al. (2000) and Pavan et al. (2011). We obtained
constraints on distance, reddening, spectral type, and X--ray luminosity 
shown in the table by considering the absolute magnitudes of early-type 
stars and by applying the method described in Papers III and IV for the 
classification of this type of X--ray sources. In particular, we exclude the possibility
that the source is of luminosity class I or III otherwise it would be 
placed well beyond the Galactic disk ($>$30 kpc from Earth). The lack of 
further detailed photometric optical information and of higher-resolution 
spectroscopy does not allow us to further refine our spectral 
classification for this object.

In addition, using the empirical formula of Predehl \& Schmitt 
(1995) to obtain the hydrogen column value from our reddening 
determination, we find $N_{\rm H} \approx$ 7$\times$10$^{21}$ cm$^{-2}$: 
this is slightly higher than the upper limit ($<$6$\times$10$^{21}$ 
cm$^{-2}$) obtained by Pavan et al. (2011) for this source; however, given 
the large uncertainties affecting the measurements involved in our 
calculation, we can consider the two values as basically compatible with 
each other.

\begin{table*}%[th!]
\caption[]{Main results for the HMXB IGR J19173+0747 (see 
Fig. 7, lower right panel) identified in the present sample of {\it 
INTEGRAL} sources.}
\hspace{-1.2cm}
\scriptsize
\vspace{-.5cm}
\begin{center}
\begin{tabular}{lccccccccccr}
\noalign{\smallskip}
\hline
\hline
\noalign{\smallskip}
\multicolumn{1}{c}{Object} & \multicolumn{2}{c}{H$_\alpha$} & 
\multicolumn{2}{c}{H$_\beta$} &
\multicolumn{2}{c}{He {\sc ii} $\lambda$4686} &
 $R$ & $A_V$ & $d$ & Spectral & \multicolumn{1}{c}{$L_{\rm X}$} \\
\cline{2-7}
\noalign{\smallskip} 
 & EW & Flux & EW & Flux & EW & Flux & mag & (mag) & (kpc) & type & \\

\noalign{\smallskip}
\hline
\noalign{\smallskip}

IGR J19173+0747 & 6.1$\pm$0.6 & 1.77$\pm$0.18 & $<$10 & $<$0.8 & $<$10 & $<$0.8 &
 16.0 & $\sim$4.2 & $\sim$10 & early B\,V & 0.027 (0.1--2.4; {\it R}) \\
 & & & & & & & & & & & 0.72 (0.5--10; {\it X}) \\
 & & & & & & & & & & & 0.67 (20--40; {\it I}) \\

\noalign{\smallskip} 
\hline
\noalign{\smallskip}
\multicolumn{12}{l}{Note: EWs are expressed in \AA, line fluxes are
in units of 10$^{-15}$ erg cm$^{-2}$ s$^{-1}$, whereas X--ray luminosities
are in units of} \\
\multicolumn{12}{l}{10$^{35}$ erg s$^{-1}$ and the reference band 
(between round brackets) is expressed in keV.} \\
\multicolumn{12}{l}{In the last column, the upper-case letter indicates the satellite 
and/or the instrument with which the} \\
\multicolumn{12}{l}{corresponding X--ray flux measurement was obtained (see text).} \\
\noalign{\smallskip}
\hline
\hline
\end{tabular} 
\end{center} 
\end{table*}

We finally note that this system has no known radio source associated. 
This implies that it is an X--ray binary that does not display collimated 
(jet-like) outflows, that is, it has not the characteristics of a 
microquasar.

\subsection{Other Galactic objects}

\begin{figure*}%[th!]
%\begin{center}
%\hspace{.1cm}
%\hspace{-.8cm}
%\centering{
\mbox{\psfig{file=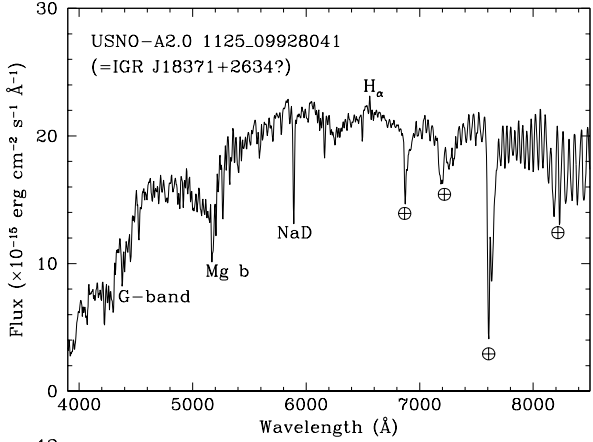,width=9cm,angle=270}}%}
%\hspace{-1.0cm}
%\centering{
\mbox{\psfig{file=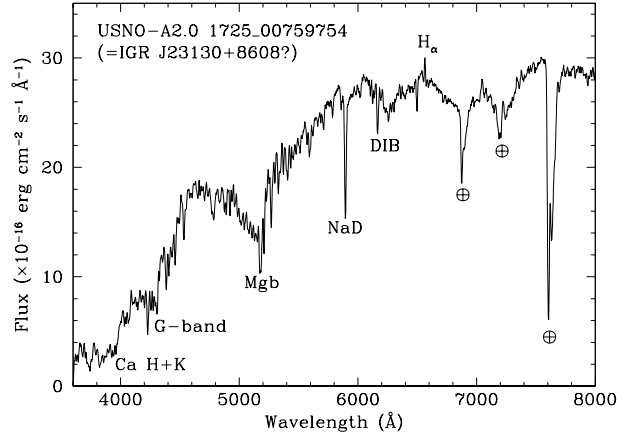,width=9cm,angle=270}}%}

\vspace{-.9cm}
\parbox{9.5cm}{
\psfig{file=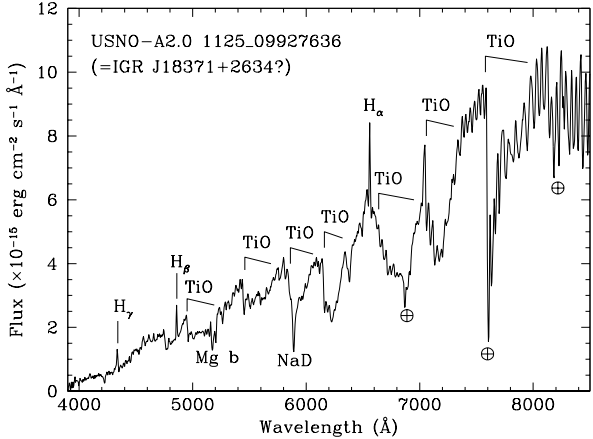,width=9cm,angle=270}
}
\hspace{0.8cm}
\parbox{8cm}{
\vspace{-.5cm}
%\hspace{-.3cm}
\caption{Spectra (not corrected for the intervening Galactic absorption) 
of the optical counterparts of the 2 chromospherically active stars (upper 
panels) belonging to the sample of {\it INTEGRAL} sources presented in 
this paper. 
For completeness we also report (in the lower left panel) the spectrum 
of an X--ray emitting symbiotic binary which lies in the 99\% confidence level 
error circle of source IGR J18371+2634 (see Landi et al. 2011). For each 
spectrum, the main spectral features are labeled. The symbol $\oplus$ 
indicates atmospheric telluric absorption bands.}}
%\end{center}
\end{figure*}

\begin{table*}%[th!]
\caption[]{Synoptic table containing the main results for the
active stars (see Fig. 8) identified in the present sample of {\it 
INTEGRAL} sources.}
\hspace{-1.2cm}
\scriptsize
\vspace{-.5cm}
\begin{center}
\begin{tabular}{lccccr}
\noalign{\smallskip}
\hline
\hline
\noalign{\smallskip}
\multicolumn{1}{c}{Object} & \multicolumn{2}{c}{H$_\alpha$} & 
$R$ & $d$ & \multicolumn{1}{c}{$L_{\rm X}$} \\
\cline{2-3}
\noalign{\smallskip} 
 & EW & Flux & mag & (pc) & \\

\noalign{\smallskip}
\hline
\noalign{\smallskip}

IGR J18371+2634(\#1) & 0.59$\pm$0.12 & 13$\pm$3 & 12.5 & 530 & 0.0057 (0.1--2.4; {\it R}) \\
 & & & & & 0.0044 (2--10; {\it X}) \\
 & & & & & 0.47 (20--40; {\it I}) \\
 & & & & & $<$0.47 (40--100; {\it I}) \\

 & & & & & \\

IGR J23130+8608 & 0.74$\pm$0.07 & 2.1$\pm$0.2 & 14.7 & 1500 & 0.019 (2--10; {\it X}) \\
 & & & & & 3.8 (20--40; {\it I}) \\
 & & & & & $<$4.6 (40--100; {\it I}) \\

\noalign{\smallskip} 
\hline
\noalign{\smallskip}
\multicolumn{6}{l}{Note: EWs are expressed in \AA, line fluxes are
in units of 10$^{-15}$ erg cm$^{-2}$ s$^{-1}$, whereas X--ray luminosities} \\
\multicolumn{6}{l}{are in units of 10$^{33}$ erg s$^{-1}$ and the 
reference band (between round brackets) is expressed in keV.} \\
\multicolumn{6}{l}{In the last column, the upper-case letter indicates the satellite and/or the 
instrument} \\
\multicolumn{6}{l}{with which the corresponding X--ray flux measurement 
was obtained (see text).} \\
\noalign{\smallskip}
\hline
\hline
\end{tabular} 
\end{center} 
\end{table*}

Two of the selected sources, IGR J18371+2634 and IGR J23130+8608, display 
a star-like continuum typical of late-G/early-K type stars, with a faint 
but statistically significant $H_\alpha$ emission at $z$ = 0 (see Fig. 8,
upper panels).
This optical spectral appearance is very similar to that of source
IGR J08023$-$6924, tentatively identified as a RS CVn star (Paper VI;
Rodriguez et al. 2010a). We thus suggest a chromospherically active 
star identification for these two sources as well.

The main observed and inferred parameters for each object are reported in 
Table 8. Luminosities are computed using the X--ray fluxes reported in
Voges et al. (2000), Bird et al. (2007, 2010), Rodriguez et al. (2010a) 
and Landi et al. (2011).

Assuming, as stated at the beginning of Section 4, that these two
objects are similar to the active star II Peg --- which has magnitude
$R \sim$ 6.9 (Monet et al. 2003) and for which we determine a 
distance of 40 pc from Earth based on its measured parallax of 25.06 
milliarcsec (van Leeuwen 2007) --- we obtain the distances reported in 
Table 8. We stress that these figures should be considered as upper limits, 
as no reddening correction on the observed magnitudes of the optical 
counterparts of IGR J18371+2634 and IGR J23130+8608 was attempted.

We point out that the counterpart of IGR J18371+2634 mentioned above is 
source \#1 of Landi et al. (2011); this is the northern source indicated 
in the upper right panel of Fig. 3. We note that we also acquired a 
spectrum of source \#2 (that is, the southern one in the upper right panel 
of Fig. 3) simultaneously with the one of source \#1. This spectrum is 
reported in Fig. 8 (lower left panel): its appearance is typical 
of a symbiotic star, similar to that of the optical counterpart of IGR 
J17197$-$3010 (see Section 3.2). However, given its position with respect 
to the IBIS error circle and the non-detection of X--ray flux above 2 keV 
from it (Landi et al. 2011), we consider this object as an unlikely 
counterpart to the X--ray source IGR J18371+2634, although one cannot 
exclude a sporadic contribution of this symbiotic star to the hard X--ray 
emission detected with {\it INTEGRAL}.

Of course, a deeper multiwavelength followup is needed to confirm (or disprove) 
the two tentative active star identifications above.

\subsection{Statistics}

As is customary for our set of papers, we here present an update of the 
statistics in our previous works by adding the results from the 
sample presented here as well as those reported in Bikmaev et al. (2010), 
Corbet et al. (2010), Pellizza et al. (2011) and Maiorano et al. (2011).

We find that the 204 {\it INTEGRAL} sources identified up until now, on the 
basis of their optical or NIR spectroscopy, are distributed in the
following manner into the main broad classes discussed in Papers VI-VIII
and in this one: 125 (61.2\%) are AGNs, 51 (25.0\%) are X--ray binaries, 
25 (12.3\%) are CVs, and 3 cases (1.5\%) possibly belong to the class of 
active stars.

When we consider the AGN subclasses, we see that 62 sources (i.e., 50\% of 
the AGN identifications) are Seyfert 1 galaxies, 46 (37\%) are narrow-line 
AGNs (including 43 Seyfert 2 galaxies and 3 LINERS), while the QSO, XBONG, 
and BL Lac subclasses are populated by 10, 4, and 3 objects (8\%, 3\%, and 
2\%), respectively (see e.g. Paper VIII and references therein for an 
explanation of the properties of these latter subclasses).

For the Galactic objects, it is found that 40 and 11 objects (78\% and 
22\% of the X--ray binary identifications) are HMXBs and LMXBs, 
respectively; in addition, 25 sources are classified as CVs, most of which 
(20, that is 80\% of them) are definite or likely dwarf novae (mostly of 
magnetic type), and the remaining 5 are symbiotic stars.

When we compare the above figures with those of our previous papers, we
see no substantial changes in the source distribution among the various 
classes, as the main group is made of AGN, followed by X--ray binaries and 
CVs. It is apparent from the identification results presented in this 
paper that the approach used in our program, that is, the use of optical 
spectroscopy to pinpoint the nature of {\it INTEGRAL} sources strongly 
favours the discovery of AGNs against the other classes of objects. 
Moreover, the use of medium-sized and large telescopes (as TNG and GTC, in 
the present case), allows us to study the faint end of the distribution of 
putative optical counterparts of these high-energy sources.

\begin{figure}[h!]
\hspace{-0.5cm}
\psfig{file=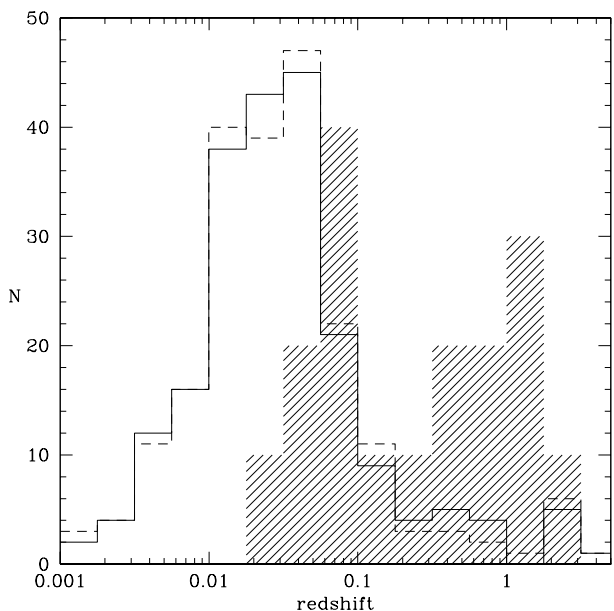,width=9.5cm}
%\vspace{9cm}
\vspace{-0.7cm}
\caption{Logarithmic histograms showing the frequency of known AGN 
redshifts in the surveys of Bird et al. (2010; continuous line) and of
Krivonos et al. (2010; dashed line), together with the logarithmic 
distribution of redshifts of AGNs identified in the present paper (shaded 
histogram; note that for this latter case the actual numbers are 
multiplied by 10 for the sake of comparison).}
\end{figure}

With regards to this issue, we note that the average redshift of the AGNs in the 
sample of the present paper is $<$$z$$>$ = 0.576. This can be compared e.g. 
with the average redshift ($<$$z$$>$ = 0.135) of the known extragalactic 
sources from the 4$^{\rm th}$ IBIS survey (Bird et al. 2010); likewise,
the value of the same parameter in the catalog of Krivonos et al. (2010)
is $<$$z$$>$ = 0.144.

This is graphically apparent in Fig. 9, where the logarithmic histograms 
of the redshift distribution of the known AGNs in the surveys of Bird et 
al. (2010) and Krivonos et al. (2010) are plotted together with that of 
the AGN identified in this paper (the numbers of the latter one have been 
multiplied by 10 in Fig. 9 to ease the comparison by eye). All histograms 
have binning of 0.25 dex. One can see that, while the redshift 
distributions of known AGNs in the two mentioned {\it INTEGRAL} surveys 
are similar to each other and are dominated by nearby objects (with $z <$ 
0.05), that of the present sample has substantially different shape and 
range: indeed, two peaks (one around $z \sim$ 0.1 and the other at $z 
\sim$ 1) are present, and no object with $z <$ 0.02 is found.

To quantify this difference between the redshift distribution of the 
present sample and those of Bird et al. (2010) and Krivonos et al. (2010), 
we applied a Kolmogorov-Smirnov nonparametric test (e.g., Kirkman 1996): 
we found that the probability that the redshift distribution of the 
present sample and those of known AGNs in either of the two above catalogs 
is less than 0.001 in both cases. This confirms that the AGNs of the 
present sample are drawn from a different distribution, of more distant 
objects, with respect to that of known AGNs in both Bird et al. (2010) and 
Krivonos et al. (2010).

The above considerations thus indicate that the deeper {\it INTEGRAL} 
observations available with the latest surveys allow one to explore the 
hard X--ray emitting sources in the far universe, at a mean distance 
$\sim$5 times larger than that of the average of such type of objects 
known up to now.

As a final comment, one may wonder whether there is a physical 
reason behind the bimodal distribution of the redshifts of the 
newly-identified AGNs in the present sample, as shown in Fig. 9. 
Specifically, are we looking at two different physical classes of AGNs in 
the nearby and distant universe, or rather is it just a selection effect 
due to the merging of a wide-field, shallow survey with a series of 
narrow-field, deeper pointed observations? Although the possibility of a 
selection bias of this kind, together with an effect of small-number statistics, 
cannot be excluded {\it a priori}, we think that (as stressed at 
the end of Section 4.1) the high-$z$ sources are indeed different from 
the other ones at lower redshift in the sense that they belong to a 
subclass of extreme blazars which emit the bulk of their power in the 
hard X--ray band; thus, they begin to be detected by wide-field surveys 
when a deep enough level of sensitivity is reached.

\section{Conclusions}

Continuing our ongoing identification program of {\it INTEGRAL} sources by 
means of optical spectroscopy (Papers I-VIII) pursued using various 
telescopes since 2004, we have identified and studied 22 objects having 
unknown or poorly explored nature and belonging to surveys of the hard 
X--ray sky (Bird et al. 2010; Krivonos et al. 2010; Pavan et al. 2011). 
This has been made possible by using 6 telescopes of different sizes (from 
1.5 to 10.4 metres of aperture) and archival data from one spectroscopic 
survey.

We found that the selected sample largely consists of AGNs; indeed, 16 
sources belong to this class: 14 are of Type 1 (6 of them lie at high 
redshift, having $z>$ 0.5) and 2 are of Type 2 (one of which is a system of
two interacting Seyfert 2 galaxies). The other objects belong to our Galaxy:
we found that two of them are (probably magnetic) CVs, one is a HMXB, one
is a symbiotic star, and two are possibly identified as active stars.

These findings confirm the absolute majority of AGNs among our 
identification program. Moreover, the observations presented here allowed 
us to increase by more than a factor of two the number of hard X-ray 
emitting high-redshift AGNs discovered through optical spectroscopy, and 
to discover and identify the source IGR J12319$-$0749 (with redshift $z$ = 
3.12) as the second farthest persistently emitting hard X--ray object 
within the {\it INTEGRAL} catalogs. Our black hole mass estimates for 
these high-$z$ AGNs also support the fact that hard X--ray surveys may 
efficiently spot extremely powerful blazars in the distant Universe. 

This identification program has also been proven to be of great 
importance in the detection of highly absorbed (and even Compton-thick) 
active galaxies, especially in the local Universe. This is fundamental 
for the estimate of their fraction among the AGN population (see Malizia 
et al. 2009 and Burlon et al. 2011); moreover, with the detections in the 
present work, this project may also start helping to resolve the hard 
X--ray background, or at least to evaluate the relative contribution of 
high-$z$ sources detected and identified in this spectral range.

All of this has been possible through the combined use of the recently 
published deep {\it INTEGRAL} surveys and of large telescopes, such as the 
most powerful ones used in this paper.

It is moreover stressed that 173 of the 204 optical and NIR spectroscopic 
identifications considered in Section 4.4 (that is, nearly 85\%) were 
obtained or refined within the framework of our spectroscopic follow-up 
program originally started in 2004 (Papers I-VIII, the present work, and 
references therein).

The results presented here once again demonstrate the high effectiveness 
of the method of catalog cross-correlation and/or follow-up observations 
(especially with soft X--ray satellites capable of providing arcsec-sized 
error boxes, such as {\it Chandra}, {\it XMM-Newton} or {\it Swift}), and 
optical spectroscopy to determine the actual nature of still unidentified 
or poorly known {\it INTEGRAL} sources. We however recall that, for 8 
objects of our present sample, only a putative albeit likely optical 
counterpart could be identified because it lies in the 99\% confidence 
level error circle of the corresponding hard X--ray source, or because 
of the lack of soft X--ray observations providing a definite arcsec-sized 
position at high energies. In the latter case, timely observations with 
soft X--ray satellites affording arcsec-sized localizations are needed to 
confirm the proposed association.

Present and future surveys at optical and NIR wavelengths, such as 
the ongoing Vista Variables in the V\'{i}a L\'actea (VVV: Minniti et al. 
2010; Saito et al. 2011) public NIR survey, will permit us to check and 
identify variable objects in the fields of the objects detected in 
published and forthcoming {\it INTEGRAL} catalogs, thus easing the search 
for putative counterparts for these high-energy sources. Indeed, this 
approach has already been tested to search for the quiescent NIR 
counterpart of hard X--ray transients detected with {\it INTEGRAL}, 
providing encouraging results and allowing one to place constraints on the 
nature of these objects (Rojas et al. 2011; Greiss et al. 2011a,b).

\begin{acknowledgements}

We thank Silvia Galleti for Service Mode observations at the Loiano 
telescope, and Roberto Gualandi and Ivan Bruni for night assistance; Giorgio 
Martorana and Mauro Rebeschini for Service Mode observations at the Asiago 
telescope and Luciano Traverso for coordinating them; Aldo Fiorenzano for 
Service Mode observations at the TNG; Manuel Hern\'andez for Service Mode 
observations at the CTIO telescope and Fred Walter for coordinating them.
NM thanks Sean Farrell for useful discussions.
We also thank the anonymous referee for useful remarks which helped us
to improve the quality of this paper. 
This research has made use of the ASI Science Data Center Multimission 
Archive; it also used the NASA Astrophysics Data System Abstract Service, 
the NASA/IPAC Extragalactic Database (NED), and the NASA/IPAC Infrared 
Science Archive, which are operated by the Jet Propulsion Laboratory, 
California Institute of Technology, under contract with the National 
Aeronautics and Space Administration.
This publication made use of data products from the Two Micron All 
Sky Survey (2MASS), which is a joint project of the University of 
Massachusetts and the Infrared Processing and Analysis Center/California 
Institute of Technology, funded by the National Aeronautics and Space 
Administration and the National Science Foundation.
This research has also made use of data extracted from the Six-degree 
Field Galaxy Survey archive; it has also made use of the SIMBAD and VIZIER
database operated at CDS, Strasbourg, France, and of the HyperLeda catalog 
operated at the Observatoire de Lyon, France.
NM acknowledges ASI-INAF financial support via grant No. I/009/10/0 and
thanks the Departamento de Astronom\'{i}a y Astrof\'{i}sica of the
Pontificia Universidad Cat\'olica de Chile in Santiago for the warm
hospitality during the preparation of this paper.
PP has been supported by the ASI-INTEGRAL grant No. I/008/07.
RL is supported by the ASI-INAF agreement No. I/033/10/0.
LM is supported by the University of Padua through grant No. CPS0204. 
VC is supported by the CONACyT research grants 54480 and 15149 (M\'exico).
DM is supported by the Basal CATA PFB 06/09, and FONDAP Center for 
Astrophysics grant No. 15010003. 
GG is supported by Fondecyt grant No. 1085267.
\end{acknowledgements}

\end{document}